\documentclass[11pt,a4paper]{article}
\usepackage[T1]{fontenc}
\usepackage[margin=1in]{geometry}
\usepackage{amsmath, amssymb}
\usepackage{hyperref}
\usepackage{mathtools}
\usepackage{amsthm}
\usepackage{enumerate}
\usepackage[shortlabels]{enumitem}
\usepackage{xspace}
\usepackage{mathtools, bm}
\usepackage{tikz}
\usetikzlibrary{snakes, positioning, arrows.meta}
\usepackage{algorithm}
\usepackage[noend]{algpseudocode}
\usepackage[capitalise]{cleveref}

\usepackage{xcolor}

\usepackage[colorinlistoftodos]{todonotes}

\newtheorem{theorem}{Theorem}
\newtheorem{definition}[theorem]{Definition}
\newtheorem{lemma}[theorem]{Lemma}
\newtheorem{corollary}[theorem]{Corollary}
\newtheorem{claim}[theorem]{Claim}
\newtheorem{proposition}[theorem]{Proposition}

\newtheorem{conjecture}[theorem]{Conjecture}

\newtheorem{remark}[theorem]{Remark}
\newtheorem{problem}[theorem]{Open Question}

\numberwithin{theorem}{section}

\newcommand{\set}[1]{\{#1\}}

\newcommand{\class}[1]{\textsf{#1}}
\newcommand{\sset}{\subseteq}
\newcommand{\N}{\mathbb{N}}
\newcommand{\NN}{\mathbb{N}}
\newcommand{\Z}{\mathbb{Z}}
\newcommand{\ZZ}{\mathbb{Z}}
\newcommand{\QQ}{\mathbb{Q}}

\newcommand{\length}[1]{\mathit{len}(#1)}

\newcommand{\Aa}{\mathcal{A}}
\newcommand{\parikh}[1]{\psi(#1)}
\newcommand{\lang}[1]{\mathcal{L}(#1)}

\newcommand{\poly}[1]{\textsf{poly}(#1)}
\newcommand{\Oh}{\mathcal{O}}
\newcommand{\abs}[1]{\lvert #1 \rvert}
\newcommand{\norm}[1]{\lVert#1\rVert}
\renewcommand{\vec}[1]{{\bf #1}}
\newcommand{\inverseAckermann}[1]{\alpha(#1)}
\newcommand{\iackermann}[1]{\mathit{Ack}^{-1}(#1)}
\newcommand{\ackermann}[1]{\mathit{Ack}(#1)}

\newcommand{\size}[1]{\mathit{size}(#1)}
\newcommand{\bitsize}[1]{\mathit{bitsize}(#1)}

\newcommand{\geqz}{\geq\!0}
\newcommand{\eqz}{=\!0}

\newcommand{\configuration}[3]{#1(\vec{#2}#3)} %some configuration p?(v?)
\newcommand{\config}[2]{\configuration{#1}{#2}{}} %a configuration p(v)
\newcommand{\iconfig}[3]{\configuration{#1_{#3}}{#2}{_{#3}}} %a configuration p(v)
\newcommand{\run}[3]{#1\xrightarrow{#2}#3}
\newcommand{\zrun}[3]{#1\xrightarrow[\ZZ]{#2}#3}
\newcommand{\Run}[4]{#1\xrightarrow{#2}_{#3}#4}
\newcommand{\eff}[1]{\mathit{eff}(#1)}
\newcommand{\step}[2]{#1\rightarrow#2}

\newcommand{\Pp}{\mathcal{P}}
\newcommand{\Uu}{\mathcal{U}}
\newcommand{\Vv}{\mathcal{V}}

\newcommand{\Ss}{\mathcal{S}}
\newcommand{\Tt}{\mathcal{T}}
\newcommand{\Nn}{\mathcal{N}}
\newcommand{\Zz}{\mathcal{Z}}

\newcommand{\var}[1]{\mathsf{#1}}
\newcommand{\LOOP}{\texttt{LOOP: }}

\newcommand{\inc}[2]{$\var{#1}$\,\texttt{+=}\,$#2$}
\newcommand{\increment}[3]{$\var{#1}_{#2}$\,\texttt{+=}\,$#3$}

\newcommand{\dec}[2]{$\var{#1}$\,\texttt{-=}\,$#2$}
\newcommand{\decrement}[3]{$\var{#1}_{#2}$\,\texttt{-=}\,$#3$}

\newcommand{\zt}[1]{\texttt{zero-test(}$\var{#1}$\texttt{)}}
\newcommand{\test}[1]{$\var{#1}\,\texttt{=?}\,0$}
\newcommand{\assert}[2]{$\var{#1}$\,\texttt{=}\,$#2$}
\newcommand{\assertt}[3]{$\var{#1}_{#2}$\,\texttt{=}\,$#3$}

\newcommand{\ugadget}[1]{\texttt{update(}$#1$\texttt{)}}

\newcommand{\nondiv}[1]{\textup{\texttt{non-divisibility(}}$#1$\textup{\texttt{)}}}

\newcommand{\satprogram}[1]{\textup{\texttt{SAT(}}$#1$\textup{\texttt{)}}}

\algblockdefx[ffor]{Ffor}{EndFfor}%
	[3]{\texttt{FOR} $#1$\,\texttt{=}\,$#2$ \texttt{TO} $#3$\,:}%
	{\texttt{END FOR}}
\algnotext{EndFfor}
\algblockdefx[fforeach]{Fforeach}{EndFforeach}%
	[1]{\texttt{FOR EACH} #1:}
	{\texttt{END FOR}}
\algnotext{EndFforeach}
\algblockdefx[biglet]{BigLet}{EndBigLet}%
	[1]{\texttt{LET} #1:}
	{\texttt{END LET}}
\algnotext{EndBigLet}

\renewcommand{\phi}{\varphi}
\renewcommand{\iota}{z} 

\newcommand{\trans}[1]{\stackrel{#1}{\longrightarrow}}

\title{The Tractability Border of Reachability in\\Simple Vector Addition Systems with States}
\author{   
    \bgroup
    \begin{tabular}{@{\hspace{0.2in}}c@{\hspace{0.2in}} @{\hspace{0.2in}}c@{\hspace{0.2in}} @{\hspace{0.2in}}c@{\hspace{0.2in}}} 
        Dmitry Chistikov\footnote{Centre for Discrete Mathematics and its Applications (DIMAP) and Department of Computer Science, University of Warwick, Coventry, UK,
        \texttt{d.chistikov@warwick.ac.uk}. During the work on this paper, DC was a visitor to the Max Planck Institute for Software Systems (MPI-SWS), Kaiserslautern and Saarbr\"ucken, Germany, a visiting fellow at St~Catherine's College and a visitor to the Department of Computer Science at the University of Oxford, UK. Supported in part by the Engineering and Physical Sciences Research Council [EP/X03027X/1].}
        & Wojciech Czerwi\'{n}ski\footnote{University of Warsaw, Poland,
        \texttt{wczerwin@mimuw.edu.pl}. Supported by the ERC grant INFSYS, agreement no. 950398.} 
        & Filip Mazowiecki\footnote{University of Warsaw, Poland,
        \texttt{f.mazowiecki@mimuw.edu.pl}. Supported by Polish National Science Centre SONATA BIS-12 grant number 2022/46/E/ST6/00230.}
        \\ {\L}ukasz Orlikowski\footnote{University of Warsaw, Poland,
        \texttt{lo418363@students.mimuw.edu.pl}. Supported by the ERC grant INFSYS, agreement no. 950398.}
        & Henry Sinclair-Banks\footnote{University of Warsaw, Poland, \texttt{hsb@mimuw.edu.pl}, \url{http://henry.sinclair-banks.com}. Supported by the ERC grant INFSYS, agreement no. 950398. Centre for Discrete Mathematics and its Applications (DIMAP) and Department of Computer Science, University of Warwick, Coventry, UK, \texttt{h.sinclair-banks@warwick.ac.uk}. Supported by EPSRC Standard Research Studentship (DTP), grant number EP/T51794X/1.}
        & Karol W\k{e}grzycki\footnote{Saarland University and MPI for
        Informatics, Saarbr\"ucken, Germany, \texttt{wegrzycki@cs.uni-saarland.de}. Supported by the ERC grant TIPEA, agreement no. 850979.}
    \end{tabular}
    \egroup
}

\date{}

\begin{document}

\maketitle
\renewcommand{\thefootnote}{\roman{footnote}}

\thispagestyle{empty}
%\vspace*{-2ex} % fits title page into a single page

\begin{abstract}
    Vector Addition Systems with States (VASS), equivalent to Petri nets, are a well-established model of concurrency. 
A $d$-VASS can be seen as directed graph whose edges are labelled by $d$-dimensional integer vectors.
While following a path, the values of $d$ nonnegative integer counters are updated according to the integer labels.
The central algorithmic challenge in VASS is the reachability problem: is there a run from a given starting node and counter values to a given target node and counter values?
When the input is encoded in binary, reachability is computationally intractable: even in dimension one, it is \class{NP}-hard.

In this paper, we comprehensively characterise the tractability border of the problem when the input is encoded in unary.
For our main result, we prove that reachability is \class{NP}-hard in unary encoded 3-VASS, even when structure is heavily restricted to be a simple linear path scheme.
This improves upon a recent result of Czerwi\'{n}ski and Orlikowski (2022), in both the number of counters and expressiveness of the considered model, as well as answers open questions of Englert, Lazi\'{c}, and Totzke (2016) and Leroux (2021). 

The underlying graph structure of a simple linear path scheme (SLPS) is just a path with self-loops at each node.
We also study the exceedingly weak model of computation that is SPLS with counter updates in $\set{-1, 0, +1}$.
Here, we show that reachability is \class{NP}-hard when the dimension is bounded by $\Oh(\inverseAckermann{k})$, where $\alpha$ is the inverse Ackermann function and $k$ bounds the size of the SLPS.

We complement our result by presenting a polynomial-time algorithm that decides reachability in 2-SLPS when the initial and target configurations are specified in binary. 
To achieve this, we show that reachability in such instances is well-structured: all loops, except perhaps for a constant number, are taken either polynomially many times or almost maximally. 
This extends the main result of Englert, Lazić, and Totzke (2016) who showed the problem is in \class{NL} when the initial and target configurations are specified in unary.

\end{abstract}

\clearpage
\setcounter{page}{1}

% !TEX root = ../main.tex
\section{Introduction}
\label{sec:introduction}

\newtheorem*{Thm1}{Theorem~\ref{thm:3-lps-hardness}}
\newtheorem*{Thm2}{Theorem~\ref{thm:unitary-hardness}}
\newtheorem*{Thm3}{Theorem~\ref{thm:mainalg}}

The reachability problem has always been of core interest in computer science.
The halting problem, which is undecidable, is an instance of reachability from some starting configuration in a Turing machine to some accepting configuration.
Based on a seminal (now folklore) result by Minsky~\cite{Minsky67}, reachability is undecidable in a much simpler model: two-counter machines.
This model can be seen as a finite automaton with two nonnegative valued integer counters in which transitions can increment, decrement, or test the counter for zero (i.e.\ the transition can only be taken if a specified counter has zero value).
It is natural to consider a model without zero tests: Vector Addition Systems with States (VASS).
VASS are a fundamental model of computation that has been studied since the 1970s~\cite{KarpM69,Greibach78a,HopcroftP79}.
They are equivalent to Petri nets, a popular and simple model of concurrency with
vast applications~\cite{ZurawskiZ94,Aalst15,BurnsKY00,BaldanCMS10,BouajjaniE13}.

Concretely, a $d$-dimensional VASS ($d$-VASS) is a directed graph with edges labelled by vectors from $\Z^d$.
Vertices of this graph are called states and edges are called transitions. 
A configuration of a VASS is a state $p$ and a vector $\vec{v} \in \NN^d$, denoted $\config{p}{v}$.
Notice that vectors in the configurations are always nonnegative, but this is not necessarily true for the transition labels.
This is because vectors labelling the transitions correspond to updates applied to counters which can only take nonnegative values.
Precisely, suppose there is a transition from $p$ to $q$ labelled by $\vec{u} \in \ZZ^d$, then one can take this transition from the configuration $\config{p}{v}$ to reach the configuration $\config{q}{v+u}$ so long as $\vec{v} + \vec{u} \in \NN^d$.
The reachability problem asks for a given VASS, initial configuration, and target configuration, whether there is a sequence of transitions from the initial to the target configuration.
The dimension of the VASS can be fixed or variable; when it is fixed, it is specified (i.e.\ reachability in $d$-VASS vis-à-vis in VASS).

The complexity of reachability in VASS has a great history and literature.
Early on, the problem was proved to be \class{EXPSPACE}-hard  by Lipton in 1976~\cite{Lipton76} and decidable by Mayr in 1981~\cite{Mayr81}.
More recently, the problem has amassed plenty of attention and its complexity was finally settled: reachability in VASS is \class{Ackermann}-complete~\cite{LerouxS15, LerouxS19, CzerwinskiLLLM21, CzerwinskiLO21, CzerwinskiO21, LerouxSTOC21}.
Finer improvements to the upper and lower bounds are an area of active research~\cite{Lasota22, CzerwinskiJLLO23}.

Reachability in fixed dimension VASS has also been studied earnestly.
When the dimension is fixed, the encoding of the VASS matters.
Integer vectors that are transition labels can be encoded in binary (``\emph{binary VASS}'') or in unary (``\emph{unary VASS}'').
Reachability is \class{NL}-complete for both unary 1-VASS~\cite{ValiantP75} and unary 2-VASS~\cite{EnglertLT16}.
On the other hand, reachability is \class{NP}-complete for binary 1-VASS (see~\cite[Theorem~3.5]{RosierY86}) and~\class{PSPACE}-complete for binary 2-VASS~\cite{BlondinFGHM15}.
Determining the complexity of reachability in 3-VASS, with either unary or binary encoding, is an intriguing and long-standing problem
(see, e.g., \cite{BlondinFGHM15,BlondinEFGHLMT21,CzerwinskiO21,Leroux21}) that can be traced back to \cite{HopcroftP79}.
In this paper, we present the first non-trivial lower bound for reachability in 3-VASS: we show that reachability in unary 3-VASS is \class{NP}-hard.
In fact, our lower bound holds for an extremely restricted kind of VASS that we will detail shortly.

Due to both its wide applicability and high complexity, reachability has been studied for numerous variants and subclasses of VASS (for a survey on the matter, see~\cite{SchmitzSurvey16}).
Flat VASS are a well-studied subclass of VASS that do not contain any nested cycles~\cite{ComonC00, LerouxS04,LerouxS05, Leroux21}; they characterise semilinear reachability relations~\cite{Leroux13}.
Each state in a flat VASS belongs to at most one cycle; the underlying graph structure can be seen as a directed acyclic graph with disjoint cycles attached to each of the nodes.
Our paper focuses on reachability in flat VASS.
Regardless of the dimension, and even when encoded in binary, reachability in flat VASS is in \class{NP}~\cite{FribourgOCONCUR97, CzerwinskiLLLM20}.
Already in dimension one, reachability in binary flat 1-VASS is \class{NP}-hard~\cite{HaaseKOW09}.
The complexity of reachability in unary flat $d$-VASS is much more delicate.
The only known upper bounds below \class{NP} are for dimensions $d=1$ and $d=2$ in which reachability is in \class{NL} (this follows from their not necessarily flat counterparts~\cite{ValiantP75, EnglertLT16}).
Negatively resolving questions in~\cite{BlondinFGHM15, EnglertLT16}, reachability in unary flat VASS was shown to be \class{NP}-hard for dimension $d=7$~\cite{CzerwinskiLLLM20}, subsequently improved to $d=5$~\cite{DubiakThesis20} and $d=4$~\cite{CzerwinskiO21}.
Dimension $d=3$ was again left open.
We close this line of research, showing that reachability in unary flat 3-VASS is \class{NP}-hard.
Perhaps surprisingly, we prove hardness by first restricting our systems even further, as follows.

Linear Path Schemes (LPS) are flat VASS that, informally speaking, do not have branching.
More precisely, an LPS is a flat VASS whose underlying graph structure is a simple path with disjoint cycles attached to the nodes.
In other words, LPS have the form $\alpha_1 \, \beta_1^{\,*} \, \alpha_2 \, \cdots \, \alpha_{n-1} \, \beta_n^{\,*} \, \alpha_n$, where each $\alpha_i$ and $\beta_i$ are sequences of transitions and the asterisk acts as a Kleene star in regular expressions (unbounded iteration).
LPS are a basic and well-known subclass of VASS; a flat VASS can be seen as a union of (possibly exponentially many) LPS~\cite{LerouxS04,BlondinEFGHLMT21}.
To go one step further, we consider an exceedingly weak model of computation: LPS with cycles of length at most one; these innocuous VASS are known as Simple Linear Path Schemes (SLPS)~\cite{EnglertLT16}.
Prior to our work, it was not known whether reachability in unary $d$-LPS was \class{NP}-hard, for any fixed dimension $d$ (and regardless of whether the LPS was simple or not).

\subsection*{Overview of Results: Contributions and Techniques}

Our first main contribution is the following theorem, on (unary flat) 3-VASS.
This result answers open questions on reachability in 3-LPS~{\cite[Section 6]{EnglertLT16}} and~{\cite[Section 5]{Leroux21}}.

\begin{theorem}
	Reachability in unary 3-SLPS is \class{NP}-complete.
	\label{thm:3-lps-hardness}
\end{theorem}

To show \class{NP}-hardness, we employ a combination of newly developed and known techniques.
Given an instance of 3-CNF SAT $\phi(x_1, \ldots, x_n)$, we associate, to the $i$-th Boolean variable $x_i$, the $i$-th prime number $p_i$.
An assignment $(a_1, \ldots, a_n) \in \set{0, 1}^n$ corresponds to an integer $v \in \NN$ so that $x_i$ is true if $v \equiv 1 \bmod p_i$ and $x_i$ is false if $v \equiv 0 \bmod p_i$.
By the Chinese remainder theorem, for every $(a_1, \ldots, a_n) \in \set{0, 1}^n$ there is a $v \in \NN$ that \emph{encodes} it: $v \equiv a_i \bmod p_i$ for each $i$.
Some $v \in \NN$ do not correspond to a valid assignment; e.g., if $v \equiv 2 \bmod p_i$.
Whilst this encoding is well known, non-deterministic branching is not available in LPS (and SLPS).
So, to ensure validity, we cannot directly ``say'' that $v \equiv 1 \bmod p_i$ or $v \equiv 0 \bmod p_i$ for every~$i$.
Because of this, our construction fits everything into a conjunction (logical \textsc{and}) of checks, which together ensure that a ``guessed'' number $v \in \NN$ encodes to a valid assignment, and moreover this assignment satisfies~$\phi$.

As it turns out, it suffices to test that $v \in \NN$ is not divisible by an appropriately chosen small integer $q$,
for polynomially many values of $q$.
This sequence of tests is expressed as a unary 1-SLPS extended with \emph{non-divisibility assertions} (\cref{lem:sat-encoding}).
Such assertions can, in turn, be realised by a novel family of unary 2-SLPS extended with zero tests (one 2-SLPS for each non-divisor $q \geq 2$).
The formal description is in~\Cref{fig:unary-non-div}, and~\Cref{clm:unary-non-div} asserts the correctness.
By combining the 3-CNF SAT implementation with this family, we now have a unary 2-SLPS with zero tests for which reachability coincides with the satisfiability of $\phi$.
Importantly, the zero tests in the 2-SLPS are heavily restricted: they only lie on transitions between the self-loops, so there are at most polynomially many of them.

Now these zero tests also need to be implemented. To this end, we review the ``controlling counter'' technique, originally developed in~{\cite[Section 4]{CzerwinskiO21}}.
Intuitively, this technique takes a VASS with zero tests as input, and outputs a VASS without zero tests.
The new VASS uses an additional counter to simulate a polynomial number of zero tests on the other counters.
In this paper, we provide a new black-box version of the ``controlling counter'' lemma, which we believe is more amenable and has wider applicability compared to the ones in the literature.
With this new \cref{lem:controlling-counter}, the proof of~\cref{thm:3-lps-hardness} is completed by applying it to the unary 2-SLPS with zero tests which we constructed above.

\bigskip

Next, we consider what is perhaps one of the weakest models of computation involving multiple counters.
These are VASS with constant updates in $\set{-u, -u+1, \ldots, u-1, u}$, where $u \geq 1$ is a constant independent of the size of the VASS.
If $u=1$, i.e.\ updates can only belong to $\set{-1, 0, 1}$, we say the VASS is unitary.
In general, one can construct a polynomial size \emph{unitary} VASS that is equivalent to a \emph{unary} VASS: for example, a transition with update \inc{x}{5} can be split into five successive transitions each with the update \inc{x}{1}.
Such a construction does not work for simple LPS, because a single-transition cycle cannot be split into several transitions.
It is not at all clear if there exists a constant dimension $d \in \NN$ for which reachability in unitary $d$-SLPS is \class{NP}-hard.
For our second main contribution, we achieve one of the closest results one could hope for.

\begin{theorem}
	\label{thm:unitary-hardness}
	Reachability in unitary $\inverseAckermann{k}$-SLPS is \class{NP}-complete.
\end{theorem}

Here, $k$ denotes the size of the reachability instance and $\alpha:\NN\to\NN$ is the inverse of the Ackermann function
$\mathit{Ack}:\NN\to\NN$.
Thus, the restriction to $\inverseAckermann{k}$-SLPS permits the dimension of the SLPS to grow, but only very slowly relative to the size.

We use the same approach to obtain \class{NP}-hardness: reduction from 3-CNF SAT.
We first reuse~\cref{lem:sat-encoding} since the aforementioned 1-SLPS with non-divisibility assertions is unitary.
However, we must work harder to implement non-divisibility.
We construct a unitary 5-SLPS to test whether $\var{x}$ (the primary counter maintaining the guessed value of the assignment) is not divisible by $q$.
Although it is more intricate than its unary counterpart, the overall idea remains: add a non-zero remainder to $\var{x}$ and then test divisibility by $q$, before restoring the original counter values.
(The construction is presented in~\Cref{fig:unitary-non-div}, and its properties are summarised in~\cref{clm:unitary-non-div}.)
Accordingly, we obtain a unitary 5-SLPS with a polynomial number of zero tests for which reachability coincides with 3-CNF satisfiability.

There is a major difference in how we simulate zero tests in unitary SLPS.
For this, we use technique known as ``multiplication triples''~{\cite[Section 3]{CzerwinskiO21}} which was initially proposed in~{\cite[Section 3]{CzerwinskiLLLMSTOC19}}.
Here, three counters with initial values $\var{b} = B$, $\var{c} = 2C$, and $\var{d} = 2BC$ can be used to simulate $C$ many zero tests on other counters whose sum is always bounded above by $B$.
The challenge for us to overcome is how to initialise a triplet of counters with the right values: we need the quantity $B$ to be at least exponential and $C$ to be polynomial.
To do this, we manipulate some ideas, in particular, from~\cite{Lasota22}, that were used to show that reachability in VASS is \class{Ackermann}-hard.
We build a unitary $d$-SLPS of size $\Oh(A(d))$\footnote{For our purposes, we define a fast-growing function $A: \NN \to \NN$, see~\cref{app:ackermann-function}. $A$ has the same asymptotic growth rate as the Ackermann function, see Claim~\ref{clm:functionA}.}
which, from an initial configuration with counter values $\vec{0}^d$, reaches a configuration with counter values $(\vec{0}^{d-3}, B, A(d), B \cdot A(d))$, where $B \in \NN$ can be arbitrarily large (\cref{clm:unitary-triple}).
By setting $d = A^{-1}(k) = \Oh(\inverseAckermann{k})$, we obtain an $\Oh(\inverseAckermann{k})$-SLPS that initialises a triplet of counters with the desired values.
Once this is achieved, the simulation only requires replacing a zero test with a constant number of transitions and self-loops, which are indeed unitary (\cref{lem:unitary-zero-tests}).

\bigskip

To complement our first two results and to trace the tractability border for reachability in unary $d$-SLPS from the other side (upper bounds), we study an extension of reachability in unary 2-SLPS where the initial and target configurations are encoded in binary.
Recall that reachability is \class{NL}-complete in unary 2-VASS~\cite{EnglertLT16} and \class{PSPACE}-complete in binary 2-VASS~\cite{BlondinFGHM15}. 
In these papers, the encoding of vectors in the VASS transitions, the initial configuration, and the target configuration are treated in the same way.
A recent line of research focuses on VASS in which the encoding of VASS is treated differently to the encoding of source and target~{\cite{DraghiciHR24}}.
Namely, the VASS itself is fixed and only the initial and target configurations form an instance of the problem.
The complexity of the reachability problem in such instances is currently unknown but conjectured to be in \class{PSPACE}, even for general VASS.
In our result, the VASS is not fixed but, in a similar spirit, the initial and target counter values are encoded differently.

\begin{theorem}\label{thm:mainalg}
    Reachability in unary 2-SLPS is in \class{P} even if the initial and target configurations are encoded in binary.
\end{theorem}

The proof builds on the result when the initial and target configurations are encoded in unary~\cite{EnglertLT16} and introduces new techniques. 
Observe that, in an SLPS $\alpha_1 \, \beta_1^{\,*} \, \alpha_2 \, \cdots \, \alpha_{n-1} \, \beta_n^{\,*} \, \alpha_n$, a run is uniquely defined by a vector $(m_1,\ldots, m_n) \in \N^n$, i.e.\ how many times each cycle is performed. We divide the proof into two parts, both interesting on their own.

\begin{figure*}[ht!]
    \centering
    \includegraphics[width=0.9\textwidth]{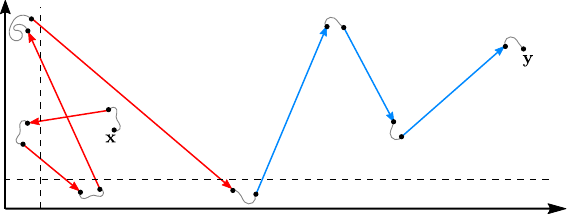}
    \caption{The figure presents the schematic view of the structure of
        reachability runs for unary 2-SLPS with binary-encoded
        $\config{s}{x}$ and $\config{t}{y}$. The grey parts of the run (little curves)
    are of polynomial length. The first four cycles (straight arrows highlighted in red) are taken $m_i$ times, and one of the
    endpoints of each of them is within polynomial distance from the axes. The last three cycles
    (straight arrows highlighted in blue) can terminate at arbitrary points.}
    \label{fig:decomposition}
\end{figure*}

First, in Theorem~\ref{thm:decomposition} we show that if there is a reachability witness then there is one of a
specific shape, where ``almost'' every $m_i$ is either small (polynomial in the size of the input SLPS) or almost as big as possible, i.e.\ such that the cycle ends not far (polynomially close) from one
of the principal axes. We write ``almost'' as we allow at most $\Oh(1)$ of
the $m_i$'s to be arbitrary (this seems to be unavoidable but is easy to handle nevertheless).
Note that in particular this means that cycles for almost all big $m_i$'s are
neither positive nor negative. They are so-called \emph{mixed vectors}, i.e.\
they strictly increase exactly one coordinate and strictly decrease the other
one.
We refer the reader to~\Cref{fig:decomposition} for a schematic overview of the characterisation.

Second, we show that such runs can be found using a dynamic programming algorithm. The main idea is simple: we maintain a subset of reachable configurations and perform consecutive cycles. 
Given our above characterisation of runs, whenever we prolong the configurations with the next cycle, we either move only within a polynomial ball of previously reachable configurations (if $m_i$ is small), or move from one of the axes to the other (if $m_i$ is big). The issue is that in the latter case, the set of reachable configurations grows multiplicatively by a polynomial factor; thus a naive implementation would result in an exponential algorithm.

Here, we rely on a new idea that allows us to maintain only a small subset
of reachable configurations. We make two big steps at once instead of one.
Recall that, for big steps, most cycles are mixed. Intuitively, a pair of mixed vector updates,
performed as much as possible, multiplies one of the coordinates by a
rational number and does not change the other much. 
For example, starting
from $(n,0)$ and performing $(-2,+1)$ followed by $(+5,-1)$ results roughly in
$(0,\frac{n}{2})$ followed by $(\frac{5}{2}n,0)$, thus multiplying the value by
$\frac{5}{2}$. Such a pair can express a vector parallel to the axis, e.g.\ in
this case $(-2,1) + (5,-1) = (3,0)$. In this case we can forget configurations
which differ by a multiple of $3$ on the first coordinate, assuming that performing $(-2,+1)$
followed by $(+5,-1)$ would not result in a counter dropping below zero. 
Intuitively, we
observe that the drop below zero occurs only if the quotient of the values in
the first coordinate is larger than $\frac{5}{2}$, which means that there are
only polynomially many regions that cannot reach one another. Altogether this
allows us to maintain a small number of configurations, as thanks to the first
part it suffices to consider only configurations polynomially close to the
axes (see~\Cref{fig:decomposition}).

We want to highlight the following observation, that most complexity upper bounds for VASS and related models are strongly dependent on the length of the shortest run witnessing reachability. 
For example, this is the case in the two above-mentioned results for 2-VASS~\cite{BlondinFGHM15,EnglertLT16}, where the shortest reachability witnesses are of polynomial and exponential size, respectively, which immediately yields the \class{NL} and \class{PSPACE} upper bounds. This is also the case for the \class{Ackermann} upper bound for general VASS~\cite{LerouxS19}. 
Our work, among few others, offers techniques going beyond length bounds of the shortest reachability witnesses. 
Under \emph{Further Related Work} below, we briefly discuss other examples of such techniques, which belong to the fundamental toolbox for VASS.

Finally and making for a secondary contribution, we study a subclass of simple linear path schemes called \emph{ultraflat VASS} \cite{Leroux21}.
Answering an open question of~{\cite[Section 5]{Leroux21}}, we prove that reachability in unary ultraflat 4-VASS is \class{NP}-complete (\cref{thm:ultraflat-hardness}).
We approach the lower bound in much the same way as for unary 3-SLPS; for a more detailed discussion, see~\cref{sec:4ultraflat}.

\subsection*{Further Related Work}

As mentioned above, reachability in flat VASS can be decided in \class{NP}, even though the shortest reachability witnesses could be exponentially long; this follows from the fact that reachability can be expressed as an integer linear program~\cite{FribourgOCONCUR97}.
Another example is \class{NP} membership of reachability in binary 1-VASS, where the shortest paths can have also have exponential length~\cite{HaaseKOW09}. 
The proof idea is to find a positive cycle followed by a negative cycle, which allows transitions to be reordered freely in the middle part of the run.

In linear path schemes, the reachability problem has also been studied with continuous semantics.
In a continuous VASS~\cite{BlondinH17}, the update of a transition can be multiplied by any real number in $(0,1]$.
Very recently, it was proven that reachability in continuous LPS is in \class{P}~{\cite[Theorem 13]{AlmagorGLP23}} and reachability in continuous 2-LPS with zero tests is \class{NP}-hard~{\cite[Theorem 21]{AlmagorGLP23}}. 
Differently from our paper (aside from the continuous semantics), in these results transitions in VASS and LPS are encoded in binary.

% !TEX root = ../main.tex
\section{Preliminaries}
\label{sec:preliminaries}

Let $\ZZ$ be the set of integers. 
Let $\NN$ be the set of natural numbers, that is, the set of nonnegative integers.
As usual, $\abs{S}$ denotes the cardinality of a set $S$.
We use $\poly{n}$ to denote $n^{\Oh(1)}$.
We denote integer intervals as $[x,y] = \set{z \in \ZZ : x \leq z \leq y}$.
We use boldface to denote vectors and we specify a vector by listing its coordinates in a tuple.
We index coordinates of vectors using square brackets: $\vec{v} = (\vec{v}[1], \ldots, \vec{v}[d])$, where $d \in \NN$ is the dimension of the vector.
For two $d$-dimensional vectors $\vec{u}$ and $\vec{v}$, we write $\vec{u} \leq \vec{v}$ if $\vec{u}[i] \leq \vec{v}[i]$ for all $i \in [1, d]$, and we also write $\vec{u} < \vec{v}$ if additionally $\vec{u} \neq \vec{v}$.
A vector $\vec{v}$ is nonnegative if $\vec{v} \geq (0, \ldots, 0)$.
The norm $\norm{\cdot}$, also denoted $\norm{\cdot}_1$, of a $d$-dimensional vector $\vec{v}$ is $\norm{\vec{v}} \coloneqq \sum_{i=1}^d\abs{\vec{v}[i]}$.
We overload this notation for sets of vectors $V$: $\norm{V} \coloneqq \sum_{\vec{v} \in V} \norm{v}$.

\paragraph*{Vector addition systems with states.}
A \emph{$d$-dimensional Vector Addition System with States} ($d$-VASS) $\Vv = (Q, T)$ consists of a non-empty finite set of \emph{states} $Q$ and a non-empty finite set of \emph{transitions} $T \subseteq Q \times \ZZ^d \times Q$.
A \emph{configuration} of a $d$-VASS, denoted $\config{q}{v}$, consists of the current state $q \in Q$ and the current counter values $\vec{v} \in \NN^d$.
Given two configurations $\config{p}{u}$ and $\config{q}{v}$, we write $\step{\config{p}{u}}{\config{q}{v}}$ if there exist a transition $t = (p, \vec{x}, q) \in T$ where $\vec{x} = \vec{v} - \vec{u}$; we refer to $\vec{x}$ as the \emph{counter updates} of $t$, and we may also write $\run{\config{p}{u}}{t}{\config{q}{v}}$ to highlight that transition $t$ was taken.

A \emph{path} in a VASS is a (possibly empty) sequence of transitions $( (q_0, \vec{x}_1, q_1), \ldots, (q_{\ell-1}, \vec{x}_\ell, q_\ell) )$, where $(q_i, \vec{x}_i, q_{i+1}) \in T$ for all $1 \leq i < \ell-1$.
A \emph{run} $\pi$ in a VASS is a sequence of configurations $\pi = ( \iconfig{q}{v}{0}, \ldots, \iconfig{q}{v}{\ell})$ such that $\step{\iconfig{q}{v}{i}}{\iconfig{q}{v}{i+1}}$ for all $0 \leq i \leq \ell-1$.
The \emph{length} of a path or run is the number of transitions taken, denoted $\length{\pi} = \ell$.
The \emph{effect} of a path or run is the sum of all the counter updates, denoted $\eff{\pi} = \vec{v}_\ell - \vec{v}_0$.
Given such a run $\pi$, we can also write $\run{\iconfig{q}{v}{0}}{\pi}{\iconfig{q}{v}{\ell}}$; $\pi$ is \emph{$b$-bounded} if $\norm{\vec{v}_i} \leq b$ for all $0 \leq i \leq \ell$.
We can also write $\Run{\config{p}{u}}{*}{\Vv}{\config{q}{v}}$ if there exists a
run from $\config{p}{u}$ to $\config{q}{v}$ in a VASS $\Vv$. 
It is often the case that the VASS $\Vv$ is implicit from the context, so we may use $\run{\config{p}{u}}{*}{\config{q}{v}}$ as shorthand for $\Run{\config{p}{u}}{*}{\Vv}{\config{q}{v}}$.
Furthermore, when the states $p$ and $q$ follow implicitly from the context, we use $\run{\vec{u}}{\rho}{\vec{v}}$ to denote that $\run{\config{p}{u}}{\rho}{\config{q}{v}}$.

\paragraph*{The reachability problem.}
An instance $(\Vv, \config{p}{u}, \config{q}{v})$ of \emph{the reachability problem} asks whether $\Run{\config{p}{u}}{*}{\Vv}{\config{q}{v}}$ is true for a given VASS $\Vv$, initial configuration $\config{p}{u}$, and target configuration $\config{q}{v}$.
To study the complexity of reachability, let us define the size of an instance.
The \emph{size} of a VASS $\Vv = (Q, T)$ \emph{encoded in unary} is $\size{\Vv} \coloneqq \abs{Q} + \sum_{(p, \vec{x}, q) \in T}\norm{\vec{x}}$.
The size of the instance of reachability, encoded in unary, is $\size{\Vv} + \norm{\vec{u}} + \norm{\vec{v}}$.
We will primarily focus on unary encoding, but both a VASS and a configuration can also be encoded in binary. 
In fact, later we will consider reachability in unary VASS where the initial and target configurations are encoded in binary.
The size of a configuration $\config{q}{v}$ encoded in binary is $\bitsize{\vec{v}} \coloneqq \log_2(\norm{\vec{v}}+1) + 1$.

\paragraph*{Flat VASS and linear path schemes.}
A VASS $\Vv = (Q, T)$ is \emph{flat} if for every state $q \in Q$, there is at most one simple cycle that contains $q$.
Intuitively speaking, $\Vv$ contains no nested cycles.
We study further restrictions of flat VASS, namely we focus on \emph{Linear Path Schemes} (LPS).
In an LPS, the states and transition structure are restricted so that the VASS is a simple path with disjoint cycles attached to each of the states along the path.
(Some of these cycles may well be self-loops with zero updates, which are useless.)
A \emph{Simple Linear Path Scheme} (SLPS) is an LPS in which the length of each cycle is at most one (first defined in~{\cite[Section 2]{EnglertLT16}}).
SLPS can be seen as a simple path with self-loops at each state, see~\Cref{fig:example} for an example.

\begin{figure}
	\begin{minipage}{.6\textwidth}
	  	\hspace{1.5cm}\begin{tikzpicture}[scale = 0.8]
	  		\node[circle, fill = black, draw, inner sep = 0.7mm, minimum size = 1.4mm] (q1) at (0, 0) {};
			\node[circle, fill = black, draw, inner sep = 0.7mm, minimum size = 1.4mm] (q2) at (2, 0) {};
			\node[circle, fill = black, draw, inner sep = 0.7mm, minimum size = 1.4mm] (q3) at (4, 0) {};
			\path[-Stealth, line width = 0.4mm] (q1) edge node[below] {$(+5, 0)$} (q2);
			\path[-Stealth, line width = 0.4mm] (q2) edge node[below] {$(-1, -10)$} (q3);
			\path[-Stealth, line width = 0.4mm, out = 120, in = 60, distance = 12mm] (q2) edge (q2);
			\node at (2, 1.2) {$(-1, +3)$};
	  	\end{tikzpicture}
	\end{minipage}
	\begin{minipage}{.4\textwidth}
	 	\begin{algorithmic}[1]
			\Require \assert{x}{0}, \assert{y}{0}
			\State \inc{x}{5}
			\State \LOOP \dec{x}{1}, \inc{y}{3}
			\State \dec{x}{1}, \dec{y}{10}
		\end{algorithmic}
	\end{minipage}
	\caption{An example 2-SLPS in which, from the initial state with initial values $(0, 0)$, the final state can only be reached with values $(0,2)$. On the left is a drawing of the 2-SLPS and on the right it is presented as a counter program.}
	\label{fig:example}
\end{figure}

The statement of the following theorem can be found in~{\cite[Section 1]{CzerwinskiLLLM20}}, note that the dimension does not even need to be fixed.
It follows from the fact that reachability can be described in existential Presburger arithmetic; for further details see~{\cite[Section 3]{FribourgOCONCUR97}}.
\begin{theorem}\label{thm:np}
	Reachability in binary encoded flat VASS is in \class{NP}. 
\end{theorem}

\paragraph*{Counter programs.}
We will present VASS, in particular SLPS, as \emph{counter programs} (as Esparza does in~{\cite[Section 7]{Esparza96}}).
Counter programs are numbered lists of instructions operating on variables (such as $\var{x}$) that are exactly the same as VASS counters: their value is a natural number ($\var{x} \geq 0$) and they can receive additive integer updates (such as \inc{x}{5}).
See~\Cref{fig:example} for an example of a 2-SLPS and its corresponding two-counter program.
In counter programs, we use \texttt{FOR} loop statements as shorthand for several instructions which are typically identical or similar (for example, see Lines 2, 3, and 7 in~\Cref{fig:lps-sat}).
Importantly, note that the ``variables'' that control the \texttt{FOR} loops are \emph{not} counters that are accessible to the program.
Additionally, we use \emph{gadgets} that are just named counter programs (or VASS) that can be referenced by their name for the succinctness of presentation (think function calls).
On top of that, later the \nondiv{\cdot} gadget (used in \satprogram{\phi} in~\Cref{fig:lps-sat}) is implemented in more than one way, depending on what encoding is required.

\paragraph*{Zero tests.}
We will also use counter programs (or VASS) that are enriched with zero tests.
A VASS with zero tests can zero test the counter on special zero-testing transitions; equivalently, a counter program can zero test a counter $\var{x}$ with the \zt{x} instruction.
A zero-testing transition does not update the counter value and can only be taken if indeed the value of $\var{x}$ is zero.
VASS with zero tests, more commonly known as \emph{counter machines}, are well-known to have undecidable reachability (already in dimension two~\cite{Minsky67}).

Formally, a \emph{$d$-VASS with zero tests} is a tuple $\Zz = (Q, T, Z)$ where $(Q, T)$ is a $d$-VASS and $Z \sset Q \times \set{\geqz, \eqz}^d \times Q$ is the set of \emph{zero-testing transitions}.
We will now define the semantics; let $(p, \vec{z}, q) \in Z$ be a zero-testing transition and let $\vec{v} \in \NN^d$.
Then $\Run{\config{p}{v}}{(p, \vec{z}, q)}{\Zz}{\config{q}{v}}$ is a run if, for every $i \in [1, d]$, if $\vec{z}[i]$ is ``$\eqz$'', then $\vec{v}[i] = 0$.
In words, the transition $(p, \vec{z}, q)$ can only be taken from a configuration $\config{p}{v}$ if all of the counters indexed by the components of $\vec{z}$ that are ``$\eqz$'' indeed have zero value.
The \emph{size} of a $d$-VASS with zero tests $\Zz = (Q, T, Z)$ encoded in unary is 
\begin{equation*}
	\size{\Zz} \coloneqq \abs{Q} + \abs{Z} + \sum_{(p, \vec{x}, q)} \norm{\vec{x}} .
\end{equation*}

In this paper, we always use zero tests in restricted ways.
It turns out that one additional counter can be used to simulate a polynomial number of zero tests on other counters.
This idea is known as the \emph{controlling counter technique}~{\cite[Lemma 10]{CzerwinskiO21}}.
Here, we present a more amenable ``black box'' version: it can be used to turn an instance of reachability $d$-VASS with a polynomial number of zero tests into an instance of reachability in a $(d+1)$-VASS.
Critically, this version also preserves the (simple) linear path scheme structure of a given VASS.
We provide further discussion and a proof of~\cref{lem:controlling-counter} in~\cref{app:controlling-counter}.

\begin{lemma}[Controlling Counter Technique, cf.~\cite{CzerwinskiO21}]
	\label{lem:controlling-counter}
	Let $\Zz$ be a $d$-VASS with zero tests and let $\config{s}{x}, \config{t}{y}$ be two configurations.
	Suppose $\Zz$ has the property that on any accepting run from $\config{s}{x}$ to $\config{t}{y}$, at most $m$ zero tests are performed on each counter.
	Then there exists a $(d+1)$-VASS $\Vv$ and two configurations $\config{s'}{0}, \config{t'}{y'}$ such that:
	\begin{enumerate}[(1)]
		\item $\Run{\config{s}{x}}{*}{\Zz}{\config{t}{y}}$ if and only if $\Run{\config{s'}{0}}{*}{\Vv}{\config{t'}{y'}}$,
		\item $\Vv$ can be constructed in $\Oh((\size{\Zz} + \norm{x}) \cdot (m+1)^d)$ time, and
		\item $\norm{\vec{y}'} \leq \norm{\vec{y}}$.
	\end{enumerate} 
	Moreover, if $\Zz$ is a flat VASS or a (simple) linear path scheme in which no zero-testing transition lies on a cycle, then $\Vv$ can be assumed to be a flat VASS or a (simple) linear path scheme, respectively.
\end{lemma}

% !TEX root = ../main.tex
\section{Hardness of Reachability in Linear Path Schemes}
\label{sec:hardness}

In this section, we detail the two main \class{NP} lower bounds for reachability in simple linear path schemes.
The first is when the SLPS has unary encoding and three dimensions.
The second is when the SLPS has unitary encoding and at least inverse Ackermann dimensions.

Both lower bounds are obtained by reducing from SAT.
In both reductions, we use an encoding of Boolean vectors by integers sometimes known as the Chinese remainder encoding
(see, e.g., Stockmeyer and Meyer~\cite[proof of Theorem~6.1]{StockmeyerM73}).
Let $\phi$ be an input formula in 3-CNF with variables $x_1, \ldots, x_n$ and clauses $C_1, \ldots, C_m$.
We choose positive integers $p_1, \ldots, p_n$ that are pairwise coprime, for example the first~$n$ prime numbers.
An integer $v$ encodes an assignment to $x_1, \ldots, x_n$ if and only if $v \bmod p_i$ is either 0 or 1 for all $i \in \set{1, \ldots, n}$; namely, this assignment has 
\begin{equation*}
	x_i \text{ is true } \text{if } v \equiv 1 \bmod p_i \text{\quad and \quad} x_i \text{ is false } \text{if } v \equiv 0 \bmod p_i.
\end{equation*}
By the Chinese remainder theorem, for each assignment vector $(a_1, \ldots, a_n) \in \set{0, 1}^n$, there exists a nonnegative integer~$v$ that encodes the assignment $x_i = a_i$ for all~$i \in \set{1, \ldots, n}$.
Note that this does not rule out the fact that there exist integers $v$ that do not correspond to assignments: if, for example, $v \equiv 2 \bmod p_i$ for some $i \in \set{1, \ldots, n}$, then $v$ does not correspond to an assignment.

First, we provide~\cref{lem:sat-encoding} that ``implements'' the above encoding of SAT into an instance of reachability in a simple linear path scheme that has one counter (denoted $\var{x}$) and the ability to perform non-divisibility checks.
The one-counter program implementation is presented in~\Cref{fig:lps-sat}.
In summary, an assignment value $v$ is ``guessed'' and stored in $\var{x} = v$.
To test whether the guessed value is a valid assignment, instead of testing whether 
\begin{equation*}
	v \equiv 1 \bmod p_i \; \text{ or } \; v \equiv 0 \bmod p_i,
\end{equation*}
we test whether 
\begin{equation*}
	v \not\equiv 2 \bmod p_i \; \text{ and } \; v \not\equiv 3 \bmod p_i \; \text{ and } \; \ldots \; \; v \not\equiv p_i-1 \bmod p_i.
\end{equation*}
The reason for this is due to the fact that (simple) linear path schemes do not allow for disjunction: for example, it is not possible to have transitions branching into two different states in one case \emph{or} another case.
Conjunction, on the other hand, is straightforward: one can compose two (simple) linear path schemes by joining the end of the first with the start of the second.

The ``conjunctive'' rewriting was used, e.g., by Sch\"oning~\cite{Schoning97}
to encode 3-CNF SAT in formulas of Presburger arithmetic.
The idea dates back to Stockmeyer and Meyer~\cite[proof of Theorem~6.1]{StockmeyerM73}.

We handle satisfiability of an assignment similarly to its validity, please see~\Cref{fig:lps-sat}.
The full proof of~\cref{lem:sat-encoding} can be found in~\cref{app:sat-encoding}.

\begin{lemma}
	Given an instance $\phi$ of 3-CNF SAT, let \satprogram{\phi} be the simple linear path scheme presented in~\Cref{fig:lps-sat} that uses one counter and non-divisibility assertions.
	Then $\phi$ is satisfiable if and only if the final state can be reached from the initial state (with any counter values) in \satprogram{\phi}.
	Moreover, \satprogram{\phi} can be constructed in polynomial time with respect to $\size{\phi}$.
	\label{lem:sat-encoding}
\end{lemma}

\begin{figure}[ht!]
	\begin{algorithmic}[1]
		\State \LOOP \inc{x}{1}
			\Comment{Guess a potential assignment.}
		\Ffor{i}{1}{n}
			\Comment{Check that $\var{x}$ encodes an assignment.}
			\Ffor{r}{2}{p_i-1}
				\State \inc{x}{p_i - r}
				\Comment{We would like to subtract $r$, but $\var{x}$ may be less than $r$.}
				\State \nondiv{p_i}
				\State \dec{x}{p_i - r}
			\EndFfor
		\EndFfor
		\Ffor{j}{1}{m}
			\Comment{Check that the assignment is satisfying.}
			\State \inc{x}{q_j - r_j}
				\Comment{\texttt{(*)} $q_j$ and $r_j$ are defined below.}
			\State \nondiv{q_j}
			\State \dec{x}{q_j - r_j}
		\EndFfor
		\Ensure $\phi = C_1 \wedge \cdots \wedge C_m$ has a satisfying assignment.
		\vspace{0.1in}
		\Statex \texttt{(*)} 
			Consider $C_j = \ell_1 \vee \ell_2 \vee \ell_3$ and suppose $x_a$, $x_b$, and $x_c$ are the variables of the literals. 
			Let $q_j = p_a \cdot p_b \cdot p_c$ and let $r_j \in \set{0, 1, \ldots, q_j-1}$ be the residue such that $r_j$ modulo $p_a$ is 0 or 1 and falsifies $\ell_1$, $r_j$ modulo $p_b$ is 0 or 1 and falsifies $\ell_2$, and $r_j$ modulo $p_c$ is 0 or 1 and falsifies $\ell_c$.
			For example, if $C_j = x_1 \vee \overline{x}_2 \vee x_3$, then $q_j = p_1p_2p_3$ and $r_j$ is selected such that $r_j \equiv 0 \bmod p_1$, $r_j \equiv 1 \bmod p_2$, and $r_j \equiv 0 \bmod p_3$.
	\end{algorithmic}
	\caption{%
		A simple linear path schemes \satprogram{\phi} with one counter $\var{x}$ and non-divisibility checks; reachability from the initial state to the final state is equivalent to satisfiability of $\phi$.
		Although only one counter $\var{x}$ is explicitly presented, the non-divisibility assertions are implemented by the \nondiv{\cdot} gadgets that use ancillary counters (see~\Cref{fig:unary-non-div} for unary encoding and~\Cref{fig:unitary-non-div} for unitary encoding).
		Notation: $n$ is the number of variables in $\phi$, $m$ is the number of clauses in $\phi$, and $p_1, \ldots, p_n$ are pairwise coprime integers greater than~$1$ (e.g., the first $n$ primes).
 	}
	\label{fig:lps-sat}
\end{figure}

\cref{lem:sat-encoding} will be used to prove all of our lower bounds for reachability in linear path schemes (Theorems~\ref{thm:3-lps-hardness},~\ref{thm:unitary-hardness}, and~\ref{thm:ultraflat-hardness}).
Though, we need to implement non-divisibility checks if we wish to obtain any lower bound, this varies depending on the exact model and encoding.
However, there is one thing in common with all of the implementations: they require some zero-tests.
How these zero tests are simulated also depends on the exact model and encoding.

\subsection{Unary Encoding and Three Dimensions}
\label{subsec:3-lps-hardness}

\begin{Thm1}
	Reachability in unary 3-SLPS is \class{NP}-complete.
\end{Thm1}
For the proof of~\cref{thm:3-lps-hardness}, we will show that the \satprogram{\phi} can be implemented by a unary 3-SLPS.
To achieve this, we must construct a unary 3-SLPS auxiliary gadget for asserting non-divisibility.
In fact, we construct a unary 2-SLPS with a polynomial number of zero tests.
We then use the controlling counter technique (\cref{lem:controlling-counter}) to simulate these zero tests at the cost of adding a dimension.
On top of the reserved primary counter $\var{x}$, for this subsection, we will reserve $\var{y}$ as the ancillary counter, which is used to complete operations on the primary counter.
We first present the unary 2-SLPS for asserting non-divisibility; its correctness (\cref{clm:unary-non-div}) is proved in~\cref{app:3slps}.

\begin{figure}[ht!]
	\begin{algorithmic}[1]
		\Require \assert{x}{v}, \assert{y}{0}
		\State \inc{x}{1}, \inc{y}{p-2} \Comment{Invariant: $\var{y} + \Delta \var{x} = p-1$.}
		\State \LOOP \dec{y}{1}, \inc{x}{1}
		\Comment{Result so far: $\Delta \var{x} \in \set{1, 2, \ldots, p-1}$.}
		\State \LOOP \dec{x}{p}, \inc{y}{p}
		\State \zt{x}
		\State \LOOP \inc{x}{1}, \dec{y}{1}
		\State \zt{y}
		\State \dec{x}{p-1}
		\Comment{This breaks the invariant, but $\var{x}=v$ is restored.}
		\Ensure \assert{x}{v}, \assert y 0
	\end{algorithmic}
	\caption{%
        The $\nondiv{p}$ gadget implemented as a unary 2-SLPS with two zero tests for asserting that the initial value $v$ of counter $\var{x}$ is not divisible by the fixed positive integer $p$.
		The construction, which has linear size in $p$, uses 2~counters, the primary~$\var{x}$ and its ancillary~$\var{y}$.
		From an initial configuration with $\var{x} = v, \var{y} = 0$, the final configuration with $\var{x} = v, \var{y} = 0$ can be reached if and only if $p$ does \emph{not} divide $v$.
		Note that $\Delta\var{x}$ is the change in the counter value of $\var{x}$.
	}
	\label{fig:unary-non-div}
\end{figure}

\begin{claim}\label{clm:unary-non-div}
	Let $p$ be a positive integer.
	In the unary 2-SLPS with zero tests presented in~\Cref{fig:unary-non-div}, from an initial configuration with counter values $\var{x} = v$, $\var{y} = 0$, the final state can be reached if and only if $p$ does not divide $v$.
	Moreover, in the case of reachability, the final configuration must have counter values $\var{x} = v$, $\var{y} = 0$.
	Furthermore, the unary 2-SLPS can be constructed in $\Oh(p)$ time and uses only two zero tests (which do not belong to cycles).
\end{claim}

\begin{proof}[Proof of Theorem~\ref{thm:3-lps-hardness}]
	As previously detailed, we obtain the lower bound via a reduction from 3-CNF SAT; the \class{NP} upper bound is given by~\cref{thm:np}.

	We extend the reduction presented in~\cref{lem:sat-encoding} by implementing the non-divisibility assertions in polynomial time.
	As seen in~\Cref{fig:unary-non-div}, we can implement a non-divisibility assertion as a unary 2-SLPS that only uses two zero tests (see~\cref{clm:unary-non-div} for its correctness).
	Note that the ancillary counter $\var{y}$ is untouched between non-divisibility assertions, so its value before and after each assertion is zero.
	Accordingly, we obtain an instance of reachability in a unary 2-SLPS with zero tests that is equivalent to the satisfiability of $\phi$.

	We can use the controlling counter technique (\cref{lem:controlling-counter}) to obtain, in polynomial time, an equivalent instance of reachability in a unary 3-SLPS.
	Importantly, observe that this is only possible because the number of zero tests is polynomially bounded, which is the case since the zero tests only occur on transitions between cycles\footnote{Note, in particular, the \emph{moreover} part of~\cref{lem:controlling-counter}, for further details see~\cref{app:controlling-counter}.}.
	Precisely, there are $m + \sum_{i=1}^n (p_i-2)$ many non-divisibility assertions each with two zero tests.
	Hence, reachability in unary 3-SLPS is \class{NP}-hard.
\end{proof}

% !TEX root = ../main.tex
\subsection{Constant Updates and Inverse Ackermann Dimensions}
\label{sec:inverse-ackermann}

Typically, the choice of updates (the numbers occurring on the transitions) of a VASS, or their encoding, does not influence expressivity of the model even though that choice is likely to influence its size.
However, if we restrict our focus to VASS subclasses, this may not be true; in this section we consider a situation in which expressivity is influenced by the choice of numbers on the transitions.

A VASS has \emph{constant updates} if all counter updates on all transitions belong to $\set{-u, -u+1, \ldots, u-1, u}$ for some constant $u \in \NN$ that is independent of the size of the VASS.
We say that a VASS is \emph{unitary} if that constant is one (so all updates belong to $\set{-1, 0, 1}$).
In general, a VASS can be simulated by a VASS with constant updates (in fact by a unitary VASS) of the same dimension.
The construction is rather simple, for example, one can replace a transition with an update of $+5$ with five transitions with updates of $+1$.
The size of the resulting unitary VASS is equal to the size of the original VASS times the greatest absolute value of any of its updates.
Thus, a unary VASS can be simulated by a polynomial size unitary VASS of the same dimension.
The same cannot be said for \emph{simple} linear path schemes; the aforementioned construction does not respect the structural requirement that the cycles can only consist of one transition.

We show that the choice of the number to fix the updates does not affect the expressivity of SLPSs.
The following lemma (proved in~\cref{app:fixed-updates}) allows us to focus on
unitary SLPS, noting that we may need to increase the dimension by a constant
factor if we require constant updates of magnitude greater than 1.

\begin{lemma}\label{lem:fixed-updates}
	A $d$-SLPS with constant updates in $\set{-u, -u+1, \ldots, u-1, u}$ can be simulated by a unitary $(d \cdot u)$-SLPS.
\end{lemma}

Unitary SLPS is an \emph{exceedingly weak} model of computation.
Prior to this work, it was not at all clear whether reachability for unitary SLPS is \class{NP}-complete when the dimension is not fixed, however one can obtain this lower bound by combining~\cref{thm:3-lps-hardness} with~\cref{lem:fixed-updates}.
Besides, we will present a much stronger result: reachability in unitary SLPS with inverse Ackermann dimensions is \class{NP}-complete.

To be as clear as possible, one can view the reachability problem in inverse Ackermann dimension unitary SLPS as follows.
The input is: 
	a number $k$ specified in unary,
	a unitary simple linear path scheme $\Vv$ of dimension at most $c\cdot\inverseAckermann{k}$ (some constant $c$) and of size at most $k$, an initial configuration $\config{p}{u}$ where $\norm{\vec{u}} \leq k$, and
	a target configuration $\config{q}{v}$ where $\norm{\vec{v}} \leq k$.
The question is: does there exist a run in $\Vv$ from $\config{p}{u}$ to $\config{q}{v}$\,?
The number $k$, given in unary, only serves to explicitly relate the size of the reachability instance and the dimension of the given SLPS.
For the remainder of this subsection and for our convenience, we let $k$ be an arbitrary natural number.

\begin{Thm2}
	Reachability in unitary $\inverseAckermann{k}$-SLPS is \class{NP}-complete.
\end{Thm2}

The reduction for this lower bound, from a high-level perspective, is approached in the same way to~\cref{thm:3-lps-hardness}.
We will reduce from SAT and, again, we will use the Chinese remainder encoding of assignments to the variables.
\cref{lem:sat-encoding} is reused; checking that a guessed assignment is both valid and satisfying is achieved through a conjunction of non-divisibility assertions.
Note that the one-counter program with non-divisibility assertions (\Cref{fig:lps-sat}) can be implemented using unitary updates: the updates on Lines 4, 6, 8, and 10 can be unrolled into several $+1$ or $-1$ updates.

The major difference is how non-divisibility is checked, though we will still use zero tests for these checks.
On top of that, we also require a different technique for simulating zero tests compared to unary SLPS, for which we employed the controlling counter technique (\cref{lem:controlling-counter}).
That is because there is no guarantee that given a unitary VASS, the additional counter prescribed the controlling counter technique is unitary; it receives updates that linearly depend on the number of zero tests remaining on the other counters.

In~\cref{app:3slps}, we present a constant dimension unitary SLPS for asserting non-divisibility (see~\cref{app:unitary-non-div}), along with the proof of~\cref{clm:unitary-non-div} that is its correctness (see~\cref{app:unitary-non-div-correctness}).

\begin{claim}
	\label{clm:unitary-non-div}
	Let $p$ be a positive integer.
	In the unitary 5-SLPS with zero tests presented in~\Cref{fig:unitary-non-div}, from an initial configuration with counter values $\var{x} = v$, $\var{y} = \var{a_1} = \var{a_2} = \var{a_3} = 0$, the final state can only be reached if and only if $p$ does not divide $v$.
	Moreover, in the case of reachability, the final configuration must have counter values $\var{x} = v$, $\var{y} = \var{a_1} = \var{a_2} = \var{a_3} = 0$.
	Furthermore, the unitary 5-SLPS can be constructed in $\Oh(p)$ time and uses $2p+5$ zero tests (which do not belong to cycles).
\end{claim}

We will now give a sketch of the proof of~\cref{thm:unitary-hardness}; the full proof can be found in~\cref{app:unitary-hardness}.
We reduce from SAT by extending the reduction for~\cref{lem:sat-encoding} by implementing the non-divisibility assertions as unitary 5-SLPSs with zero tests (see~\cref{clm:unitary-non-div} and~\Cref{fig:unitary-non-div}).
Therefore, the satisfiability of $\phi$ is equivalent to an instance reachability in a unitary 5-SLPSs with polynomially many zero tests.
We conclude by using~\cref{lem:unitary-zero-tests} and~\cref{clm:unitary-triple}, which simulate zero tests with a multiplication triple and initialise said triple, respectively.

More concretely, to simulate the zero tests we make use of a technique for simulating zero tests known as \emph{multiplication triples}~\cite{CzerwinskiLLLM21,CzerwinskiO21, Lasota22, CzerwinskiO22}.
We wish to set up three counters $\var{b}$, $\var{c}$, and $\var{d}$ with initial values $\var{b} = B$, $\var{c} = 2C$, and $\var{d} = 2BC$ that can perform $C$ many zero tests on some (other) counters whose value are bounded above by $B$.
The following lemma, which is proved in~\cref{app:unitary-zero-tests}, is an SLPS analogue of~{\cite[Lemma 2.6]{CzerwinskiO22}}.
We say that the \emph{aggregate effect} of a transition is the sum of its updates, for example the aggregate effect of the transition on Line 13 of~\Cref{fig:unitary-non-div} is $-1$.
We note that the bounding counter $\var{b}$ needs to be split into several counters in order to maintain the fact that the resulting SLPS is unitary (this is done in a way mirroring the proof of~\cref{lem:fixed-updates}).
We will use~\cref{lem:unitary-zero-tests} only when the magnitude of the largest aggregate effect of any transition is bounded by a constant.

\begin{lemma}\label{lem:unitary-zero-tests}
	Let $\Uu$ be a unitary $d$-SLPS with at most $C$ zero tests, where the absolute value of the aggregate effect of any transition is at most $a \in \NN$, and such that the sum of the $d$ counters is always bounded by $B$.
	Then, there exists a unitary $(d+a+2)$-SLPS $\Vv$ where the absolute value of the aggregate effect of any transition is at most one, that can be constructed in $\Oh(a\cdot\norm{\Uu} + d\cdot C)$ time, such that 
	\begin{equation*}
		\Run{\config{s}{x}}{*}{\Uu}{\config{t}{y}} \;\text{ if and only if }\; \Run{s'(\vec{x}, \vec{0}^{a-1}, B, 2C, 2BC)}{*}{\Vv}{t'(\vec{y}, \vec{0}^{a-1}, 0, 0, 0)}.
	\end{equation*}
\end{lemma}

In the following claim, the fast-growing function $A:\NN\to\NN$ is defined in~\cref{app:ackermann-function}; it has the same asymptotic growth rate as the Ackermann function (which is also defined in~\cref{app:ackermann-function}), see~\cref{clm:functionA}.
We provide detailed commentary and a proof of~\cref{clm:unitary-triple} in~\cref{app:ackermann-generator}.

\begin{claim}\label{clm:unitary-triple}	
	There exists a unitary $\Oh(d)$-SLPS $\Tt$ of size $\Oh(A(d))$ such that, if, from an initial configuration with counter values $\vec{0}^d$ and for any $C \in \NN$, the final state can only be reached with counter values $(\vec{0}^{d-3}, x, y, z)$ where $x = A(d)$, $y \geq C$, and $z = A(d) \cdot y$.
\end{claim}

\newcommand{\lps}{\mathcal{L}}
\newcommand{\cone}{cone}
\newcommand{\Q}{\mathbb{Q}}
\newcommand{\Qplus}{\mathbb{Q}_{\ge 0}}
\newcommand{\C}{{\bf C}}
\newcommand{\reach}{\trans{*}}
\newcommand{\cycles}[1]{Cycles(#1)}
\newcommand{\axis}{Axis}

\newcommand{\epath}[1]{\mathit{path}(#1)}

\definecolor{thinfill}{RGB}{110, 100, 230}
\definecolor{smallfill}{RGB}{244, 180, 120}
\definecolor{thincolour}{RGB}{70, 80, 150}
\definecolor{pathcolour}{RGB}{20,20,20}
\definecolor{firstcolour}{RGB}{160, 20, 70}
\definecolor{freecolour}{RGB}{20, 130, 30}
\definecolor{zigzagcolour}{RGB}{30, 33, 200}
\definecolor{smallcolour}{RGB}{80,100,80}
\definecolor{incolour}{RGB}{200,95, 10}
\definecolor{greyout}{RGB}{221,222,223}
\newcommand{\shortminus}{\scalebox{0.5}[1.0]{$-$}}

\section{Decomposing Runs in Unary 2-SLPS Between Binary-Encoded Initial and Target Configurations}
\label{sec:polynomial}

In this section, we will present a structural result about the shape of the runs that are sufficient to consider as witnesses of reachability in unary 2-SLPS where the initial and target configurations are encoded in binary.
Intuitively speaking, a cycle can be exhausted if the cycle has a negative effect on (at least) one counter.
We will argue that cycles are either iterated a ``small'' number of times or are ``nearly exhausted'', except for a constant number of cycles that can be taken many times.
We will then leverage our structural result about runs to obtain a polynomial time algorithm for reachability in unary 2-SLPS with binary encoded initial and target configurations in~\cref{sec:algorithm}.

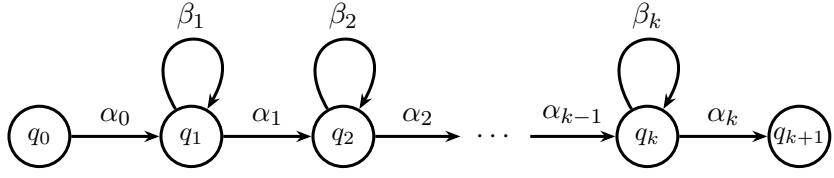
\begin{figure*}
    \centering
    \begin{tikzpicture}
	\node[circle, draw=black, line width = 0.4mm, minimum size = 8mm] (q0) at (0,0) {\small$q_0$};
	\node[circle, draw=black, line width = 0.4mm, minimum size = 8mm] (q1) at (2,0) {\small$q_1$};
	\node[circle, draw=black, line width = 0.4mm, minimum size = 8mm] (q2) at (4,0) {\small$q_2$};
	\node[circle, draw=white, line width = 0.4mm, minimum size = 8mm] (q3) at (6,0) {$\cdots$};
	\node[circle, draw=black, line width = 0.4mm, minimum size = 8mm] (qk) at (8,0) {\small$q_k$};
	\node[circle, draw=black, line width = 0.4mm, minimum size = 8mm] (q) at (10,0) {};
	\node at (10,0) {\small$q_{k+1}$};

	\path[draw=black, -{Stealth[width=1.5mm, length=2mm]}, line width = 0.4mm] (q0) -- node[above]{$\alpha_0$} (q1);
	\path[draw=black, -{Stealth[width=1.5mm, length=2mm]}, line width = 0.4mm] (q1) -- node[above]{$\alpha_1$} (q2);
	\path[draw=black, -{Stealth[width=1.5mm, length=2mm]}, line width = 0.4mm] (q2) -- node[above]{$\alpha_2$} (q3);
	\path[draw=black, -{Stealth[width=1.5mm, length=2mm]}, line width = 0.4mm] (q3) -- node[above]{$\alpha_{k-1}$} (qk);
	\path[draw=black, -{Stealth[width=1.5mm, length=2mm]}, line width = 0.4mm] (qk) -- node[above]{$\alpha_k$} (q);

	\path[draw=black, -{Stealth[width=1.5mm, length=2mm]}, line width = 0.4mm] (q1) edge[loop above, looseness = 30, distance = 40, out = 120, in = 60] node[above] {$\beta_1$} (q1);
	\path[draw=black, -{Stealth[width=1.5mm, length=2mm]}, line width = 0.4mm] (q2) edge[loop above, looseness = 30, distance = 40, out = 120, in = 60] node[above] {$\beta_2$} (q2);
	\path[draw=black, -{Stealth[width=1.5mm, length=2mm]}, line width = 0.4mm] (qk) edge[loop above, looseness = 30, distance = 40, out = 120, in = 60] node[above] {$\beta_k$} (qk);

\end{tikzpicture}
    \caption{A drawing of the SLPS $\Vv = \alpha_0 \beta_1^* \alpha_1 \cdots \alpha_{k-1} \beta_k^* \alpha_k$.}
    \label{fig:slps}
\end{figure*}

Throughout the section fix our attention on the 2-dimensional SLPS $\Vv = \alpha_0 \beta_1^* \alpha_1 \cdots \alpha_{k-1} \beta_k^* \alpha_k$ which is illustrated in~\Cref{fig:slps}.
Namely, for each $i$, $\beta_i$ is a cycle that can be taken an arbitrary amount of times and $\alpha_i$ is a transition leading between $\beta_i$ and $\beta_{i+1}$ (that can only be taken once).
To avoid extra notation, in this section and in \Cref{sec:algorithm} we will identify $\alpha_i$ and $\beta_i$ with their effects.
Furthermore, without loss of generality, we assume that $\beta_i \neq (0,0)$.

For further convenience, we will omit the state when specifying configurations.
In every scenario, the state is either implicit (because there is only one underlying path in this VASS) or specified a priori.
We will fix an instance of reachability in 2-SLPS $(\Vv, \vec{s}, \vec{t})$, where $\vec{s} \in \NN^2$ and $\vec{t} \in \NN^2$ are shorthand for $\config{q_0}{s}$ and $\config{q_{k+1}}{t}$, respectively.

To make it clear in this context, the size of $\Vv$ encoded in unary is
\begin{equation*}
    \size{\Vv} = k+1 + \sum_{i=0}^k \norm{\alpha_i} + \sum_{i=1}^k\norm{\beta_i}.
\end{equation*}
Note that $\size{\Vv} > k$ and $\size{\Vv} > \max_{\vec{v} \in \set{\alpha_0, \beta_1, \alpha_1, \ldots, \alpha_{k-1}, \beta_k, \alpha_k}} \norm{\vec{v}}$.
 
\paragraph*{Paths and runs.}
Paths in $\Vv$ have the form:
\begin{align}\label{eq:run}
    \pi = \alpha_0 \, \beta_1^{n_1} \, \alpha_1 \, \cdots \, \alpha_{k-1} \, \beta_k^{n_k} \, \alpha_k,
\end{align}
where $n_1, \ldots, n_k \in \NN$ are the \emph{exponents}.
A path in $\Vv$ can be uniquely represented by the vector of its exponents, a vector in $\NN^k$.
We define the function $\mathit{path}$ that accordingly maps vectors in $\NN^k$ to paths in $\Vv$.
For example, $\epath{n_1, \ldots, n_k}$ represents $\pi$ in~\Cref{eq:run}. 
For the remainder of this section, we fix $\pi = \epath{n_1, \ldots, n_k}$.

A \emph{subpath} is an infix of a path.
Similarly, a \emph{subrun} is an infix of a run.
Both subpaths and subruns can be prefixes and suffixes of paths and runs, respectively.

We will now introduce some notation to easily specify useful subpaths of an SLPS.
For a path $\pi = \epath{n_1, \ldots, n_k} = \alpha_0 \, \beta_1^{n_1} \, \alpha_1 \, \cdots \, \alpha_{k-1} \, \beta_k^{n_k} \, \alpha_k$ and two indices $i, j \in \set{0, 1, \ldots, k}$, we define
\begin{center}
    \begin{tabular}{r l}
        for $i \geq 1$, $j \geq i$, 
            & $\pi[i,j] \coloneqq \beta_i^{n_i} \alpha_i \cdots \beta_j^{n_j} \alpha_j$; \\[2mm]
        for $i \geq 1$, $j \geq i$,
            & $\pi[i,j) \coloneqq \beta_i^{n_i} \alpha_i \cdots \beta_j^{n_j}$; \\[2mm]
        for $i \geq 0$, $j \geq i$, 
            & $\pi(i,j] \coloneqq \alpha_i \cdots \beta_j^{n_j} \alpha_j$; \text{ and} \\[2mm]
        for $i \geq 0$, $j > i$,
            & $\pi(i,j) \coloneqq \alpha_i \cdots \beta_j^{n_j}$.
    \end{tabular}
\end{center}
For absolute clarity, we define $\pi(i,i)$ to be the empty path for every $i$.

At times, it will be useful to allow the counters to take negative values.
A \emph{$\ZZ$-configuration} is a tuple consisting of a state $q \in Q$ and current (potentially negative) integer counter values $\vec{v} \in \ZZ^2$, denoted $q(\vec{v})$.
A \emph{$\ZZ$-run} $(\iconfig{q}{v}{i})_{i=0}^k$ is sequence of $\ZZ$-configurations such that, for every $1 \leq i \leq k$, $(q_{i-1}, \vec{v}_i - \vec{v}_{i-1}, q_i) \in T$.
We use $\zrun{\config{p}{u}}{*}{\config{q}{v}}$ to denote a $\ZZ$-run from $\config{p}{u}$ to $\config{q}{v}$.

\subsection{Midpoint Configurations}
\label{subsec:midpoints}

Suppose, for a moment, that $\zrun{\vec{s}}{\pi}{t}$.
A \emph{midpoint} is a $\ZZ$-configuration that is reached after traversing $\pi(0, i]$ or $\pi(0,i)$ from $\vec{s}$; in other words, a midpoint is a $\ZZ$-configuration reached just before iterating $\beta_i$ or just after all iterations of $\beta_i$ are taken. 
We denote the midpoints using $\vec{a}_0^\pi, \vec{b}_0^\pi, \ldots, \vec{a}_k^\pi, \vec{b}_k^\pi$ ($\pi$ explicitly written in order to differentiate midpoints of different runs).
Precisely, for every $i$, $\vec{a}_i^\pi$ is the $\ZZ$-configuration reached after traversing $\pi(0,i)$ from $\vec{s}$ and $\vec{b}_i^\pi$ is the $\ZZ$-configuration reached after traversing $\pi(0,i]$ from $\vec{s}$.
Here, we note that (again) the current state is implicit: $\vec{a}_i^\pi$ is shorthand for $q_i(\vec{a}_i^\pi)$ and $\vec{b}_i^\pi$ is shorthand for $q_{i+1}(\vec{b}_i^\pi)$.
Altogether,
\begin{equation*}
        \vec{s} = \vec{a}_0^\pi 
        \xrightarrow{\alpha_0} \vec{b}_0^\pi 
        \xrightarrow{\beta_1^{n_1}} \vec{a}_1^\pi
        \xrightarrow{\alpha_1} \vec{b}_1^\pi 
        \xrightarrow{\hspace{0.1in}} \cdots \xrightarrow{\hspace{0.1in}} \vec{a}_{k-1}^\pi
        \xrightarrow{\alpha_{k-1}} \vec{b}_{k-1}^\pi
        \xrightarrow{\beta_k^{n_k}} \vec{a}_k^\pi 
        \xrightarrow{\alpha_k} \vec{b}_k^\pi = \vec{t}.
\end{equation*}

It turns out that, for a given path $\pi$, in order to verify that following $\pi$ from $\vec{s}$ yields a valid run, it suffices to check the nonnegativity of the midpoints.
The proof of~\Cref{lem:checkrun} can be found in~\cref{app:checkrun}

\begin{lemma}\label{lem:checkrun}
    In $\Vv$, $\run{\vec{s}}{\pi}{\vec{t}}$ if and only if $\zrun{\vec{s}}{\pi}{\vec{t}}$ and $\vec{a}_0^\pi, \vec{b}_0^\pi, \ldots, \vec{a}_k^\pi, \vec{a}_b^\pi \geq \vec{0}$.
\end{lemma}

Thanks to~\cref{lem:checkrun}, we shall limit the discussion from all configurations to just midpoints.

\subsection{Shifting Paths}
\label{sec:shifting}

Recall that we fixed the path $\pi = \epath{n_1, \ldots, n_k}$.

\begin{definition}[Shifting] \label{def:shifting}
    Let $i_1, i_2, i_3 \in \set{1, \ldots, k}$ be three indices such that $i_1 < i_2 < i_3$ and $e_1, e_2, e_3 \in \ZZ$. 
    We say that the path $\rho = \epath{m_1, \ldots, m_k}$ is $\pi$ \emph{shifted} at $i_1, i_2, i_3$ by coefficients $e_1, e_2, e_3$ if
    \begin{enumerate}[(i)]
        \item $m_{i_1} = n_{i_1} + e_1 \geq 0$, $m_{i_2} = n_{i_2} + e_2 \geq 0$, and $m_{i_3} = n_{i_3} + e_3 \geq 0$; and
        \item for all $j \in \set{1, \ldots, k}\setminus\set{i_1, i_2, i_3}$, $m_j = n_j$.
    \end{enumerate}
\end{definition}

For convenience, if some coefficients are zero, then we may drop them when describing a shifted path.
For example, if $c_{i_3} = 0$ in the definition of a shifted path, then we can drop the third index and coefficient and say that $\rho$ is $\pi$ shifted at $i_1, i_2$ by $e_1, e_2$.

The following is an auxiliary claim that is used multiple times later in this section; it is proved in~\cref{app:zero-effect-shifting}.

\begin{claim}\label{clm:zero-effect-shifting}
    Let $\pi$ be a path and let $\rho$ be the path that is $\pi$ shifted at $i_1, i_2, i_3$ by $e_1, e_2, e_3$.
    Suppose $e_1 \cdot \beta_{i_1} + e_2 \cdot \beta_{i_2} + e_3 \cdot \beta_{i_3} = \vec{0}$, then 
    \begin{enumerate}[(1)]
        \item $\zrun{\vec{s}}{\pi}{\vec{t}}$ if and only if $\zrun{\vec{s}}{\rho}{\vec{t}}$; and
        \item $\vec{a}_j^\pi = \vec{a}_j^\rho$ and $\vec{b}_j^\pi = \vec{b}_j^\rho$ for all $j \in \set{1, \ldots, i_1-1}\cup\set{i_3, \ldots, k}$.
    \end{enumerate}
\end{claim}

\subsection{Bounds for Configurations}
\label{sec:configuration-bounds}

We will now introduce some bounds that will be convenient to use throughout the rest of this section.
Consider three vectors $\vec{v}_1, \vec{v}_2, \vec{v}_3 \in \set{\beta_1, -\beta_1, \ldots, \beta_k, -\beta_k}$ such that the equation
\begin{align}\label{eq:cramer}
    x_1 \cdot \vec{v}_1 + x_2 \cdot \vec{v}_2 + x_3 \cdot \vec{v}_3 = \vec{0}
\end{align}
has a solution in natural numbers (nonnegative integers). 
Since $\vec{v}_1, \vec{v}_2, \vec{v}_3$ are two-dimensional vectors, if there is a solution to~\Cref{eq:cramer}, then there exists a solution $x_1, x_2, x_3 \leq N$ where $N$ is polynomial in $\size{\Vv}$. 
Indeed, this follows by Cramer's rule and Hadamard's inequality;
in fact such a bound also holds in arbitrary dimension, see, e.g., Domenjoud~\cite{Domenjoud91}.

We now define a bound that will be used repeatedly on midpoint configurations:
\begin{equation}\label{eq:N}
    B \coloneqq N\cdot\size{\Vv}^2 + 2N\cdot\size{\Vv}.
\end{equation}
Note that since $N$ is polynomial in $\size{\Vv}$, $B$ is also polynomial in $\size{\Vv}$.

Intuitively, one should think of $B$ as follows. 
Consider a path $\pi = \epath{n_1, \ldots, n_k}$ in which $n_i < N$, for every $1 \leq i \leq k$ and consider an initial configuration $\vec{s} \geq (N\cdot\size{\Vv}^2, N\cdot\size{\Vv}^2)$. 
It is easy to observe that, from $\vec{s}$, following $\pi$ yields a run; it is not possible for either counter to drop below zero at any point since $k + 1 \leq \size{\Vv}$ and $\alpha_i, \beta_i \geq (-\size{\Vv}, -\size{\Vv})$.
The additional $+2N\cdot\size{\Vv}$ in the definition of $B$ will be used to shift runs and maintain the nonnegativity of the midpoints of the shifted run.
This idea is formalised by the subsequent definition and claim.

\begin{definition}[Shiftable midpoint]\label{def:shiftable}
    Let $\pi = \epath{n_1, \ldots, n_k}$ be a path.
    We say that a midpoint $\vec{a}^\pi_i$ is \emph{shiftable} if $n_i \geq N$ and either
    \begin{enumerate}[(a)]
        \item $\vec{a}^\pi_i \geq (B, B)$; or
        \item for some $\iota \in \set{1,2}$, $\beta_i[\iota] = 0$ and $\vec{a}^\pi_i[3-\iota] \geq B$. 
    \end{enumerate}
\end{definition}

\begin{claim} \label{clm:shifting-subclaim}
    Let $\pi = \epath{n_1, \ldots, n_k}$ be a path, let $i$ and $j$ be an indices such that $i < j$, and let $x \in \ZZ$ such that
    \begin{enumerate}[(i)]
        \item $\vec{a}_i^\pi$ is shiftable;
        \item for all $i < s < j$, $n_s < N$; and
        \item $\abs{x} \leq N$.
    \end{enumerate}
    Suppose $\rho$ is the path obtained by shifting $\pi$ at $i$ by $x$.
    If $\run{\vec{s}}{\pi}{\vec{t}}$, then $\run{\vec{s}}{\rho(0,j-1]}{\vec{v}}$ for some $\vec{v} \geq \vec{0}$.
\end{claim}

The proof of~\cref{clm:shifting-subclaim} can be found in~\cref{app:shifting-subclaim}.

\begin{definition}[$C$-Safety] \label{def:c-safe}
    Let $C \in \NN$ and let $i, j \in \set{1, \ldots, k}$.
    We say that, from $\vec{s}$, path~$\pi$ is $C$-\emph{safe} over $[i,j)$ if there does not exist $i \leq s < j$ such that $\vec{a}_s \leq (C,C)$. 
    In other words, at least one coordinate of every $\vec{a}_i, \vec{a}_{i+1}, \ldots, \vec{a}_{j-1}$ is at least $C$.
\end{definition}

Note that, in the definition of $C$-safety, we do not bound the final midpoint $\vec{a}_j$.
It is possible that $\vec{a}_j \leq (C, C)$; this fact will be convenient in some of the following proofs.

\subsection{Simplifying the Runs}
\label{sec:simplifying-runs}

\begin{definition}[Essential subpaths] \label{def:essential}
    Let $1 \leq i \leq j \leq k$ be two indices and consider a subpath $\pi[i,j) = \beta_i^{n_i}\,\alpha_i\,\cdots\,\alpha_{j-1}\,\beta_j^{n_j}$.
    Let $a$ be the first shiftable midpoint after $i$, in other words $a \coloneqq \min\set{s \in (i,j] : \vec{a}_s^\pi \text{ is shiftable}}$.
    We say that $\pi[i,j)$ is \emph{essential} if there exists an index $b$ such that
    \begin{equation*}
        \sum_{s \in \set{a+1, \ldots, j}\setminus\set{b}} n_s \leq B + 2N\cdot\size{\Vv}^2 + N\cdot\size{\Vv}+\size{\Vv}. 
    \end{equation*}
\end{definition}

In other words, the subpath $\pi[i,j)$ is essential if there is at most one cycle $\beta_b$ among $\beta_{a+1}, \ldots, \beta_j$ which is taken more than a fixed polynomial number of times. 
Note that the index~$b$ in the definition does not necessarily belong to the set~$\set{a+1, \ldots, j}$.
Also note that there is no lower bound on the number of times that $\beta_b$ is taken, it could very well be the case that all cycles are taken at a fixed polynomial number of times; in other words, it could true that $\sum_{s \in \set{a+1, \ldots, j}} n_s \leq  B + 2N\cdot\size{\Vv}^2 + N\cdot\size{\Vv}+\size{\Vv}$. 

For convenience, given a subpath $\pi[i,j)$, we define a set that contains indices of cycles that are taken a ``large'' number of times:
\begin{equation*}
    I_{\pi[i,j)} \coloneqq \set{ s \in \set{i, \ldots, j} : n_s \geq N}.
\end{equation*}

We now provide a claim that will be used to simplify runs that contain shiftable midpoints.
In the following claim, one can think that $C = B$, but later in this section it will be convenient to enlarge the set of ``unsafe'' configurations.
We will compare two subpaths $\pi[i,j) = \beta_i^{n_i}\,\alpha_i\,\cdots\,\alpha_{j-1}\,\beta_j^{n_j}$ and $\rho[i,j) = \beta_i^{m_i}\,\alpha_i\,\cdots\,\alpha_{j-1}\,\beta_j^{m_j}$ using the reverse lexicographic ordering (over the vectors containing the cycle iterations).
We shall denote the order using $\prec$; it is defined by:
\begin{equation*}
    \rho[i,j) \prec \pi[i,j)  \iff \text{ there exists } s \in \set{i, \ldots, j} \text{ such that } m_s < n_s,\, m_{s+1} = n_{s+1},\,\ldots,\, m_j = n_j.
\end{equation*}

\begin{lemma}\label{lem:iteration}
    Let $\pi$ be a path in $\Vv$ such that $\run{\vec{s}}{\pi}{\vec{t}}$, let $i, j \in \set{1, \ldots, k}$ be two indices, and let $C = \size{\Vv}^2\cdot(B + (2N+3)\cdot\size{\Vv})$.
    If there exists $a \in \set{i, \ldots, j}$ such that $\vec{a}_a^\pi$ is a shiftable midpoint, then there exists a path $\rho$ such that $\run{\vec{s}}{\rho}{\vec{t}}$, $\rho(0,i-1] = \pi(0,i-1]$, $\rho(j,k] = \pi(j,k]$, and either
    \begin{enumerate}[(a)]
        \item $\rho[i,j)$ is essential; 
        \item $\rho$ is not $C$-safe over $[i,j)$; or
        \item $\rho[i,j) \prec \pi[i,j)$.
    \end{enumerate}
\end{lemma}

\begin{proof}
    Firstly, if $\pi[i,j)$ is essential, then we immediately conclude with (a) by setting $\rho=\pi$.
    Similarly, if $\pi$ is not $C$-safe over $[i,j)$, then we conclude with (b) by setting $\rho=\pi$.
    Therefore, for the remainder of the proof, we assume that $\pi[i,j)$ is not essential and that $\pi$ is $C$-safe over $[i,j)$.

    Let $a \in \set{i, \ldots, j}$ be the minimal index such that $\vec{a}_a^\pi$ is a shiftable midpoint (we know that such an index exists by the assumptions in this claim).
    We will now argue that there exists $b \in I_{\pi[i,j)}$ such that $b > a$.
    Suppose, on the contrary, that there does not exist $b \in I_{\pi[i,j)}$ such that $b > a$.
    In this case, we know that $I_{\pi[i,j)} \cap \set{a+1, \ldots, j} = \emptyset$.
    Thus, for all $s \in \set{a+1, \ldots, j}$, $n_s < N$.
    So, by selecting $b=a$ (or any $b \notin \set{a+1, \ldots, j}$) in the definition of an essential subpath (\cref{def:essential}), one observes
    \begin{equation*}
        \sum_{s \in \set{a+1, \ldots, j}\setminus\set{a}} n_s 
            = \sum_{s \in \set{a+1, \ldots, j}} n_s
            < \sum_{s \in \set{a+1, \ldots, j}} N 
            < N(j-a) \leq Nk \leq N\cdot\size{\Vv}.
    \end{equation*} 
    This implies that $\pi[i,j)$ is essential.
    However, this contradicts an opening assumption that $\pi[i,j)$ is not essential.
    Therefore, there exists $b \in I_{\pi[i,j)}$ such that $b > a$.
    In fact, let us define $b$ to be the minimum index in $I_{\pi[i,j)}$ such that $b > a$.

    Thus far, we know that $n_a, n_b \geq N$ and, for all $s \in \set{a+1, \ldots, b-1}$, $n_s < N$.
    We shall now consider three cases depending on whether or not $\beta_a$ and $\beta_b$ are linearly dependent and whether or not $\vec{a}_b \geq (B,B)$.

    \paragraph*{Case 1.} 
        Suppose $\beta_a$ and $\beta_b$ are linearly dependent. 
        This means that there exists $x_1 \in \ZZ$ and $x_2 \in \NN$ such that
        \begin{equation}\label{eq:case1}
            x_1\cdot\beta_a + (-x_2)\cdot\beta_b = \vec{0}.
        \end{equation}
        As discussed around~\Cref{eq:cramer}, there exists a nonzero solution to~\Cref{eq:case1} such that $0 < \abs{x_1}, x_2 \leq N$.

        The idea is to modify the path $\pi$ by decreasing $n_b$ (by $x_2$).
        Precisely, we shift $\pi$ in $a,b$ by $x_1, -x_2$ to obtain a path $\rho$. 
        The following claim, that is proved in~\cref{app:case1-shifting}, argues that $\run{\vec{s}}{\rho}{\vec{t}}$; following we argue that $\rho[i,j) \prec \pi[i,j)$.

        \begin{claim}\label{clm:case1-shifting}
            In the case that $\beta_a$ and $\beta_b$ are linearly dependent, if $\rho$ is the path that is obtained by shifting $\pi$ at $a,b$ by $x_1, -x_2$, then $\run{\vec{s}}{\rho}{\vec{t}}$.
        \end{claim}

        To conclude Case 1, we show that $\rho[i,j) \prec \pi[i,j)$.
        Recall that $\pi[i,j) = \beta_i^{n_i}\,\alpha_i\,\cdots\,\alpha_{j-1}\,\beta_j^{n_j}$ and suppose that $\rho = \beta_i^{m_i}\,\alpha_i\,\cdots\,\alpha_{j-1}\,\beta_j^{m_j}$.
        Recall also that $x_2 > 0$, and $m_b = n_b - x_2$; we deduce that $m_b < n_b$.
        Additionally, recall that $a < b$ and that $\rho$ was defined to be $\pi$ shifted at $a,b$ by $x_1, -x_2$; we therefore know that $s \in \set{b+1, \ldots, j},\, m_s = n_s$.
        Hence, $\rho[i,j) \prec \pi[i,j)$.

    \paragraph*{Case 2.} Suppose $\beta_a$ and $\beta_b$ are linearly independent and $\vec{a}^\pi_b \geq (B,B)$.

        On top of $b \in I_{\pi[i,j)}$ such that $b > a$, we claim that there exists $c \in I_{\pi[i,j)}$ such that $c > b$.
        \cref{clm:case2-later-cycle} is proved in~\cref{app:case2-later-cycle}.
        \begin{claim}\label{clm:case2-later-cycle}
            There exists $c \in I_{\pi[i,j)}$ such that $c > b$.
        \end{claim}

        In fact, given that such a $c$ exists, let $c \in I_{\pi[i,j)}$ be the minimal index such that $c > b$.
        Now, consider the equation
        \begin{equation}\label{eq:case2}
            x_1\cdot \beta_a + x_2 \cdot \beta_b + (-x_3) \cdot \beta_c = \vec{0}
        \end{equation}
        where $x_1,x_2 \in \ZZ$ and $x_3 \in \NN$.
        Since $\beta_a$, $\beta_b$, and $\beta_c$ are two-dimensional vectors, and the pair of vectors $\beta_a, \beta_b$ are linearly independent, there exists a solution to~\Cref{eq:case2} such that $x_3 > 0$.
        Furthermore, as discussed around~\Cref{eq:cramer}, there exists a solution such that $\abs{x_1}, \abs{x_2}, x_3 \leq N$.

        The idea is to modify the path $\pi$ by decreasing $n_c$ (by $x_3$).
        Precisely, this time, we shift $\pi$ at $a,b,c$ by $x_1,x_2,-x_3$ to obtain a path $\rho$.
        The following claim, that is proved in~\cref{app:case2run}, argues that $\run{\vec{s}}{\rho}{\vec{t}}$; following we argue that $\rho[i,j) \prec \pi[i,j)$.

        \begin{claim}\label{clm:case2run}
            In the case that $\beta_a$ and $\beta_b$ are linearly independent and that $\vec{a}^\pi_b \geq (B,B)$, if $\rho$ is the path that is obtained by shifting $\pi$ at $a,b,c$ by $x_1, x_2, -x_3$, then $\run{\vec{s}}{\rho}{\vec{t}}$.
        \end{claim}

        To conclude Case 2, we show that $\rho[i,j) \prec \pi[i,j)$.
        Recall that $\pi[i,j) = \beta_i^{n_i}\,\alpha_i\,\cdots\,\alpha_{j-1}\,\beta_j^{n_j}$ and suppose that $\rho = \beta_i^{m_i}\,\alpha_i\,\cdots\,\alpha_{j-1}\,\beta_j^{m_j}$.
        Recall also that $x_3 > 0$, and $m_c = n_c - x_3$; we deduce that $m_c < n_c$.
        Additionally, recall that $c > b > a$ and that $\rho$ was defined to be $\pi$ shifted at $a,b,c$ by $x_1, x_2, -x_3$; we therefore know that $s \in \set{c+1, \ldots, j},\, m_s = n_s$.
        Hence, $\rho[i,j) \prec \pi[i,j)$.

    \paragraph*{Case 3.} 
        Suppose $\beta_a$ and $\beta_b$ are linearly independent and $\vec{a}^\pi_b \not\geq (B,B)$.
        Since $\pi$ is assumed to be $C$-safe over $[i,j)$, we know that either $\vec{a}^\pi_b[1] < B$ and $\vec{a}^\pi_b[2] \geq C$ or $\vec{a}^\pi_b[1] \geq C$ and $\vec{a}^\pi_b[2] < B$.
        Without loss of generality, assume that $\vec{a}^\pi_b[1] < B$ and $\vec{a}^\pi_b[2] \geq C$.
        The following claim is proved in~\cref{app:case3-later-cycle}.
        \begin{claim}\label{clm:case3-later-cycle}
            There exists $c \in I_{\pi[i,j)}$ such that $c > b$ and $\beta_c[1] \geq 0$.
        \end{claim}
        
        In fact, given that such a $c$ exists, let $c \in I_{\pi[i,j)}$ be the minimal index such that $c > b$ and $\beta_c[1] \geq 0$.
        Now, like in Case 2, consider the equation
        \begin{equation}\label{eq:case3}
            x_1 \cdot \beta_a + x_2 \cdot \beta_b + (-x_3)\cdot \beta_c = \vec{0},
        \end{equation}
        where $x_1, x_2 \in \ZZ$ and $x_3 \in \NN$.
        Again, since $\beta_a$, $\beta_b$, and $\beta_c$ are two-dimensional vectors, and the pair of vectors $\beta_a, \beta_b$ are linearly independent, there exists a solution to~\Cref{eq:case2} such that $x_3 > 0$.
        Furthermore, as discussed around~\Cref{eq:cramer}, there exists a solution such that $\abs{x_1}, \abs{x_2}, x_3 \leq N$.

        The idea is to modify the path $\pi$ by decreasing $n_c$ (by $x_3$).
        Precisely, this time, we shift $\pi$ at $a,b,c$ by $x_1,x_2,-x_3$ to obtain a path $\rho$.
        The following claim, that is proved in~\cref{app:case3-shifting}, argues that $\run{\vec{s}}{\rho}{\vec{t}}$; following we argue that $\rho[i,j) \prec \pi[i,j)$.

        \begin{claim}\label{clm:case3-shifting}
            In the case that $\beta_a$ and $\beta_b$ are linearly independent and that $\vec{a}^\pi_b \not\geq (B,B)$, if $\rho$ is the path that is obtained by shifting $\pi$ at $a,b,c$ by $x_1, x_2, -x_3$, then $\run{\vec{s}}{\rho}{\vec{t}}$.
        \end{claim}

        To conclude Case 3, we show that $\rho[i,j) \prec \pi[i,j)$.
        Recall that $\pi[i,j) = \beta_i^{n_i}\,\alpha_i\,\cdots\,\alpha_{j-1}\,\beta_j^{n_j}$ and suppose that $\rho = \beta_i^{m_i}\,\alpha_i\,\cdots\,\alpha_{j-1}\,\beta_j^{m_j}$.
        Recall also that $x_3 > 0$, and $m_c = n_c - x_3$; we deduce that $m_c < n_c$.
        Additionally, recall that $c > b > a$ and that $\rho$ was defined to be $\pi$ shifted at $a,b,c$ by $x_1, x_2, -x_3$; we therefore know that $s \in \set{c+1, \ldots, j},\, m_s = n_s$.
        Hence, $\rho[i,j) \prec \pi[i,j)$.

        This concludes the proof of~\cref{lem:iteration} because, in each of the three cases, the path $\rho$ that is obtained from $\pi$ by shifting at two ($a$ and $b$) or three indices ($a$, $b$, and $c$) where $i \leq a,b,s \leq j$. 
        This means that $\rho(0,i-1] = \pi(0,i-1]$, $\rho(j,k] = \pi(j,k]$.
        We proved that in each case, $\run{\vec{s}}{\rho}{\vec{t}}$ and $\rho[i,j) \prec \pi[i,j)$, so we conclude with (c).
\end{proof}

Given that the (sub)paths are well-ordered by $\prec$, the following lemma is a straightforward corollary of~\cref{lem:iteration}.
The proof of~\cref{lem:nobigpoints} can be found in~\cref{app:nobigpoints}. 
Recall that $C = \size{\Vv}^2\cdot(B + (2N+3)\cdot\size{\Vv})$ (see~\cref{lem:iteration}).

\begin{lemma}\label{lem:nobigpoints}
    Let $\pi$ be a path in $\Vv$ such that $\run{\vec{s}}{\pi}{\vec{t}}$, let $i, j \in \set{1, \ldots, k}$ be two indices.
    If there exists $a \in \set{i, \ldots, j}$ such that $\vec{a}_a^\pi$ is a shiftable midpoint, then there exists a path $\rho$ such that $\run{\vec{s}}{\rho}{\vec{t}}$, $\rho(0,i-1] = \pi(0,i-1]$, $\rho(j,k] = \pi(j,k]$, and either
    \begin{enumerate}[(a)]
        \item $\rho[i,j)$ is essential; or
        \item $\rho$ is not $C$-safe over $[i,j)$.
    \end{enumerate}
\end{lemma}

We will need to invoke a result by Englert, Lazi\'{c}, and Totzke~\cite{EnglertLT16} (see also~\cite{BlondinEFGHLMT21}) that, roughly speaking, states that for a given 2-VASS, initial configuration, and target configuration all encoded in unary, if reachability holds, then there is a polynomial-length run from the initial configuration to the target configuration.
We will invoke~\cref{claim:ranko} for runs in the ``unsafe part''; this will be helpful for finding runs between configurations that are bounded above by $(C,C)$.

\begin{corollary}[see {\cite[Theorem~16]{EnglertLT16}} or {\cite[Theorem~3.2 and Corollary~3.3]{BlondinEFGHLMT21}}]\label{claim:ranko}
    Let $C \in \NN$, $\vec{s}, \vec{t} \leq (C,C)$ and suppose $\run{\vec{s}}{*}{\vec{t}}$.
    Then there exists a path $\pi$ such that $\run{\vec{s}}{\pi}{\vec{t}}$; $\length{\pi} = \poly{C+\size{\Vv}}$; and if $\vec{v}$ is a configuration in $\run{\vec{s}}{\pi}{\vec{t}}$, then $\vec{v} \leq (\poly{n+C}, \poly{n+C})$.
    Moreover, one can determine the existence of such a path $\pi$ in polynomial time.
\end{corollary}

Let $\vec{v} \in \Q^2 \setminus\set{\vec{0}}$. 
We say that $\vec{v}$ is \emph{positive} if $\vec{v} > \vec{0}$; \emph{negative} if $\vec{v} < \vec{0}$; and \emph{mixed} otherwise.
Furthermore, let $\vec{u}, \vec{v}$ be two mixed vectors; we say that the pair $\vec{u}, \vec{v}$ are \emph{opposite} if $\vec{u}$ and $\vec{v}$ have swapped signs, precisely if either $\vec{u}[1] < 0$ and $\vec{v}[1] > 0$ or $\vec{u}[1] > 0$ and $\vec{v}[1] < 0$.

We are now ready to state and prove the main theorem of~\cref{sec:polynomial}.
After the formal statement, we provide a supplementary discussion that explains~\cref{thm:decomposition}.

\begin{theorem}\label{theorem:decomposition}\label{thm:decomposition}

    There exists $M \le \poly{\size{\Vv}}$ with the following property.
    Whenever $\run{\vec{s}}{*}{\vec{t}}$, there is a path $\pi = \epath{n_1,\ldots,n_k}$ such that $\run{\vec{s}}{\pi}{\vec{t}}$ and such that:
    
    \begin{enumerate}[label=\textup{(\Alph*)}, ref={(\Alph*)}]

        \item\label{con:a} either there are indices $1 \leq b' < a' < i_1 < \ldots < i_x < p < q < j_1 < \ldots < j_y < a < b \leq k$ such that

        \begin{enumerate}[label=\textup{(\arabic*)},ref={(A\arabic*)}]
            \item\label{subcon:a-not-small} for all $s \in \set{0, 1, \ldots, p-1} \cup \set{q+1, q+2, \ldots, k}$, $\vec{a}^\pi_s \not\leq (M,M)$;
            \item\label{subcon:a-slither} for all $s \in \set{a'+1, a'+2, \ldots, p-1} \cup \set{q+1, q+2, \ldots, a-1}$, $\vec{a}^\pi_s \not \geq (M,M)$;
            \item\label{subcon:a-unary} for all $s \in \set{p, p+1, \ldots, q}$, $\vec{a}^\pi_s \leq (M,M)$;         
            \item\label{subcon:a-opposite} $\beta_{i_1}, \ldots, \beta_{i_x}, \beta_{j_1}, \ldots, \beta_{j_y}$ are mixed, for all $t \in \set{1, 2, \ldots, x-1}$, $\beta_{i_t}$ and $\beta_{i_{t+1}}$ are opposite, and for all $t \in \set{1, 2, \ldots, y-1}$, $\beta_{j_t}$ and $\beta_{j_{t+1}}$ are opposite; and
            \item\label{subcon:a-short} for all $s \in \set{1, 2, \ldots, k}\setminus\set{b', a', i_1, \ldots, i_x, j_1, \ldots, j_y, a, b}$, $n_s < M$;
        \end{enumerate}

        \item\label{con:b} or there are indices $1 \leq f < i_1 < i_2 < \ldots < i_x < a < b\leq k$ such that

        \begin{enumerate}[label=\textup{(\arabic*)},ref={(B\arabic*)}]
            \item\label{subcon:b-not-small} for all $s \in \set{i_1, i_1 + 1, \ldots, k}, \vec{a}^\pi_s \not \leq (M,M)$;
            \item\label{subcon:b-slither} for all $s \in \set{f, f + 1, \ldots, a-1}, \vec{a}^\pi_s \not\geq (M,M)$; 
            \item\label{subcon:b-first-cycle} $\beta_f$ is not positive;
            \item\label{subcon:b-opposite} $\beta_{i_1}, \beta_{i_2}, \ldots, \beta_{i_x}$ are mixed and, for all $t \in \set{1, 2, \ldots, x-1}$, $\beta_{i_t}$ and $\beta_{i_{t+1}}$ are opposite; and
            \item\label{subcon:b-short} for all $s \in \set{1, 2,\ldots,k}\setminus\set{f, i_1, i_2, \ldots, i_x, a, b}$, $n_s < M$.
        \end{enumerate}
    
    \end{enumerate}

\end{theorem}

Before introducing some terms that will be used in the proof of~\cref{thm:decomposition}, we will explain Conditions~\ref{con:a} and~\ref{con:b}, and their respective subconditions, in basic terms. 
Overall, the theorem tells us that whenever there is a run between the initial configuration $\vec{s}$ and the target configuration $\vec{t}$, then there must exist a path $\pi$ that has one of two specific `shapes'. 
Either $\pi$ satisfies Condition~\ref{con:a}, in this scenario the `shape' is illustrated in~\Cref{fig:decomposition-a}, or $\pi$ satisfies Condition~\ref{con:b}, in this scenario the `shape' is illustrated in~\Cref{fig:decomposition-b}.
We will obtain a path $\pi$ that satisfies Condition~\ref{con:a} when, while on a run from $\vec{s}$ to $\vec{t}$, a `small' configuration is observed; specifically, when there is a configuration in which counter values are bounded by $(C, C)$. 
Here, $C$ is the value, polynomial in $\size{\Vv}$, that is defined in~\cref{lem:iteration}.
On the other hand, we obtain a path $\pi$ that satisfies Condition~\ref{con:b} when, while on a run from $\vec{s}$ to $\vec{t}$, such a `small' configuration is not observed.

In order for the path $\pi$ to satisfy Condition~\ref{con:a}, there must exist a certain collection of distinguished indices $(b', a', i_1, \ldots, i_x, p, q, j_1, \ldots, j_y, a, b)$.
These indices are used to specify notable features in the path; some indices indicate cycles that may be iterated greater than a small number of times and some indices indicate where `small' configurations are observed.
Notice that in this case, there are three sections in the run: the prefix until $p$-th midpoint, the infix between the $p$-th midpoint and the $q$-th midpoint, and the suffix from the $q$-th midpoint.
The subconditions are symmetric on the prefix and the suffix; so we shall speak about the infix and the suffix here.

Subcondition~\ref{subcon:a-not-small} specifies that all midpoint configurations in the suffix are `not small'; specifically, at least one coordinate is at least $M$.
In~\Cref{fig:decomposition-a}, this corresponds to midpoints $\vec{a}_{q+1}^\pi, \vec{a}_{j_1}^\pi, \vec{a}_{j_2}^\pi, \vec{a}_{j_3}^\pi, \vec{a}_a^\pi, \vec{a}_b^\pi$ being outside of the pale orange region bounded by $(M,M)$, in the lower-left corner.
Subcondition~\ref{subcon:a-slither} specifies that all midpoint configurations until the $a$-th midpoint are not greater than $(M, M)$; together with Subcondition~\ref{subcon:a-not-small}, this means that the midpoints $\vec{a}_{q+1}^\pi, \vec{a}_{q+1}^\pi, \ldots, \vec{a}_{p-1}^\pi$ have one coordinate that is at least $M$ and one coordinate that is at most $M$.
In~\Cref{fig:decomposition-a}, this corresponds to midpoints $\vec{a}_{q+1}^\pi, \vec{a}_{j_1}^\pi, \vec{a}_{j_2}^\pi, \vec{a}_{j_3}^\pi$ being inside one of the two pale blue regions near each of the axes.
Subcondition~\ref{subcon:a-unary} simply states that all of the midpoints observed during the infix of the run are bounded above by $(M,M)$.
In~\Cref{fig:decomposition-a}, this corresponds to the fact that the black squiggly subrun ($\tau$) remains inside the pale orange region.
Jumping ahead, Subcondition~\ref{subcon:a-short} specifies that all cycles that are not indexed by one of the distinguished indices can only be taken at most a fixed polynomial number of times (at most $M$).
Finally, Subcondition~\ref{subcon:a-opposite} is just used to specify that the cycles $\beta_{j_1}, \beta_{j_2}, \ldots, \beta_{j_y}$, which can be taken more than a fixed polynomial number of times, must alternate between being positive on the first coordinate and negative on the second coordinate and being negative on the first coordinate and positive on the second coordinate; hence the zigzag shape observed in~\Cref{fig:decomposition-a}.

In much the same way, in order for the path $\pi$ to satisfy Condition~\ref{con:b}, there must exist a certain collection of distinguished indices $(f, i_1, \ldots, i_x, a, b)$.
Again, these indices are used to specify notable features in the path; some indices indicate cycles that may be iterated greater than a small number of times and some indices indicate where `small' configurations are observed.
In this case, we shall consider two parts of the run: the first is from the start until the $f$-th midpoint configuration and the second is from after the $f$-th midpoint configuration to the end of the run.

One can notice that second part of the run (after the $f$-th midpoint) must have the same properties as the suffix of a run that satisfies Condition~\ref{con:a}. 
To see why this is the case, align Subcondition~\ref{subcon:b-not-small} with~\ref{subcon:a-not-small}, Subcondition~\ref{subcon:b-slither} with~\ref{subcon:a-slither}, Subcondition~\ref{subcon:b-opposite} with~\ref{subcon:a-opposite}, and Subcondition~\ref{subcon:b-short} with~\ref{subcon:a-short}.
We remark here that, later, we will leverage this commonality between runs satisfying Condition~\ref{con:a} and runs satisfying Condition~\ref{con:b} in the process of obtaining a polynomial time algorithm for deciding reachability in unary 2-SLPS with binary encoded initial and target configurations (see, in particular, Subsection~\ref{subsec:algorithm-preliminaries}).
 
There is, however, a noticeable difference between runs that satisfy Condition~\ref{con:b} (compared to runs that satisfy Condition~\ref{con:a}).
That is the midpoint $\vec{b}^\pi_{f-1}$, that is observed just before the first cycle that may be iterated more than a fixed polynomial number of times ($\beta_f$), may not satisfy $\vec{b}^\pi_{f-1} \not\geq (M,M)$.
In~\Cref{fig:decomposition-a}, this corresponds to the point $\vec{b}^\pi_{f-1}$ that does not belong to either of the two pale blue regions close to the axes.
Fortunately, we are able to obtain the following useful property about $\beta_f$: at least one coordinate of $\beta_f$ is negative (Subcondition~\ref{subcon:b-first-cycle}).
We therefore know that from $\vec{b}_{f-1}^\pi$, after iterating $\beta_f$ (potentially more than a fixed polynomial number of times), the subsequent midpoint reached satisfies $\vec{a}_f^\pi \not\geq (M,M)$.
From here, as previously discussed, we know that the shape of the remainder of the run (after $\vec{a}_f^\pi$) is the same as the shape of the suffix of a run (after $\vec{a}_{q+1}^\pi$) that satisfies Condition~\ref{con:a}.

In the proof of~\cref{thm:decomposition}, we will decompose runs into parts. 
Let $\pi = \epath{n_1, \ldots, n_k}$ such that $\run{\vec{s}}{*}{\vec{t}}$. 
We say that $\pi$ is decomposed into $\pi = \pi_1\,\tau\,\pi_2$ if there exist $p, q \in \set{1, \ldots, k}$ such that $p \leq q$ and
\begin{equation*}
    \vec{s} \xrightarrow{\pi_1} \vec{a}^\pi_p \xrightarrow{\tau} \vec{b}^\pi_{q} \xrightarrow{\pi_2} \vec{t}.
\end{equation*}

One can think that $\pi_1 = \pi(0,p)$, $\tau = \pi(p,q]$, and $\pi_2 = \pi[q+1,k]$.
Furthermore, we say that the decomposition $\pi_1\,\tau\,\pi_2$ is a $C$-\emph{safe} if $\pi$ is $C$-safe over $[1, p)$ and $[q+1, k)$ and $\vec{a}^\pi_p, \vec{b}^\pi_q \leq (C,C)$. 
Such a decomposition is unique for $\pi$ as $p$ and $q$ are the indices of the first and the last midpoints bounded by $(C,C)$, respectively. 
We allow $\tau$ and $\pi_2$ to be empty runs (if there is at most one midpoint bounded above by $(C,C)$). 
We also say that $\pi$ is $C$-\emph{unsafe} over $[p,q)$.

\begin{proof}[Proof of \cref{theorem:decomposition}]
    Recall $C = \size{\Vv}^2\cdot(B + (2N+3)\cdot\size{\Vv})$ from the statement of~\cref{lem:nobigpoints}.
    Let $\pi$ be the path such that $\run{\vec{s}}{\pi}{\vec{t}}$ and such that the interval $[p,q)$, for which $\pi$ is unsafe over, is maximal.
    Consider the $C$-safe decomposition $\pi = \pi_1 \, \tau \, \pi_2$.
    We will split this proof into the cases: one for when $\tau$ is nonempty (here, we will argue that $\pi$ satisfies Condition~\ref{con:a} of this theorem) and one for when $\tau$ is empty (here, we will argue that $\pi$ satisfies Condition~\ref{con:b} of this theorem).

    \paragraph*{Case 1.} $\tau$ is nonempty. 
    See~\Cref{fig:decomposition-a} for a picture sketching the run $\run{\vec{s}}{\pi}{\vec{t}}$ in this case.

    \begin{figure}
        \begin{center}
        \begin{minipage}{.5\textwidth}
            \input{figures/decomposition-a1}
        \end{minipage}\begin{minipage}{.5\textwidth}
            \input{figures/decomposition-a2}
        \end{minipage}
        \end{center}
        \caption{A drawing of the run $\run{\vec{s}}{\pi}{\vec{t}}$ when the path $\pi$ satisfies Condition~\ref{con:a} in~\cref{thm:decomposition}.}
        \label{fig:decomposition-a}
    \end{figure}

    We shall consider $\pi_1 = \pi(0,p)$, $\tau = \pi(p,q]$, and $\pi_2 = \pi[q+1,k]$ separately.
    First, we will consider the segment $\pi_2$.
    Among the midpoints $\vec{a}^\pi_{q+1}, \ldots, \vec{a}^\pi_k$ observed in the run over the segment $\pi_2$, we shall identify the first one that is shiftable.
    Let $a \in \set{q+1, q+2, \ldots, k}$ be the minimal index such that $\vec{a}^\pi_a$ is shiftable; if no such index exists, we set $a = k+1$.
    If there does exists $a \in \set{q+1, q+2, \ldots, k}$ such that $\vec{a}^\pi_a$ is shiftable, we will use~\cref{lem:nobigpoints} (with $i=q+1$ and $j=k$) to obtain a path $\rho$.
    Since $\pi_2 = \pi[q+1,k]$ is $C$-safe, then we know that the path $\rho$ must satisfy (a) of~\cref{lem:nobigpoints}; precisely, $\rho(0,q] = \pi(0,q]$ and $\rho[q+1,k]$ is an essential subpath.
    We will replace the path segment $\pi_2 = \pi[q+1,k]$ with $\rho[q+1,k]$.

    Let $I \coloneqq \set{s \in \set{q+1, \ldots, a-1} : n_s \geq N}$.
    By definition of $a$, for all $s \in I$, we know that $\vec{a}^\pi_s$ is not shiftable; in particular $\vec{a}^\pi_s \not\geq (B,B)$.
    Recall that $\pi$ is $C$-safe over $[q+1,k)$.
    Altogether, for all $s \in I$, either 
    \begin{equation*}
        \vec{a}^\pi_s[1] > C \text{ and } \vec{a}^\pi_s[2] < B  \text{\; or \;} \vec{a}^\pi_s[1] < B \text{ and } \vec{a}^\pi_s[2] > C.
    \end{equation*}
    If $\vec{a}^\pi_s[1] > C$ and $\vec{a}^\pi_s[2] < B$, then we say that $\vec{a}^\pi_s$ is \emph{horizontal}, otherwise if $\vec{a}^\pi_s[1] < B$ and $\vec{a}^\pi_s[2] > C$, then we say that $\vec{a}^\pi_s$ is \emph{vertical}.
    Let $i$ and $j$ be two consecutive indices in $I$; in other words, consider $i, j \in I$ such that $i < j$ and there does not exist $i' \in I$ such that $i < i' < j$.
    We call the pair of vectors $\vec{a}^\pi_i$ and $\vec{a}^\pi_j$ \emph{progressive} if one is horizontal and the other is vertical.
    We will now make some claims about midpoints indexed by $I$; the proof of~\cref{clm:progressive} and~\ref{clm:mixed} can be found in~\cref{app:progressive} and~\ref{app:mixed}, respectively.

    \begin{claim}\label{clm:progressive}
        Let $i, j \in I$ be consecutive indices ($i < j$ and there does not exist $i' \in I$ such that $i < i' < j$).
        If $n_j > B + (N+1)\cdot\size{\Vv}$, then $\vec{a}^\pi_i$ and $\vec{a}^\pi_j$ are progressive.
    \end{claim}

    \begin{claim}\label{clm:mixed}
        Let $j \in I$ and suppose $n_j \geq B + (N+1)\cdot{\size{\Vv}}$.
        \begin{enumerate}[(1)]
            \item Then $\beta_j$ is not a positive vector. 
            \item If, additionally, there exists $i \in I$ such that $i < j$, then $\beta_j$ is not a negative vector.
        \end{enumerate}
    \end{claim}

    Since $\pi[q+1,k]$ is an essential subpath, there exists at most one index $b > a$ such that $n_b > B + 2N\cdot\size{\Vv}^2 + (N+1)\cdot\size{\Vv}$.
    We shall define $J \coloneqq \set{ j_1, \ldots, j_y }$, where $j_1 < \ldots < j_y$ as defined as follows: 
    \begin{equation*}
    j_i = 
        \begin{cases}
            \min(I) 
                & \text{ if } i = 1 \text{ and } \\
            \min\set{s \in I : s > j_{i-1} \text{ and } \vec{a}^\pi_{j_{i-1}},\vec{a}^\pi_s\text{ are progressive} } 
                & \text{ if } i > 1.
        \end{cases}
    \end{equation*}
    \begin{itemize}
        \item For all $i \in \set{1, \ldots, y-1}$, the pair of vectors $\vec{a}^\pi_{j_i}$ and $\vec{a}^\pi_{j_{i+1}}$ are progressive (by definition of $J$).
        \item For all $i \in \set{1, \ldots, y}$, $\vec{a}^\pi_{j_i} \not\geq
            (B, B)$ (because $J \sset I$).
        \item For all $i \in \set{1, \ldots, y-1}$ and for all $s \in \set{ j_i+1, \ldots, j_{i+1} - 1 }$, $n_s < B + (N+1)\cdot\size{\Vv}$ (By~\Cref{clm:progressive} and~\Cref{clm:mixed}).
        \item $\beta_{j_1}, \beta_{j_2}, \ldots, \beta_{j_y}$ are mixed and for all $t \in \set{1, 2, \ldots, y-1}$, $\beta_{i_t}$ and $\beta_{i_{t+1}}$ are opposite.
        \item For all $s \in \set{q+1, \ldots, j_1-1}$, $n_s < N$ (by definition of $I$).
    \end{itemize}
    For the segment $\pi_1$, the analysis is the same as for $\pi_2$, with a tweak that, one needs to consider the automaton in reverse.
    Notice that there is a symmetry between indices $b', a', i_1, \ldots, i_x, p$ and indices $q, j_1, \ldots, j_y, a, b$ in Condition~\ref{con:a}.

    For the segment $\tau$, we will use~\cref{claim:ranko}.
    Let $M = \poly{\size{\Vv} + C}$ be a bound on both the length of the run and on the counter values of the configurations observed in the run between $\vec{a}^\pi_p$ and $\vec{b}^\pi_{q}$.
    From this, we can deduce that not only, for all $s \in \set{p, p+1, \ldots, q}, \vec{a}^\pi_s \leq (M,M)$, but for all $s \in \set{p+1, \ldots, q-1, q}$, $n_s < M$.
    Altogether this yields the desired conditions of Condition~\ref{con:a} in~\cref{thm:decomposition}.

    \paragraph*{Case 2.} $\tau$ is empty. 
    See~\Cref{fig:decomposition-b} for a picture sketching the run $\run{\vec{s}}{\pi}{\vec{t}}$ in this case.

    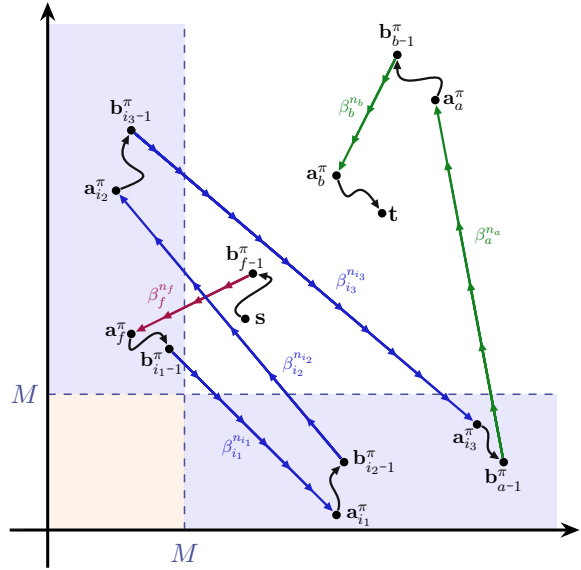
\begin{figure}
        \begin{center}
            \begin{tikzpicture}

	\fill[thinfill, opacity = 0.15] (0,1.8) rectangle (1.8, 6.7);
	\fill[thinfill, opacity = 0.15] (1.8,0) rectangle (6.7, 1.8);
	\fill[smallfill, opacity = 0.15] (0,0) rectangle (1.8, 1.8);
	\draw[thincolour, line width = 0.2mm, dashed] (0, 1.8) -- (6.7, 1.8);
	\draw[thincolour, line width = 0.2mm, dashed] (1.8, 0) -- (1.8, 6.7);
	\node[thincolour] at (1.8, -0.3) {\small $M$};
	\node[thincolour] at (-0.3, 1.8) {\small $M$};

	\node[circle, fill = black, inner sep = 0.4mm] (s) at (2.6, 2.8) {};
	\node[scale=0.8] at (2.8, 2.8) {$\vec{s}$};

	\node[circle, fill = black, inner sep = 0.4mm] (b0) at (2.7, 3.4) {};
	\node[scale=0.7] at (2.6, 3.6) {$\vec{b}^\pi_{f\shortminus1}$};

	\draw[pathcolour, -{Stealth[inset=0pt,length=1.25mm,width=1mm]}, line width = 0.3mm] (s) .. controls (2.2, 3) and (3.2,3.3) .. (b0); 

	\node (a11) at (2.3, 3.2) {};
	\node (a12) at (1.9, 3) {};
	\node (a13) at (1.5, 2.8) {};
	
	\node[circle, fill = black, inner sep = 0.4mm] (a1) at (1.1, 2.6) {};
	\node[scale=0.7] at (0.9, 2.6) {$\vec{a}^\pi_f$};

	\draw[firstcolour, -{Stealth[inset=0pt,length=1.25mm,width=1mm]}, line width = 0.3mm] (b0) -- (a11.center);
	\draw[firstcolour, -{Stealth[inset=0pt,length=1.25mm,width=1mm]}, line width = 0.3mm] (b0) -- (a12.center);
	\draw[firstcolour, -{Stealth[inset=0pt,length=1.25mm,width=1mm]}, line width = 0.3mm] (b0) -- (a13.center);
	\draw[firstcolour, -{Stealth[inset=0pt,length=1.25mm,width=1mm]}, line width = 0.3mm] (b0) -- (a1);
	\node[firstcolour, scale = 0.6] at (1.5, 3.1) {$\beta_f^{n_f}$};

	\node[circle, fill = black, inner sep = 0.4mm] (b1) at (1.6, 2.4) {};
	\node[scale=0.7] at (1.5, 2.2) {$\vec{b}^\pi_{i_1\shortminus1}$};

	\draw[pathcolour, -{Stealth[inset=0pt,length=1.25mm,width=1mm]}, line width = 0.3mm] (a1) .. controls (1, 2.1) and (1.4, 2.8) .. (b1); 

	\node (a21) at (2, 2) {};
	\node (a22) at (2.3, 1.7) {};
	\node (a23) at (2.6, 1.4) {};
	\node (a24) at (2.9, 1.1) {};
	\node (a25) at (3.2, 0.8) {};
	\node (a26) at (3.5, 0.5) {};

	\node[circle, fill = black, inner sep = 0.4mm] (a2) at (3.8, 0.2) {};
	\node[scale=0.7] at (4.1, 0.2) {$\vec{a}^\pi_{i_1}$};

	\draw[zigzagcolour, -{Stealth[inset=0pt,length=1.25mm,width=1mm]}, line width = 0.3mm] (b1) -- (a21.center);
	\draw[zigzagcolour, -{Stealth[inset=0pt,length=1.25mm,width=1mm]}, line width = 0.3mm] (b1) -- (a22.center);
	\draw[zigzagcolour, -{Stealth[inset=0pt,length=1.25mm,width=1mm]}, line width = 0.3mm] (b1) -- (a23.center);
	\draw[zigzagcolour, -{Stealth[inset=0pt,length=1.25mm,width=1mm]}, line width = 0.3mm] (b1) -- (a24.center);
	\draw[zigzagcolour, -{Stealth[inset=0pt,length=1.25mm,width=1mm]}, line width = 0.3mm] (b1) -- (a25.center);
	\draw[zigzagcolour, -{Stealth[inset=0pt,length=1.25mm,width=1mm]}, line width = 0.3mm] (b1) -- (a26.center);
	\draw[zigzagcolour, -{Stealth[inset=0pt,length=1.25mm,width=1mm]}, line width = 0.3mm] (b1) -- (a2);
	\node[zigzagcolour, scale = 0.6] at (2.5, 1.1) {$\beta_{i_1}^{n_{i_1}}$};
	
	\node[circle, fill = black, inner sep = 0.4mm] (b2) at (3.9, 0.9) {};
	\node[scale=0.7] at (4.3, 0.9) {$\vec{b}^\pi_{i_2\shortminus1}$};

	\draw[pathcolour, -{Stealth[inset=0pt,length=1.25mm,width=1mm]}, line width = 0.3mm] (a2) .. controls (3.9, 0.5) and (3.6, 0.6) .. (b2);

	\node (a31) at (3.4, 1.5) {};
	\node (a32) at (2.9, 2.1) {};  
	\node (a33) at (2.4, 2.7) {}; 
	\node (a34) at (1.9, 3.3) {}; 
	\node (a35) at (1.4, 3.9) {}; 

	\node[circle, fill = black, inner sep = 0.4mm] (a3) at (0.9, 4.5) {}; 
	\node[scale=0.7] at (0.65, 4.5) {$\vec{a}^\pi_{i_2}$};

	\draw[zigzagcolour, -{Stealth[inset=0pt,length=1.25mm,width=1mm]}, line width = 0.3mm] (b2) -- (a31.center);
	\draw[zigzagcolour, -{Stealth[inset=0pt,length=1.25mm,width=1mm]}, line width = 0.3mm] (b2) -- (a32.center);
	\draw[zigzagcolour, -{Stealth[inset=0pt,length=1.25mm,width=1mm]}, line width = 0.3mm] (b2) -- (a33.center);
	\draw[zigzagcolour, -{Stealth[inset=0pt,length=1.25mm,width=1mm]}, line width = 0.3mm] (b2) -- (a34.center);
	\draw[zigzagcolour, -{Stealth[inset=0pt,length=1.25mm,width=1mm]}, line width = 0.3mm] (b2) -- (a35.center);
	\draw[zigzagcolour, -{Stealth[inset=0pt,length=1.25mm,width=1mm]}, line width = 0.3mm] (b2) -- (a3);
	\node[zigzagcolour, scale = 0.6] at (3.3, 2.2) {$\beta_{i_2}^{n_{i_2}}$};

	\node[circle, fill = black, inner sep = 0.4mm] (b3) at (	1.1, 5.3) {};
	\node[scale=0.7] at (1.1, 5.55) {$\vec{b}^\pi_{i_3\shortminus1}$};

	\draw[pathcolour, -{Stealth[inset=0pt,length=1.25mm,width=1mm]}, line width = 0.3mm] (a3) .. controls (1.7, 4.8) and (0.8, 4.7) .. (b3);

	\node (a41) at (1.45, 5.0) {};
	\node (a42) at (1.8, 4.7) {};
	\node (a43) at (2.15, 4.4) {};
	\node (a44) at (2.5, 4.1) {};
	\node (a45) at (2.85, 3.8) {};
	\node (a46) at (3.2, 3.5) {};
	\node (a47) at (3.55, 3.2) {};
	\node (a48) at (3.9, 2.9) {};
	\node (a49) at (4.25, 2.6) {};
	\node (a410) at (4.60, 2.3) {};
	\node (a411) at (4.95, 2.0) {};
	\node (a412) at (5.30, 1.7) {};

	\node[circle, fill = black, inner sep = 0.4mm] (a4) at (5.65, 1.4) {};
	\node[scale=0.7] at (5.5, 1.2) {$\vec{a}^\pi_{i_3}$};

	\draw[zigzagcolour, -{Stealth[inset=0pt,length=1.25mm,width=1mm]}, line width = 0.3mm] (b3) -- (a41.center);
	\draw[zigzagcolour, -{Stealth[inset=0pt,length=1.25mm,width=1mm]}, line width = 0.3mm] (b3) -- (a42.center);
	\draw[zigzagcolour, -{Stealth[inset=0pt,length=1.25mm,width=1mm]}, line width = 0.3mm] (b3) -- (a43.center);
	\draw[zigzagcolour, -{Stealth[inset=0pt,length=1.25mm,width=1mm]}, line width = 0.3mm] (b3) -- (a44.center);
	\draw[zigzagcolour, -{Stealth[inset=0pt,length=1.25mm,width=1mm]}, line width = 0.3mm] (b3) -- (a45.center);
	\draw[zigzagcolour, -{Stealth[inset=0pt,length=1.25mm,width=1mm]}, line width = 0.3mm] (b3) -- (a46.center);
	\draw[zigzagcolour, -{Stealth[inset=0pt,length=1.25mm,width=1mm]}, line width = 0.3mm] (b3) -- (a47.center);
	\draw[zigzagcolour, -{Stealth[inset=0pt,length=1.25mm,width=1mm]}, line width = 0.3mm] (b3) -- (a48.center);
	\draw[zigzagcolour, -{Stealth[inset=0pt,length=1.25mm,width=1mm]}, line width = 0.3mm] (b3) -- (a49.center);
	\draw[zigzagcolour, -{Stealth[inset=0pt,length=1.25mm,width=1mm]}, line width = 0.3mm] (b3) -- (a410.center);
	\draw[zigzagcolour, -{Stealth[inset=0pt,length=1.25mm,width=1mm]}, line width = 0.3mm] (b3) -- (a411.center);
	\draw[zigzagcolour, -{Stealth[inset=0pt,length=1.25mm,width=1mm]}, line width = 0.3mm] (b3) -- (a412.center);
	\draw[zigzagcolour, -{Stealth[inset=0pt,length=1.25mm,width=1mm]}, line width = 0.3mm] (b3) -- (a4);
	\node[zigzagcolour, scale = 0.6] at (4, 3.3) {$\beta_{i_3}^{n_{i_3}}$};

	\node[circle, fill = black, inner sep = 0.4mm] (b4) at (6, 0.9) {};
	\node[scale=0.7] at (6, 0.65) {$\vec{b}^\pi_{a\shortminus1}$};

	\draw[pathcolour, -{Stealth[inset=0pt,length=1.25mm,width=1mm]}, line width = 0.3mm] (a4) .. controls (5.9, 1.3) and (5.7, 1.1) .. (b4);

	\node (a51) at (5.85, 1.7) {};
	\node (a52) at (5.7, 2.5) {};
	\node (a53) at (5.55, 3.3) {};
	\node (a54) at (5.4, 4.1) {};
	\node (a55) at (5.25, 4.9) {};
	
	\node[circle, fill = black, inner sep = 0.4mm] (a5) at (5.1, 5.7) {};
	\node[scale=0.7] at (5.35, 5.7) {$\vec{a}^\pi_{a}$};

	\draw[freecolour, -{Stealth[inset=0pt,length=1.25mm,width=1mm]}, line width = 0.3mm] (b4) -- (a51.center);
	\draw[freecolour, -{Stealth[inset=0pt,length=1.25mm,width=1mm]}, line width = 0.3mm] (b4) -- (a52.center);
	\draw[freecolour, -{Stealth[inset=0pt,length=1.25mm,width=1mm]}, line width = 0.3mm] (b4) -- (a53.center);
	\draw[freecolour, -{Stealth[inset=0pt,length=1.25mm,width=1mm]}, line width = 0.3mm] (b4) -- (a54.center);
	\draw[freecolour, -{Stealth[inset=0pt,length=1.25mm,width=1mm]}, line width = 0.3mm] (b4) -- (a55.center);
	\draw[freecolour, -{Stealth[inset=0pt,length=1.25mm,width=1mm]}, line width = 0.3mm] (b4) -- (a5);
	\node[freecolour, scale = 0.6] at (5.8, 3.9) {$\beta_a^{n_a}$};

	\node[circle, fill = black, inner sep = 0.4mm] (b5) at (4.6, 6.3) {};
	\node[scale=0.7] at (4.6, 6.55) {$\vec{b}^\pi_{b\shortminus1}$};

	\draw[pathcolour, -{Stealth[inset=0pt,length=1.25mm,width=1mm]}, line width = 0.3mm] (a5) .. controls (5.4, 6.3) and (4.6, 5.7) .. (b5);

	\node (a61) at (4.4, 5.9) {};
	\node (a62) at (4.2, 5.5) {};
	\node (a63) at (4.0, 5.1) {};

	\node[circle, fill = black, inner sep = 0.4mm] (a6) at (3.8, 4.7) {};
	\node[scale=0.7] at (3.55, 4.7) {$\vec{a}^\pi_{b}$};

	\draw[freecolour, -{Stealth[inset=0pt,length=1.25mm,width=1mm]}, line width = 0.3mm] (b5) -- (a61.center);
	\draw[freecolour, -{Stealth[inset=0pt,length=1.25mm,width=1mm]}, line width = 0.3mm] (b5) -- (a62.center);
	\draw[freecolour, -{Stealth[inset=0pt,length=1.25mm,width=1mm]}, line width = 0.3mm] (b5) -- (a63.center);
	\draw[freecolour, -{Stealth[inset=0pt,length=1.25mm,width=1mm]}, line width = 0.3mm] (b5) -- (a6);
	\node[freecolour, scale = 0.6] at (4, 5.6) {$\beta_b^{n_b}$};

	\node[circle, fill = black, inner sep = 0.4mm] (t) at (4.4, 4.2) {};
	\node[scale=0.8] at (4.55, 4.2) {$\vec{t}$};

	\draw[pathcolour, -{Stealth[inset=0pt,length=1.25mm,width=1mm]}, line width = 0.3mm] (a6) .. controls (4.0, 4.1) and (4.0, 4.8) .. (t);

	%================
	%================
	%================
	% axis
	\draw[-{Stealth[width=2mm, length=2.5mm]}, line width = 0.4mm] (-0.5,0) -- (7,0);
	\node[gray, scale = 0.05] at (-0.4,0) {\textit{Henry Sinclair-Banks}};
	\draw[-{Stealth[width=2mm, length=2.5mm]}, line width = 0.4mm] (0,-0.5) -- (0,7);
\end{tikzpicture}
        \end{center}
        \caption{A drawing of the run $\run{\vec{s}}{\pi}{\vec{t}}$ when the path $\pi$ satisfies Condition~\ref{con:b} in~\cref{thm:decomposition}.}
        \label{fig:decomposition-b}
    \end{figure}

    This case follows from the argument presented for the segment $\pi_2 = \pi[q+1,k]$ in Case 1.
    The only difference is that in Case 1, we could assume that $\vec{b}^\pi_{q} \not\geq (C,C)$; here $\vec{b}^\pi_{q}$ is the initial configuration in the run $\run{\vec{b}^\pi_{q}}{\pi_2}{\vec{t}}$.
    For Case 1, we could therefore deduce that $\beta_{j_1}$, the first cycle taken many times after $\vec{b}^\pi_{q}$, was mixed.
    In this case however, it could be the case that the initial configuration $\vec{s} = \vec{a}_0^\pi \geq (C, C)$. 
    Moreover, it is possible that $\beta_f$, the first cycle taken many times, is negative.
    From~\cref{lem:nobigpoints}, we can conclude that $\vec{a}^\pi_{i_1} \not\geq (C,C)$, and therefore $\vec{a}^\pi_{i_1} \not\geq (M, M)$, and start the analysis from here.

    To this end, we will prove that Subcondition~\ref{subcon:b-first-cycle} is true by contradiction. 
    Let $1 \leq f \leq k$ be the least index such that $n_f \geq M$.
    First, suppose that $\beta_f[1], \beta_f[2] > 0$.
    In this case, we deduce that  $\vec{a}^\pi_f \geq (M,M)$; this immediately contradicts Subcondition~\ref{subcon:b-slither}. 
    Now, suppose that $\beta_f[1] = 0$ and $\beta_f[2] > 0$ (symmetrically $\beta_f[1] > 0$ and $\beta_f[2] = 0$), then the midpoint $\vec{a}_f^\pi$ is shiftable.
    Given that $\tau$ is empty, we know that $\pi$ is $C$-safe over $[1,k)$; in particular, it is $C$-safe over $[f,k)$.
    Therefore, by~\cref{lem:nobigpoints}, it must be true that $\pi[f,k)$ is essential.
    However, recall that $x \geq 1$ and that $n_{i_1}, \ldots, n_{i_x}, n_a, n_b \geq M$.
    So, regardless of the choice of $c \in \set{f+1, \ldots, k}$, we deduce that 
    \begin{align*}
        \sum_{s \in \set{f+1, \ldots, k}\setminus\set{c}} n_s 
        & \geq n_{i_1} + \cdots + n_{i_x} + n_a + n_b  \\
        & \geq n_{i_1} + n_a + n_b 
        \geq 3\cdot M 
        \geq B + 2N\cdot\size{\Vv}^2 + N\cdot\size{\Vv}+\size{\Vv}.
    \end{align*}
    This contradicts the fact that $\pi(f,k]$ is essential.
    Therefore, we deduce that $\beta_f$ is not positive; that is exactly Subcondition~\ref{subcon:b-first-cycle}.
\end{proof}

\newcommand{\dist}[2]{\mathit{dist}(#1, #2)}
\newcommand{\ball}[2]{\mathit{Ball}_{#1}(#2)}
\newcommand{\polyflat}{B'}
\newcommand{\polyimportant}{P_\text{dom}}
\newcommand{\polydec}{N' \cdot n}
\newcommand{\polyexp}{N'}

\section{Polynomial Time Algorithm for Reachability in Unary 2-SLPS\linebreak Between Binary-Encoded Initial and Target Configurations}
\label{sec:algorithm}

Throughout this section, we fix our attention on the 2-SLPS $\Vv = \alpha_0\beta_1^* \alpha_1\ldots \alpha_{n-1}\beta_k^*\alpha_k$ shown in~\Cref{fig:slps} on~\cpageref{fig:slps}.
Just like in~\cref{sec:polynomial}, we will identify $\alpha_i$ and $\beta_i$ with their effects, as well as omit the state when specifying a configuration.
In this section, we will focus on deciding reachability from an initial configuration $\vec{s} \in \NN^2$ to a target configuration $\vec{t} \in \NN^2$.
We encode the initial and target configurations in binary; recall that, for a vector $\vec{v} \in \ZZ^2$, $\bitsize{\vec{v}} = \log_2(\norm{\vec{v}} + 1) + 1$.
Accordingly, in this section, our goal is to obtain an algorithm for reachability that has polynomial running time with respect to $\size{\Vv}$, $\bitsize{\vec{s}}$, and $\bitsize{\vec{t}}$.

\subsection{Preliminary Results and Simpler Reachability Witnesses}
\label{subsec:algorithm-preliminaries}

Before setting up proving that reachability in unary 2-SLPS with binary-encoded initial and target configurations can be decided in polynomial time, we shall provide some preliminary results that will be used later this this section.

\begin{remark}\label{rem:bounds}
    Since reachability can be expressed in existential Presburger arithmetic (see Fribourg and Ols{\'{e}}n~\cite[Section~3]{FribourgOCONCUR97}), and by upper bounds on (minimal) solutions to systems of linear equations over~$\NN$ (see, e.g., Domenjoud~\cite[Section~5]{Domenjoud91}), we can also assume that the counters are exponentially bounded.
    Specifically, all configurations have counter values at  most $2^{c(n + \bitsize{\vec{s}} + \bitsize{\vec{t}})^3}$, for some constant $c$. 
\end{remark}

The following lemma is similar to~\cref{claim:ranko}; the difference is that here, the initial configuration $\vec{s}$ and the target configuration $\vec{t}$ need not be polynomially bounded, but their difference is polynomially bounded.
The proof of~\cref{lem:close-reachability} is in~\cref{app:close-reachability}.

\begin{lemma}\label{lem:close-reachability}
    Let $P \in \NN$ be at most polynomial in $\size{\Vv}$ and let $\vec{s}, \vec{t} \in \NN^2$ be configurations such that $\norm{\vec{t}-\vec{s}} \leq P$.
    Then, in time polynomial in $P+\size{\Vv}$, one can decide whether $\run{\vec{s}}{*}{\vec{t}}$. 
\end{lemma}

Let $M$ be the bound, polynomial in $\size{\Vv}$, from~\cref{thm:decomposition}. 
First, we show that it suffices to consider slightly simpler witnesses that satisfy Condition~\ref{con:a} or Condition~\ref{con:b} in~\cref{thm:decomposition}.

\begin{definition}\label{def:simple-witness}
    We call $\pi = \epath{n_1, \ldots, n_k}$ a \emph{simplified reachability witness} if there exists indices $1 \leq i_1 < \ldots < i_x < a < b \leq k$ such that:
    \begin{enumerate}[label=\textup{(\arabic*)},ref={(\arabic*)}]
        \item\label{subcon:c-not-small} for all $s \in \set{i_1, i_1+1, \ldots, k}$ $\vec{a}^\pi_s \not\leq (M,M)$;
        \item\label{subcon:c-slither} for all $s \in \set{1, \ldots, a-1}, \vec{a}^\pi_s \not\geq (M,M)$; 
        \item $\beta_{i_1}, \ldots, \beta_{i_x}$ are mixed, and for all $t \in \set{1, 2, \ldots, x-1}$, $\beta_{i_t}$ and $\beta_{i_{t+1}}$ are opposite; and
        \item\label{subcon:c-short} for all $s \in \set{1, \ldots, k} \setminus \set{i_1, \ldots, i_x, a, b}, n_s < M$;
    \end{enumerate}
\end{definition}

We note that~\cref{pro:simplify} is important when reasoning (inductively) about the runs.
Namely, if there is a run $\run{\vec{a}_s}{*}{\vec{t}}$ from some midpoint configuration $\vec{a}_s$ (for $s \geq 1$), then we can assume that the path underlying this run satisfies~\cref{def:simple-witness}.

\begin{proposition}\label{pro:simplify}
    In other to verify whether $\run{\vec{s}}{*}{\vec{t}}$ in polynomial time, it suffices to verify whether there exists a simplified reachability witnesses in polynomial time.
    %By~\cref{thm:decomposition}, we can assume that for all $s \in \set{1, \ldots, k}$, $\vec{a}_s \not\leq (M,M)$.
\end{proposition}

\begin{proof}
    By~\cref{thm:decomposition}, if there exists a run $\run{\vec{s}}{*}{\vec{t}}$, then there exists a path $\pi$ such that either $\pi$ satisfies Condition~\ref{con:a} or Condition~\ref{con:b}.
    In both cases, we will argue that the original instance of reachability from $\vec{s}$ to $\vec{t}$ is implied by at most polynomially many instance of reachability from (potentially) new initial configurations $\vec{s}'$ to (potentially) new target configurations $\vec{t}'$ that have simplified witnesses.

    \paragraph{For paths $\boldsymbol{\pi}$ that satisfy Condition~\ref{con:a}.}
    Given Subcondition~\ref{subcon:a-unary}, the configurations in the run $\run{\vec{s}}{\pi}{\vec{t}}$ between indices $p$ and $q$ are bounded above by $(M,M)$.
    Precisely, the configurations in $\run{\vec{a}_p^\pi}{\tau}{\vec{b}_q^\pi}$ lie inside $[0, M]^2$. 
    Since $M$ is polynomial in $\size{\Vv}$, there are polynomially many choices for the pair of configurations $\vec{a}_p^\pi$ and $\vec{b}_q^\pi$.
    To verify whether $\run{\vec{a}_p^\pi}{*}{\vec{b}_q^\pi}$, one can use the non-deterministic log-space\footnote{Recall that \class{NL} $\sset$ \class{P}.} algorithm by Englert, Lazi\'{c} and Totzke~\cite{EnglertLT16}; see~\cref{claim:ranko}.
    From this, we conclude that it suffices only to consider polynomially many instances of reachability: for every $\vec{a}_p^\pi, \vec{b}_q^\pi \leq (M,M)$ such that $\run{\vec{a}_p^\pi}{*}{\vec{b}_q^\pi}$, one needs to determine whether $\run{\vec{s}}{*}{\vec{a}_p^\pi}$ and $\run{\vec{b}_q^\pi}{*}{\vec{t}}$.

    It remains to observe that there are simplified reachability witnesses that imply the instance of reachability $\run{\vec{s}}{*}{\vec{a}_p^\pi}$ and $\run{\vec{b}_q^\pi}{*}{\vec{t}}$.
    The existence of a simplified reachability witness for $\run{\vec{b}_q^\pi}{*}{\vec{t}}$ immediately follows from the fact that $\vec{b}_q^\pi \leq (M,M)$ implies that $\vec{b}_q^\pi \not\geq (M,M)$ and Subconditions ~\ref{subcon:a-not-small},~\ref{subcon:a-slither},~\ref{subcon:a-opposite}, and~\ref{subcon:a-short}.
    As for $\run{\vec{s}}{*}{\vec{a}_p^\pi}$: one needs to consider the automaton in reverse.
    Again, notice that there is a symmetry between indices $b', a', i_1, \ldots, i_x, p$ and indices $q, j_1, \ldots, j_y, a, b$ in Condition~\ref{con:a}.
    Like before, the fact that $\vec{a}_p^\pi \leq (M,M)$ implies that $\vec{a}_p^\pi \not\geq (M,M)$ and Subconditions~\ref{subcon:a-not-small},~\ref{subcon:a-slither},~\ref{subcon:a-opposite}, and~\ref{subcon:a-short} directly imply the existence of a simplified reachability witness for $\run{\vec{a}_p^\pi}{*}{\vec{s}}$ in the reverse automaton (in which the new first and last states are the $p$-th and first  states in the original automaton, respectively).

    \paragraph{For paths $\boldsymbol{\pi}$ that satisfy Condition~\ref{con:b}.}
    In the case when $x = 0$, then there are at most three cycles $\beta_f$, $\beta_a$, and $\beta_b$ that may be taken more than $M$ times.
    Given that $1 \leq f < a < b \leq k$, there are at most $k^3 \leq \size{\Vv}^3$ many combinations of said cycles.
    For each combination, reachability amounts to solving an integer linear program that uses a constant number of variables (see Case 2 in the proof of~\cref{lem:important-sets} for more details).

    In the case when $x \geq 1$, consider the first cycle $\beta_f$ that may be iterated more than $M$ times; there are at most $k$ many choices for $f$.
    Recall that for all $i$, $\beta_i \neq (0,0)$ and we know that $\beta_f$ is not positive given Subcondition~\ref{subcon:b-first-cycle}, so there exists $\iota \in \set{1,2}$ such that $\beta_f[\iota] < 0$.
    Given Subcondition~\ref{subcon:b-slither}, we also know that $\vec{a}_f^\pi \not \geq (M,M)$.
    It remains to argue that there are a polynomial number of midpoints $\vec{a}_f^\pi$ such that $\run{\vec{s}}{*}{\run{\vec{a}_f^\pi}{*}{\vec{t}}}$, there exists a simplified reachability witness for $\run{\vec{a}_f^\pi}{*}{\vec{t}}$, and $\run{\vec{s}}{*}{\vec{a}_f^\pi}$ can be verified in polynomial time.
    
    Given Subcondition~\ref{subcon:b-short}, observe that the effect of the path $\pi(0,f)$ is polynomially bounded; precisely, $\abs{\eff{\pi(0,f)}[z]} \leq M\cdot \size{\Vv} + \size{\Vv}$ for $z \in \set{1, 2}$.
    Therefore, there are at most $2\cdot(M+1)\cdot\size{\Vv}+1)^2$ many points $\vec{b}_{f-1}^\pi$ such that $\run{\vec{s}}{\pi(0,f)}{\vec{b}_{f-1}^\pi}$.
    For each such point $\vec{b}_{f-1}^\pi$, since $\norm{\vec{t} - \vec{b}_{f-1}^\pi} \leq (M+1)\cdot\size{\Vv}$, one can use~\cref{lem:close-reachability} to check whether $\run{\vec{s}}{*}{\vec{b}_{f-1}^\pi}$ in polynomial time.
    Now, for any such point $\vec{b}_{f-1}^\pi$, notice that there are at most $M$ many values $n_f$ such that 
    \begin{equation*}
        \run{\vec{b}_{f-1}^\pi}{\beta_f^{n_f}}{\vec{a}_f^\pi}
    \end{equation*}
    and $\vec{a}_f^\pi \not\geq (M,M)$.
    Accordingly, for each $\vec{b}_{f-1}^\pi$ there are at most $M$ many such points $\vec{a}_f^\pi$.

    By iterating over all such $f$, $\vec{b}_{f-1}^\pi$, and $\vec{a}_f^\pi$, we observe that there is a polynomial number of midpoints $\vec{a}_f^\pi$ such that $\run{\vec{s}}{*}{\run{\vec{a}_f^\pi}{*}{\vec{t}}}$.
    Lastly, it immediately follows from Subconditions~\ref{subcon:b-not-small},~\ref{subcon:b-slither},~\ref{subcon:b-opposite}, and~\ref{subcon:b-short} that $\run{\vec{a}_f^\pi}{*}{\vec{t}}$ has a simplified reachability witness.
\end{proof}

\subsection{Dynamic Programming Tuples and Transitions}
\label{sec:dynamic-programming-tuples}

For the remainder of this section, thanks to~\cref{pro:simplify}, we shall focus on verifying whether $\vec{t}$ can be reached from $\vec{s}$ via a simplified reachability witness.

Since we are attempting to verify whether reachability holds, we will not be given a path a priori, thus we will drop the superscript reference to a path $\pi$ when specifying (midpoint) configurations; so, for example, we write $\vec{a}_s$ instead of $\vec{a}^\pi_s$.
To give more detail to the choice of notation for midpoint configurations, the subscript index of a midpoint configuration represents the state of that configuration.
Precisely, if $\vec{a}_i$ is reachable, then $\vec{a}_i$ is reachable by a run with an underlying path of the form $\alpha_0 \, \beta_1^{n_1}\, \alpha_1\, \cdots \, \alpha_{i-1} \, \beta_i^{n_i}$, or in short $\epath{n_1, \ldots, n_i}$. 
Given $i, j \in \set{1, \ldots, k}$ such that $i < j$, when we consider runs $\run{\vec{a}_i}{*}{\vec{a}_j}$, we are considering runs of the form $\alpha_i \, \beta_{i+1}^{n_{i+1}} \, \alpha_{i+1} \, \cdots \, \alpha_{j-1} \, \beta_j^{n_j}$, in short $\epath{n_{i+1}, \ldots, n_j}$.

Notice that, for the majority of midpoints $\vec{a}_s$ observed along a run whose underlying path is a simplified reachability witness (\cref{def:simple-witness}), it is true that both $\vec{a}_s \not\leq (M,M)$ and\linebreak $\vec{a}_s \not\geq (M,M)$. 
Roughly speaking, these are configurations with one large and one small coordinate. 
To ease the notation, we say that a configuration $\vec{a}_s$ is $M$\emph{-slim} if $\vec{a}_s \not\leq (M,M)$ and $\vec{a}_s \not\geq (M,M)$.

\begin{definition}[Dynamic programming tuple]
    The \emph{configuration tuple} of our dynamic programming algorithm\footnote{In the dynamic programming nomenclature, the word ``state'' typically refers to the current configuration of the dynamic program. However, this word also appears in the nomenclature of vector addition systems with \emph{states}. So, for the sake of clarity, we have instead decided to adopt the word ``tuple'' (for the state of a dynamic programming algorithm).} is a tuple that consists of:
    \begin{enumerate}[(1)]
        \item an index $0 \le i \le k$ of the 2-SLPS $\Vv$;
        \item counter values $\vec{a}_i \in \N^2$ such that if $i > 0$ then $\vec{a}_i$ is $M$-slim; and
        \item two indices $i_1$ and $i_2$ such that either $i < i_1 < i_2 \leq k$, in which case $\beta_{i_1}$ and $\beta_{i_2}$ are opposite, or $i_1 = i_2 = \infty$.
    \end{enumerate}
    We denote such a tuple as $(\vec{a}_i, \beta_{i_1}, \beta_{i_2})$, and if $i_1, i_2 = \infty$, then we denote the tuple as $(\vec{a}_i, \emptyset, \emptyset)$.
\end{definition}

One should think that our dynamic procedure builds, in an online sense, a run where $\vec{a}_i$ is the midpoint configuration reached.
The cycles $\beta_{i_1}, \beta_{i_2}$ (following $\vec{a}_i$) are opposite. 
One should think that other cycles between $i$ and $i_2$ are performed a ``small'' number of times; precisely, $n_j < M$ for $j \in \set{i+1, \ldots, i_1-1}\cup\set{i_1+1, \ldots, i_2-1}$. 
The range of $i_1$ and $i_2$ is extended to $\infty$ (beyond $k$) to consider the case when there does not exist a pair of later occurring cycles that have opposing effects. 
Note that~\cref{def:simple-witness} (in turn from~\cref{thm:decomposition}) allows for two cycles, $\beta_a$ and $\beta_b$, that are iterated many times; so eventually it will be useful (in fact necessary) to consider tuples with $i_1, i_2 = \infty$.

Configuration tuples naturally lead to a definition of transitions between them. 

\begin{definition}[Dynamic programming transition]\label{def:transition}
    In our dynamic program, we say that a tuple $S_1 = (\vec{a}_i, \beta_{i_1}, \beta_{i_2})$ traverses to a tuple $S_2 = (\vec{a}_{j}, \beta_{j_1}, \beta_{j_2})$ if $i < j$ and there exists a run $\run{\vec{a}_i}{\pi}{\vec{a}_j}$, where $\pi = \epath{n_{i+1}, \ldots, n_j}$ is a path with one of the two following properties.
    \begin{enumerate}[(1)]
        \item If $i < i_1 < i_2 \leq j$, then 
        \begin{enumerate}[(a)]
            \item $n_{i_1}, n_{i_2} \geq M$; for all $s \in \set{i+1,\ldots,j} \setminus \set{i_1,i_2}$, $n_s < M$; and
            \item for every $\iota \in \set{1,2}$, $\beta_{i_1}[\iota] > 0$ if and only if for all $s \in \set{i, i+1,\ldots, i_1 - 1} \cup \set{i_2, i_2 + 1, \ldots j}$, $n_s < M$ and $\beta_{i_2}[\iota] > 0$ if and only if for all $s \in \set{i_1, i_1 +1, \ldots, i_2-1}$, $n_s < M$. 
        \end{enumerate}
        \item Otherwise, if $i_1 = i_2 = j_1 = j_2 = \infty$, then $n_s < M$ for all $s \in \set{i+1,\ldots,j}$.
    \end{enumerate}
    We write $\run{S_1}{}{S_2}$ to denote this transition.
\end{definition}

Roughly speaking, subproperty (1b) states that the midpoint configurations $\vec{a}_{i_1}$ and $\vec{a}_{i_2}$ must be opposite.
In other words, $\vec{a}_{i_1}$ and $\vec{a}_{i_2}$ swap from horizontal to vertical, or from vertical to horizontal.

We say that a tuple $T$ is \emph{reachable} from a tuple $S$, denoted $\run{S}{*}{T}$, if there exists a sequence of tuples connected by a sequence of transitions: $S \xrightarrow{} S_1 \xrightarrow{} \ldots \xrightarrow{} S_n \xrightarrow{} T$ (for some $n \geq 0$).
Furthermore, we say that $S_1$ \emph{dominates} $S_2$ if every configuration tuple reachable from $S_2$ is also reachable from $S_1$.
Intuitively speaking, the dynamic programming algorithm will need to forget dominated tuples in order to be efficient.

Transitions between tuples require runs of specific shape that exploit the opposite pair of cycles $\beta_{i_1}$ and $\beta_{i_2}$.
In the following lemma, that is proved in~\cref{app:anyrun}, will show that, actually, the existence of any run suffices for one tuple to transition to another.

\begin{lemma}\label{lem:anyrun}
    Let $S = (\vec{a}_i, \beta_{i_1}, \beta_{i_2})$ be a tuple such that $i < i_1 < i_2 \leq k$, and let $\vec{a}_j \in \NN^2$ be a configuration. 
    If $\run{\vec{a}_i}{*}{\vec{a}_j}$, then there exist tuples $S' = (\vec{a}_i, \beta_{i_1'}, \beta_{i_2'})$ and $T = (\vec{a}_j, \beta_{j_1}, \beta_{j_2})$ such that $\run{S'}{*}{T}$.
\end{lemma}

We say that a set of configuration tuples $\Ss$ is \emph{complete} if $(\vec{a}_i, \emptyset, \emptyset) \in \Ss$ and, for every tuple $(\vec{a}_i, \beta_{i_1}, \beta_{i_2}) \in \Ss$ and every pair of opposite cycles $\beta_{j_1}, \beta_{j_2}$ such that $i_1 < i_2 < j_1 < j_2$, then $(\vec{a}_i,\beta_{j_1}, \beta_{j_2}) \in \Ss$.

The next bound can be used to assert that we need only maintain a complete set of configuration tuples of polynomial size.

\begin{lemma}\label{lem:bound-on-configurations}
    Let $\Ss$ be a complete set of configuration tuples such that if $(\vec{a}_i, \beta_{i_1}, \beta_{i_2}) \in \Ss$ such that $\beta_{i_1}$, $\beta_{i_2}$ are opposite ($i < i_1 < i_2 \leq k$) and such that $\beta_{i_1} \neq c \cdot \beta_{i_2}$ for every $c \in \QQ$ ($\beta_{i_1}$ and $\beta_{i_2}$ are not parallel).
    There exists $\polyimportant$, polynomially bounded in $\size{\Vv}$, such that if $\abs{\Ss} > \polyimportant$, then there exist $S_1, S_2 \in \Ss$ such that $S_1$ dominates $S_2$. Moreover, such a pair of tuples can be found in polynomial time.
\end{lemma}
\begin{proof}
    Assume $\abs{\Ss} > \polyimportant$. We will start with a series of pigeonhole arguments to be able to reason just about vectors.
    
    Let $\Ss(i,i_1,i_2) \subseteq \Ss$ be the subset of configuration tuples of the form $(\vec{a}_i, \beta_{i_1}, \beta_{i_2})$. 
    There are only $k^3 \leq \size{\Vv}^3$ possible choices of $i$, $i_1$, and $i_2$, so there exists a subset such that $\abs{\Ss(i, i_1, i_2)} \ge \frac{\polyimportant}{\size{\Vv}^3}$.
    Since all other parts of the tuple are now fixed, let $C \subseteq \NN^2$ be the set of counter values of tuples in $\Ss(i,\beta_{i_1},\beta_{i_2})$. 
    In other words, $\vec{a}_i \in C$ if and only if $(\vec{a}_i, \beta_{i_1}, \beta_{i_2}) \in \Ss(i,\beta_{i_1},\beta_{i_2})$.

    For $\iota \in \set{1, 2}$, let $C_\iota \subseteq C$ such that every $\vec{v} \in C_\iota$ has $\vec{v}[\iota] \leq M$. 
    Recall that, by assumption, all configurations are $M$-slim, so $C = C_1 \cup C_2$. 
    Without loss of generality, assume that $\abs{C_2} \geq \abs{C_1}$ (otherwise the following argument is symmetric).
    Let $Y = C_2$.
    We get $\size{Y} > \frac{\polyimportant}{2\size{\Vv}^3}$.
    For every $x \in [0, M]$, let
    \begin{displaymath}
        Y_x \coloneqq \set{ y : (x, y) \in Y }. 
    \end{displaymath}
    By the pigeonhole principle, there exists $x \in [0, M]$ such that $\abs{Y_x} > \frac{\polyimportant}{2\size{\Vv}^3\cdot(M+1)}$. 

    For this particular value $x$, we will find a pair of tuples of the dynamic programming where one dominates the other. 
    Recall that $\beta_{i_1} \neq c \cdot \beta_{i_2}$ for every $c \in \QQ$.
    Let $h$ be an integer with the minimal absolute value such that $x_1\cdot\beta_{i_1} + x_2\cdot\beta_{i_2} = (h, 0)$ for some nonzero $x_1, x_2 \in \NN$.
    We argue that such an $h$ exists and that $h$, $x_1$ and $x_2$ are all polynomially bounded.
     
    Given that $\beta_{i_1}$ and $\beta_{i_2}$ are opposite, we know that both $\beta_{i_1}[1] \cdot \beta_{i_1}[2] < 0$ and $\beta_{i_2}[1] \cdot \beta_{i_2}[2] < 0$. 
    We define $x_1 = \abs{\beta_{i_2}[2]}$ and $x_2 = \abs{\beta_{i_1}[2]}$ and observe that
    \begin{align*}
        \beta_{i_1}^{x_1} + \beta_{i_2}^{x_2} 
        & = (\abs{\beta_{i_2}[2]}\cdot\beta_{i_1}[1] + \abs{\beta_{i_1}[2]} \cdot \beta_{i_2}[1], \abs{\beta_{i_2}[2]}\cdot\beta_{i_1}[2] + \abs{\beta_{i_1}[2]}\cdot\beta_{i_2}[2]) \\
        & = (\abs{\beta_{i_2}[2]}\cdot\beta_{i_1}[1] + \abs{\beta_{i_1}[2]} \cdot \beta_{i_2}[1], 0).
    \end{align*} 
    Hence, $h = \abs{\beta_{i_2}[2]}\cdot\beta_{i_1}[1] + \abs{\beta_{i_1}[2]} \cdot \beta_{i_2}[1]$ and observe that $\abs{h} \leq 2\cdot\size{\Vv}^2$ and $\abs{x_1}, \abs{x_2} \leq \size{\Vv}$.  
    By the assumptions of this lemma, we know that $\abs{h} > 0$, otherwise there exists a rational number $c$ such that $\beta_{i_1} = c \cdot \beta_{i_2}$. Without loss of generality, we assume that $h > 0$, otherwise the argument is symmetric.
 
    We stratify the values in $Y_x$ based on their value modulo $h$; for every $s \in [0, h-1]$, let
    \begin{displaymath}
        Z_r \coloneqq \set{ y \in Y_x : y \equiv r \bmod h}.
    \end{displaymath}
    Now, let $r \in \set{0, 1, \ldots, h-1}$ be the index for which $\abs{Z_r}$ is the largest. 
    By the pigeonhole principle, we know that $\abs{Z_r} \geq \abs{Y_x} / h$.

    The proof of~\cref{clm:short} and~\ref{clm:long} can be found in~\cref{app:short} and~\ref{app:long}, respectively. 
    \Cref{clm:short} and its proof are accompanied by~\Cref{fig:clm:short} on~\cpageref{fig:clm:short}.

    \begin{claim}\label{clm:short}
        If there exist $u, v \in Z_r$ such that 
        \begin{equation}\label{eq:dominated}
            2M\cdot\size{\Vv}^3 + M \leq v < u < v\left(1 + \frac{1}{2\cdot\size{\Vv}^2}\right) - M\cdot\size{\Vv}^2,
        \end{equation}
        then there exist indices $j_1, j_2$ such that $((r,v)_i, \beta_{i_1}, \beta_{i_2}) \in \Ss$ is dominated by $((r,u)_i, \beta_{i_1'}, \beta_{i_2'}) \in \Ss$, for some $\beta_{i_1'}$, $\beta_{i_2'}$.
    \end{claim}

    \begin{claim}\label{clm:long}
        There exist $P_1, P_2$, polynomially bounded in $\size{\Vv}$, such that if, for every $u, v \in Z_r$ such that $u > v \geq P_1$ we have $u \ge v (1 + \frac{1}{2\cdot\size{\Vv}^2}) - M\cdot\size{\Vv}^2$, then $\abs{Z_r} \leq P_2 + P_1$.
    \end{claim}

    Now, we can conclude the proof of~\cref{lem:bound-on-configurations}.
    Observe that the number of tuples that do not satisfy conditions of~\cref{clm:short} and~\cref{clm:long} is polynomial in $\size{\Vv}$.
    Now, given the lower bound on the size of $Z_r$ and~\cref{clm:long}, we determine (via~\cref{clm:short}) that there exists a pair of tuples $S_1, S_2 \in \Ss$ such that $S_1$ dominates $S_2$.
    
    Now, in order to find a pair of tuples $S_1, S_2 \in \Ss$, in polynomial time, such that $S_1$ dominates $S_2$, it suffices to compute:
    \begin{enumerate}[(1)]     
        \item $i, i_1, i_2 \in \set{1, \ldots, k}$ such that $\abs{\Ss(i,i_1,i_2)} \geq \frac{\polyimportant}{\size{\Vv}^3}$; 
        \item $\iota \in \set{1,2}$ such that $\abs{C_z} \geq \frac{\polyimportant}{2\size{\Vv}^3}$;
        \item $x \in [0,M]$ such that $\abs{Y_x} > \frac{\polyimportant}{2\size{\Vv}^3\cdot(M+1)}$;
        \item $r \in \set{0, 1, \ldots, h-1}$ such that $\abs{Z_r} \geq \abs{Y_x}/h$; and
        \item $u, v \in Z_r$ that satisfy~\Cref{eq:dominated}.
    \end{enumerate}
    We note that the computations (1) -- (4) are simple counting exercises that are polynomial in $\abs{\Ss}$ and computation (5) can be achieved by enumerating all such $u$ and $v$ in $Z_r$ and checking the three inequalities specified in~\Cref{eq:dominated}.
    Finally, by~\cref{clm:short} and by enumerating over all $1 \leq i_1' < i_2' \leq k$, we know that $S_1 = ((r,u)_i, \beta_{i_1'}, \beta_{i_2'})$ dominates $S_2 = ((r,v)_i, \beta_{i_1}, \beta_{i_2})$.
\end{proof}

\subsection{Polynomial Time Algorithm} \label{sec:alg}
\begin{Thm3}
    Reachability in unary 2-SLPS can be decided in polynomial time even if the initial and target configurations are encoded in binary.
\end{Thm3}

\newcommand{\erun}[2]{\run{#1}{\textup{ess}}{#2}} 

Recall that we are focussing on an instance of reachability in $\Vv$ from an initial configuration $\vec{s}$ to a target configuration $\vec{t}$.
Before we prove~\cref{thm:mainalg}, we will introduce some notation.
Let us denote the initial configuration as $\vec{a}_0 = \vec{s}$ and the target configuration as $\vec{a}_{k+1} = \vec{t}$ (to emphasise, with the indices, that these configurations are at the first and last state of the 2-SLPS, respectively).
We denote $\Ss_0 = \set{(\vec{a}_{0}, \beta_{i_1}, \beta_{i_2}) : 1 \leq i_1 < i_2 \leq k \text{ or } i_1 = i_2 = \infty}$ and $T = (\vec{a}_{k+1}, \emptyset, \emptyset)$.
We note that $\Ss_0$ contains all of the tuples with index $0$; and $T$ is the only possible tuple with index $k+1$.

We will also allow for an additional, special, kind of transition: $\erun{(\vec{a}_i, \emptyset, \emptyset)}{T}$ if there exists a run $\rho = \epath{n_{i+1}, \ldots, n_k}$ such that there are three indices $i < i_1 < i_2 < i_3 \leq k$ where $n_s < M$ for all $s \in \set{i+1, \ldots, k} \setminus \set{i_1, i_2, i_3}$. 
In other words, at most three loops (after $\vec{a}_i$ is reached) are taken at least $M$ times. 
Note that this transition can only be applied once and at the end. 
Let $\Ss$ be a configuration tuple; we extend $\run{S}{*}{T}$ to mean that either there exist $S_1, \ldots, S_n$ (for some $n \geq 0$) such that $S \xrightarrow{} S_1 \xrightarrow{} \ldots \xrightarrow{} S_n \xrightarrow{} T$ or $S \xrightarrow{} S_1 \xrightarrow{} \ldots \xrightarrow{} \erun{S_n}{T}$.
We will also extend the notion of transitions to sets. 
Let $\Ss$ be a set of configuration tuples, and let $T$ be a configuration tuple.
We write $\run{\Ss}{}{T}$ if there exists $S \in \Ss$ such that $\run{S}{}{T}$; similarly, we write $\run{\Ss}{*}{T}$ if there exists $S \in \Ss$ such that $\run{S}{*}{T}$.
The following is a simple corollary of~\cref{pro:simplify} and, more specifically, of the runs considered in this section (\cref{def:simple-witness}).

\begin{corollary}\label{cor:runs}
    In $\Vv$, $\run{\vec{s}}{*}{\vec{t}}$ if and only if $\run{\Ss_0}{*}{T}$.
\end{corollary}

Let $\Ss$ be a complete set of configuration tuples. 
Let $\Ss_i \subseteq \Ss$ be all of the configuration tuples of the form $(\vec{a}_i, \beta_{i_1}, \beta_{i_2})$; in other words, $\Ss_i$ contains all the tuples at index $i$.
We say that $\Ss_i$ is \emph{important} if the following two conditions are satisfied:
\begin{enumerate}[(1)]
    \item For every $S = (\vec{a}_i, \beta_{i_1}, \beta_{i_2})$ such that $\run{\Ss_0}{*}{\run{S}{*}{T}}$, there exists $S' \in \Ss_i$ such that $\run{\Ss_0}{*}{\run{S'}{*}{T}}$. 
 In other words, if there is a run witnessing reachability through a configuration tuple at index $i$, then there is a run witnessing reachability through a tuple at index $i$ that belongs to $\Ss_i$.
    \item $\Ss_0 \trans{} S_i$ for every $S \in \Ss$. In other words, we restrict to reachable tuples.
\end{enumerate}

It will also be convenient to introduce the notion of distance between pairs of configuration tuples. 
In short, it is a natural distance between the two vectors of counter values in the tuples (and all other components in the tuples are ignored).
Let $S_1 = (\vec{a}_i, \beta_{i_1}, \beta_{i_2})$ and $S_2 = (\vec{a}_j, \beta_{j_1}, \beta_{j_2})$ be two configuration tuples.
The \emph{distance} between them is defined to be $\dist{S_1}{S_2} \coloneqq \norm{\vec{a}_j-\vec{a}_i}$.
Let $\Tt$ be the set of all dynamic programming tuples; for a set of tuples $\Ss$ and a \emph{radius} $R \in \NN$ we write
\begin{equation*}
    \ball{R}{\Ss} \coloneqq \set{ S_1 \in \Tt : \text{ there exists } S_2 \in \Ss \text{ such that } \dist{S_1}{S_2} \leq R }.
\end{equation*}

\begin{figure}[ht!]
    \centering
    \includegraphics[width=0.8\textwidth]{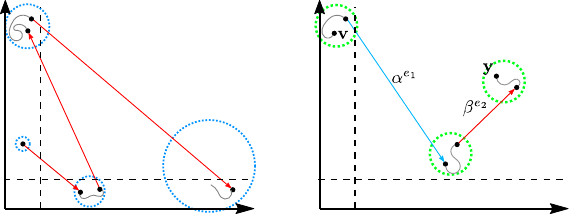}
    \caption{
        This figure presents the two main cases considered in the proof of~\cref{lem:important-sets}. 
        On the left, for the first case, we have sketched one of the runs; here we will keep track of tuples close to the axis. 
        We argue, for the sake of the running time, that all considered configurations inside balls (highlighted blue) whose radius increases only additively by a polynomial factor after every zig and zag, which is when a cycle is taken many times. 
        On the right, for the second case, we know that the run may take three cycles $\alpha,\beta$ and $\gamma$ a large number of times (on the picture we have $\gamma = \emptyset$ for visibility, but keep in mind that there could be three cycles). 
        In that case, we argue that paths in between are short (hence the green highlighted balls), so that we can guess them efficiently. 
        Then, to find the number of iterations of these cycles we can encode it as an integer linear program with a constant number of variables.}
    \label{fig:alg}
\end{figure}

We will utilise the following subprocedure in the efficient algorithm for~\cref{thm:mainalg}.
For a brief sketch of the intuition behind~\Cref{lem:important-sets}, see~\Cref{fig:alg}; for the full proof, see~\cref{app:important-sets}.

\begin{lemma}\label{lem:important-sets}
    Fix $1 \le i \le k+1$ and suppose we have $\Ss_0, \ldots,\Ss_{i-1}$ complete and important set of tuples.
    Let $\Ss_i' = \set{S : \run{\Ss_j}{}{S} \text{ for some } j < i}$. 
    In polynomial time, one can compute a complete and important subset of tuples $\Ss_i \subseteq \Ss_i'$.
    Moreover, there exists $R$, polynomially bounded in $\size{\Vv}$, such that 
    \begin{equation*}%\label{eq:ball}
        \left|\Ss_i' \setminus \ball{R}{\Ss_i}\right| \leq \polyimportant.
    \end{equation*}
\end{lemma}

We can now conclude this section with a proof that reachability in unary 2-SLPS can be decided in polynomial time, even when the initial and target configurations are encoded in binary.

\begin{proof}[Proof of~\cref{thm:mainalg}]
    By \cref{lem:important-sets}, we can compute consecutive sets of tuples $\Ss_i$ that are both complete and important.
    Therefore, $\Ss_{k+1}$ can be computed in polynomial time.
    In the end, it suffices to check whether $\Ss_{k+1}$ is empty or not.
    If $\Ss_{k+1}$ is not empty, then $\Ss_{k+1} = \set{T}$ and we know that $\run{\Ss_0}{*}{T}$.
    By~\cref{cor:runs}, this means that indeed $\run{\vec{s}}{*}{\vec{t}}$.
\end{proof}

% !TEX root = ../main.tex
\section{Further research}
\label{sec:conclusion}

We think that the following questions are interesting open problems worth further research.

Theorem~\ref{thm:unitary-hardness} shows that reachability is \class{NP}-hard for SLPS already in an ``almost fixed'' dimension.
We conjecture that even a stronger version of this statement is true.

\begin{conjecture}\label{conj:unitary-lps}
There exists a number $d \in \N$ such that the reachability problem for unitary $d$-dimensional SLPS is \class{NP}-hard.
\end{conjecture}

In the above conjecture the term ``unitary'' can be equivalently changed into ``fixed-updates'',
as by Lemma~\ref{lem:fixed-updates} changing a unary SLPS with fixed updates to a unitary SLPS results only
in multiplication of the dimension by a constant. This change, however, makes a difference if we fix the dimension.
Therefore, the following is an interesting open problem. We deliberately chose to ask it for SLPS with constant updates instead of for unitary SLPS, as we believe that for fixed dimension this is a more robust class of VASS.

\begin{problem}\label{problem:fixed-3-slps}
Is reachability in $3$-dimensional SLPS with constant updates in \class{P}?
\end{problem}

Also, to extend our upper bound, it is natural to ask whether this result extends from unary 2-SLPS to general unary 2-VASS.
We believe that the main challenge lies in first extending Theorem~\ref{thm:mainalg} to unary flat 2-VASS.

\begin{problem}\label{problem:2-vass}
Is reachability in unary 2-VASS with initial and target configurations encoded in binary in \class{P}?
\end{problem}

\subsubsection*{Acknowledgements}
We would like to thank Georg Zetzsche for helping us prove part of~\cref{lem:close-reachability}.

\bibliographystyle{plainurl}
\bibliography{bib}

\begin{appendix}
    \section{Further Details of~\cref{sec:preliminaries}}
\label{app:prelims}
\subsection{A Simpler Proof of Hardness of Reachability in Unary Flat 3-VASS}
\label{app:flat-3vass}

Here, we present a relatively simple argument for why reachability in flat unary 3-VASS is \class{NP}-hard.
Indeed, simple linear path schemes are a restricted variant of flat VASS, so this is subsumed by~\cref{thm:3-lps-hardness}.
Nevertheless, we outline the proof as this paper marks the conclusion of the line of research concerning the complexity of reachability in flat VASS.
The main high-level difference between linear path schemes and flat VASS is that non-deterministic branching is not a feature of linear path schemes; this simpler argument leverages non-deterministic branching.
Here, we focus our attention to a reduction from 3-CNF SAT. 
Consider an instance $\phi$ with $n$ Boolean variables $x_1, \ldots, x_n$ and $m$ clauses $C_1, \ldots, C_m$.
We will create an instance of reachability in unary flat 2-VASS with zero tests that do not belong to the cycles.
\cref{lem:controlling-counter} can then be used to simulate polynomially many zero tests with just one additional counter instead; giving us an instance of reachability in unary flat 3-VASS that is equivalent to the satisfiability of $\phi$.
We call the primary counter $\var{x}$ and its ancillary counter $\var{y}$.

We associate, to the $j$-th clause $C_j$, the $j$-th prime $p_j$.
For each variable $x_i$, let $P_i$ be the set of indices of clauses containing the positive literal $x_i$ and $N_i$ be the set of indices of clauses containing the negative literal $\overline{x}_i$.
On the counter $\var{x}$, we shall ``guess'' a value corresponding to an assignment in a nonstandard way. 
Initially $\var{x} = 1$.
Now, suppose $x_i$ is set to true, we multiply the value of $\var{x}$ by all of the primes $p_j$ such that $j \in P_i$.
Similarly, if $x_i$ is set to false, then we will multiply the value of $\var{x}$ by all of the primes $p_j$ such that $j \in N_i$.
Each multiplication of $\var{x}$ by a number $n \in \NN$ is realised by first transferring the value of $\var{x}$ to $\var{y}$
by a loop
\begin{algorithmic}[1]
	\State \LOOP \dec{x}{1}, \inc{y}{1}
\end{algorithmic}
then zero testing counter $\var{x}$ to check whether the value was fully transferred. Next, the value is transferred back from $\var{y}$ to $\var{x}$ but multiplied by $n$, namely
\begin{algorithmic}[1]
	\State \LOOP \dec{y}{1}, \inc{x}{n}
\end{algorithmic}
and finally counter $\var{y}$ is zero tested to verify whether also this transfer was fully realised.
To control the guessed assignment, we will have a series of non-deterministic branches, one for each variable.
Note that clause $C_j$ is satisfied if $p_j$ divides $\var{x}$.

Therefore, to check the satisfiability, we can divide $\var{x}$ by every prime $p_1, \ldots, p_m$. Division is implemented analogously to multiplication by a few zero tests. To summarise, the satisfiability of $\phi$ can be reduced to the instance of reachability in unary, flat 2-VASS with polynomially many zero tests.
Thanks to the prime number theorem, we know that the $m$-th prime is at most $\Oh(m\log(m))$, so this instance of reachability in a unary 2-VASS with zero tests can be constructed in polynomial time.

\subsection{The Controlling Counter Technique}
\label{app:controlling-counter}

\paragraph*{Context.} 
In essence, the \emph{controlling counter technique} of Czerwi\'{n}ski and Orlikowski~\cite{CzerwinskiO21} is a technique for simulating a fixed number of zero tests in a VASS. The technique uses just one additional counter; the magnitude of the updates that the counter receives is proportional to the number of zero tests that need to be simulated. Therefore, a polynomial size unary $(d+1)$-VASS can be used to simulate a unary $d$-VASS that uses a polynomial number of zero tests.

To this end, the original lemma supporting the controlling counter technique~{\cite[Lemma 10]{CzerwinskiO21}} offers a ``hindsight'' view for verifying that zero tests were faithfully simulated. 
That is, so long as a run in a given $(d+1)$-VASS has some distinguished properties, then one can guarantee at certain points in the run, some of the first $d$ counters must have had value zero.

However, we instead offer a ``foresight'' view.
If a given $d$-VASS uses a fixed and small number of zero tests, then there exists a $(d+1)$-VASS that can implicitly simulate the zero tests.
This view is more amenable; for an instance of reachability in a $d$-VASS with a small number of zero tests, we can produce an equivalent instance of reachability in a $(d+1)$-VASS (without zero tests).

Our assumptions and notation deviate somewhat from~{\cite[Lemma 10]{CzerwinskiO21}}.
Namely, we assume that a transition in a $d$-VASS can only zero test one counter at a time.
Such an assumption can be made without loss of generality: a transition with more than one zero test can be split into at most $d$ transitions that zero test just one counter each.
We also show that our construction preserves, in some way, the structure of the graph underlying the VASS; we can apply the controlling counter technique to linear path schemes with zero tests and obtain a linear path scheme.

\paragraph*{Intuition.}
Before we present the construction of $\Vv$ and provide a full proof, we will explain the core idea behind the controlling counter techniques via a basic example. 
This example has been presented similarly before~{\cite[Example 9]{CzerwinskiO21}}.

Consider a scenario in which we need to perform exactly three zero tests on a given counter $\var{x}$ in a run $\rho$. 
We shall assume, for simplicity, that the initial value of $\var{x}$ is zero.
We shall cut the run into four about the three moments in which $\var{x}$ is zero-tested.
Call these runs $\rho_1$, $\rho_2$, $\rho_3$ and $\rho_4$; so $\rho = \rho_1\,\rho_2\,\rho_3\,\rho_4$.
Let us denote by $x_i \in \NN$ the effect of $\rho_i$ on $\var{x}$, for each $i \in \set{1, 2, 3, 4}$.
The first zero test asserts that $x_1 = 0$, the second asserts $x_1 + x_2 = 0$, and the third asserts $x_1 + x_2 + x_3 = 0$. 
Given the semantics of VASS, we also know that $\var{x} \geq 0$ at all times, so $x_1, x_1 + x_2, x_1 + x_2 + x_3 \geq 0$ must hold (it is possible for $x_2$ and $x_3$ to be negative).
Thus, to check whether $x_1$, $x_1 + x_2$, and $x_1 + x_2 + x_3$ are equal to zero, it suffices to check whether their sum equals zero. 
In other words, we only need to check whether $3x_1 + 2x_2 + x_3 = 0$. 

To actually perform this check, we add one additional counter $\var{c}$, which we call \emph{the controlling counter}.
The role of $\var{c}$ is to maintain the value $3x_1 + 2x_2 + x_3$, so an instance of reachability can assert that at the end of a run $\var{c} = 3x_1 + 2x_2 + x_3 = 0$.
Initially, set $\var{c}$ to have zero value.
In the first part of the run $\rho_1$, $\var{c}$ receives updates that are three times the updates that $\var{x}$ receives; after $\rho_1$ we know that $\var{c} = 3x_1$ holds. 
Similarly, in the second part of the run $\rho_2$, $\var{c}$ receives updates that are two times the updates that $\var{x}$ receives; after $\rho_2$, we know that $\var{c} = 3x_1 + 2x_2$ holds.
Then, in the third part of the run $\rho_3$, $\var{c}$ receives updates that are equal to the updates that $\var{x}$ receives; after $\rho_3$, we know that $\var{c}= 3x_1 + 2x_2 + x_3$ holds.
For the remainder of the run after $\rho_3$, so throughout $\rho_4$, $\var{c}$ is not updated.
Therefore, we can just add one counter to simulate the three zero tests on $\var{x}$, so long as $\var{c} = 0$ is asserted in the target configuration of the reachability instance.

Observe that this technique can be generalised to handle $m$ zero tests. 
In the section of the run in which $\var{x}$ still has $m$ many zero tests to be performed, the controlling counter $\var{c}$ receives updates that are $m$ times the updates that $\var{x}$ receives. 
Furthermore, this technique can be extended to handle zero tests over multiple counters; the values are just independently added to $c$, and $c$ is checked for being zero at the end of the run. 
Concretely speaking, in a $d$-VASS (with counters $\var{x}_1, \ldots, \var{x}_d$), the controlling counter $\var{c}$ can be added to handle zero tests on all of the $d$ counters.
Suppose at some moment in a run through the VASS, $\var{x}_i$ still has $m_i$ many zero tests to be performed, for each $i \in \set{1, \ldots, d}$.
Suppose also that the following transition updates $\var{x}_i$ by $u_i$, for each $i \in \set{1, \ldots, d}$.
Then $\var{c}$ should be updated by $a_1 \cdot m_1 + a_2 \cdot m_2 + \ldots + a_d \cdot m_d$, or in other words
by $(a_1, \ldots, a_d) \cdot (m_1, \ldots, m_d)$.

\begin{proof}[Proof of~\cref{lem:controlling-counter}]
    We create $\Vv$ by taking $(m+1)^d$ copies of $\Zz$ with all zero test transitions removed.
    Namely, for each vector $\vec{a} \in [0,m]^d$ we create a copy $\Zz_\vec{a}$ of $\Zz$.
    Additionally, we create a new initial state $s'$.
    For the ease of notation, for each $\vec{a}$, the state $q_\vec{a}$ in $\Zz_\vec{a}$ is the copy of the original state $q$ in $\Zz$.
    We also add a $(d+1)$-st counter to $\Vv$. 
    For the instance of reachability, the initial configuration is $\config{s'}{\vec{0}}$ and the target configuration is $\config{t_\vec{0}}{y'}$, where $\vec{y}'[i] = \vec{y}[i]$ for all $i \in [1, d]$ and $\vec{y}'[d+1] = 0$. 
    Notice that $t_\vec{0}$ is the target state $t$ of $\Zz$ in $\Zz_\vec{0}$, that is the copy of $\Zz$ corresponding to vector $\vec{0} = (0, \ldots, 0) \in [0,m]^d$.

    As the intuition before this proof indicated, the first $d$ counters of $\Vv$ just mimic the original counters in $\Zz$ and the $(d+1)$-st counter in $\Vv$ takes care of the faithful simulation of all of the zero tests.
    Precisely, we ensure that a run from $\config{s'}{\vec{0}}$ to $\config{t_0}{y'}$ in $\Vv$ corresponds to a run from $\config{s}{x}$ to $\config{t}{y}$ in $\Zz$ where all of the zero tests were correct.
    It is realised in such a way that the final value of the $(d+1)$-st counter equals the combined sum of all the values of the counters at each moment when they were zero-tested. 
    Thus, $\vec{y'}[d+1]$ is set to be $0$, with the aim that then all the zero-tested counters were indeed zero at the moment when they were supposed to have zero value.
    
    To implement this idea, we need to track the number of zero tests remaining on each counter, the updates that the $(d+1)$-st counter receives depends on these values. 
    It is for this reason why we have the $(m+1)^d$ many copies of $\Zz$. 
    The copy $\Zz_\vec{a}$ corresponds to a situation in which exactly $\vec{a}[1]$ zero tests remain on the first counter, exactly $\vec{a}[2]$ zero tests remain on the second counter, and so on.
    Precisely and in general, exactly $\vec{a}[i]$ zero tests will be performed on the $i$-th counter, for each $i \in [1,d]$.

    To fully define $\Vv$, we must specify its transitions; there are three kinds of transitions.
    The first kind of transitions are between the copies of $\Zz$, they correspond to the zero-testing transitions of $\Zz$.
    Suppose there is a transition in $\Zz$ from state $p$, to state $q$, and that zero tests the $i$-th counter.
    Then, there will be several transitions in $\Vv$, one from $p_\vec{a}$ to $q_\vec{b}$ for every $\vec{a}, \vec{b} \in [0, m]^d$ such that $\vec{a} - \vec{e}_i = \vec{b}$ (recall that $\vec{e}_i$ is the $i$-th standard basis vector).
    Note that these transitions do not have zero tests because $\Vv$ indeed does not have zero tests.
    Instead of zero tests, we will have transitions in $\Vv$ that update the $(d+1)$-st counter depending on the number of zero tests remaining.
    
    The second kind of transitions in $\Vv$ are transitions inside the copies, they correspond to the transitions of $\Zz$ that do not zero test any counters.
    Suppose $(p, \vec{u}, q)$ is a transition in $\Zz$. 
    Then there is a transition $(p_\vec{a}, \vec{u}_\vec{a}, q_\vec{a})$ in $\Vv$ for each $\vec{a} \in [0,m]^d$ in $\Zz$.
    The first $d$ counters of $\Vv$ receive the same updates at the $d$ counters in $\Zz$, so $\vec{u}_\vec{a}[i] = \vec{u}[i]$ for each $i \in [1, d]$.
    However, the update given the $(d+1)$-st counter depends on a linear combination of the updates the other $d$ counters receive times the number of zero tests remaining on each of those $d$ counters.
    \begin{equation*}
    \vec{u}_\vec{a}[d+1] = \vec{a}[1]\cdot\vec{u}[1] + \ldots + \vec{a}[d]\cdot\vec{u}[d] = \vec{a}\cdot\vec{u}
    \end{equation*}
    Where, to be clear, $\vec{u}[i]$ is the update of this transition on the $i$-th counter and $\vec{a}[i]$ is the number of zero tests remaining on the $i$-th counter.

    The third and final kind of transitions in $\Vv$ are initialisation transitions that are outgoing from the new initial state $s'$.
    For each $\vec{a} \in [0,m]^d$ we add the transition $(s', \vec{x}_\vec{a}, s_\vec{a})$ where $\vec{x}_\vec{a}[i] = \vec{x}[i]$ for all $i \in [1, d]$, and $\vec{x}_\vec{a}[d+1] = \vec{a}\cdot\vec{x}$.
    Intuitively, one can view these transitions as ``guessing'' how many zero tests will be performed on each counter by transitioning to the initial state of the corresponding copy $\Zz_\vec{a}$, by adding the initial counter values to the first $d$ counters, and by updating the additional counter according to the initial values.

    Now we show that indeed $\config{s}{x} \reach \config{t}{y}$ in $\Zz$ if and only if $\config{s'}{\vec{0}} \reach \config{t_\vec{0}}{y'}$ in $\Vv$. 
    We start with the only if direction.

    Suppose $\run{\config{s}{x}}{\pi}{\config{t}{y}}$ in $\Zz$.
    We will assume, for each $i \in [1, d]$, that $\pi$ zero-tests the $i$-th counter $a_i$ many times.
    Let $\vec{a} = (a_1, a_2, \ldots, a_d) \in \N^d$.
    Consider a run $\rho$ in $\Vv$ that first uses the transition from the initial state $s'$ to the state $s_\vec{a}$.
    In other words, $\rho$ correctly guesses the correct copy $\Zz_\vec{a}$ of $\Zz$ for the run to start from. 
    Then $\rho$ follows $\pi$. 
    The only difference being that when $\pi$ takes a transition with a zero test, $\rho$ instead takes the corresponding transition between copies of $\Zz$.
    Since the first $d$ counters in $\Vv$ mimic the $d$ counters in $\Zz$, we only need to check that reachability holds for the $(d+1)$-st counter.

    To achieve this, consider any non-empty prefix $\rho'$ of $\rho$, and suppose $\run{\config{s}{0}}{\rho'}{\config{p_\vec{a}}{v}}$.
    We will argue that the following invariant holds:
    \begin{equation}\label{eq:invariant}
        \vec{v}[d+1] = \vec{a}\cdot\vec{v}[1;d],
    \end{equation}
    where by $\vec{v}[1;d]$ we denote the projection of $\vec{v} \in \NN^{d+1}$ onto the first $d$ components.
    In other words, let $\vec{v} \in \NN^{d+1}$, then we define $\vec{v}[1;d] \coloneqq (\vec{v}[1], \ldots, \vec{v}[d])$.
    Note that if the invariant holds, then the $(d+1)$-st counter will remain nonnegative.

    If the invariant holds, then consider the situation when all the zero tests have been performed (that is when $\rho' = \rho$).
    The run $\rho$ is in the last copy $\Zz_\vec{0}$; so $\vec{a}[i] = 0$ for all $i \in [1,d]$. 
    The configuration reached is $\vec{y}'$ since the first $d$ counters have values equal to $\vec{y}$ and the value of the $(d+1)$-st counter is $\vec{0}\cdot\vec{y} = 0$.
    Indeed, this means $\config{s'}{\vec{0}} \reach \config{t_\vec{0}}{y'}$ in $\Vv$, as required.

    We show that~\Cref{eq:invariant} holds by induction on the length of the run $\rho'$ leading to $\config{p_\vec{a}}{v}$.
    As $\rho'$ is non-empty, we shall consider $\rho'$ of length one.
    The first transition of $\rho'$ was set to be the transition $s = (s', \vec{x}_\vec{a}, s_\vec{a})$.
    In $\Vv$, $\run{\config{s'}{0}}{s}{\config{s}{x_a}}$, and we know that $\vec{x_a}[d+1] = \vec{a}\cdot\vec{x} = \vec{a}\cdot\vec{x_a}[1;d]$, which satisfies the invariant.

    Now, for the inductive step, suppose that $\rho'$ leads to $\run{\config{s}{0}}{\rho'}{\config{p_\vec{a}}{v}}$ and that $\config{p_\vec{a}}{v}$ satisfies the invariant~\Cref{eq:invariant}.
    Suppose that the next transition is $t$ and let us denote $\config{q_\vec{b}}{w}$ to be the next configuration that is reached; so we know that $\run{\config{q_\vec{a}}{v}}{t}{\config{q_\vec{b}}{w}}$.
    We need to show that $\config{q_\vec{b}}{w}$ also satisfies the invariant~\Cref{eq:invariant}.

    There are two cases: either $t$ is a transition that remains inside the copy $\Zz_\vec{a}$, or $t$ is a transition between copies $\Zz_{\vec{a}}$ and $\Zz_\vec{b}$, where $\vec{b} \neq \vec{a}$.
    In the first case, $t$ corresponds to a transition that does not zero test a counter in the original VASS $\Zz$, and in the second case, $t$ corresponds to a transition that does zero test a counter in the original VASS $\Zz$.

    In the first case, we know that as $t$ does not leave $\Zz_\vec{a}$, it is true that $\vec{b} = \vec{a}$.
    Let $\vec{u}$ be the effect of the transition; by definition, we know that 
    \begin{equation*}
        \vec{u}[d+1] = \vec{a} \cdot \vec{u}[1;d].
    \end{equation*}
    Observe that $\vec{w} = \vec{v} + \vec{u}$; the invariant~\Cref{eq:invariant} is satisfied.
    \begin{align*}
        \vec{w}[d+1] 
            = \vec{v}[d+1] + \vec{u}[d+1] 
            & = \vec{a}\cdot\vec{v}[1;d] + \vec{a}\cdot\vec{u}[1;d] \\
            & = \vec{a}\cdot(\vec{v}[1;d] + \vec{u}[1;d]) 
            = \vec{a}\cdot\vec{w}[1;d]
    \end{align*}

    In the second case, we know that $t$ transitions from $\Zz_\vec{a}$ to $\Zz_\vec{b}$ for $\vec{b} \neq \vec{a}$.
    We use $\vec{e}_j$ to denote the $j$-th standard basis vector in $\NN^d$.
    Suppose $\vec{b} = \vec{a} - \vec{e}_j$, for some $j \in [1,d]$.
    This means that $t$ corresponds to a transition that zero-tested the $j$-th counter in $\Zz$.
    Therefore, in the configuration $\config{p_\vec{a}}{v}$, we know that $\vec{v}[j] = 0$ and this implies $\vec{e}_j\cdot\vec{v}[1;d] = \vec{0}$.
    Since this transition zero-tests a counter, its effect is $\vec{0}$. 
    In particular, we know that $\vec{w} = \vec{v}$ holds. 
    We can now show that the invariant~\eqref{eq:invariant} holds for $\config{q_\vec{b}}{w}$:
    \begin{align*}
        \vec{w}[d+1] = \vec{v}[d+1]
        = \vec{a}\cdot\vec{v}[1;d]
        & = \vec{a}\cdot\vec{v}[1;d] - \vec{e}_j\cdot\vec{v}[1;d] \\
        & = (\vec{a} - \vec{e}_j)\cdot\vec{v}[1;d]
        = \vec{b}\cdot\vec{v}[1;d]
        = \vec{b}\cdot\vec{w}[1;d].
    \end{align*}

    For the converse, suppose $\run{\config{s'}{0}}{\rho}{\config{t_\vec{0}}{y'}}$ in $\Vv$.
    Let $\vec{a} \in [0, m]^d$ be the vector such that $\config{s'}{0}\xrightarrow{}\config{s_\vec{a}}{\vec{x}}$ is the configuration observed after the first transition in the run.
    After the initial configuration in the separate initial state, the run immediately starts from the copy $\Zz_\vec{v}$ and will traverse through various copies of $\Zz$ before finally reaching $\Zz_\vec{0}$.
    Consider the sequence of copies visited in $\rho$: $(\Zz_{\vec{a}_0}, \Zz_{\vec{a}_1}, \ldots, \Zz_{\vec{a}_k})$, where $k \in \NN$, $\vec{a}_0 = \vec{a}$, $\vec{a}_k = \vec{0}$.
    
    For this direction, we need to argue that $\config{s}{\vec{x}} \reach \config{t}{\vec{y}}$ in $\Zz$; we will show that the ``projection'' of run $\rho$ from $\Vv$ onto just first $d$ counters (ignoring the last counter) suffices. 
    Precisely, we shall construct a run $\pi$, from $\rho$, such that $\run{\config{s}{\vec{x}}}{\pi}{\config{t}{\vec{y}}}$ in $\Zz$.
    Whenever $\rho$ uses a transition inside a copy $\Zz_\vec{a}$, then the original transition is taken and whenever $\rho$ uses a transition between copies $\Zz_\vec{a}$ and $\Zz_\vec{b}$, then the transition with the corresponding zero test is taken.
    Indeed, all the transitions of $\rho$ that are inside a copy $\Zz_\vec{a}$ have the same effect on the first $d$ components, so $\pi$ captures their behaviour in a straightforward way.
    The only challenge is to show that when $\rho$ uses a transition between two copies, then, the corresponding counter that is supposed to be zero-tested is actually equal to zero so that the run $\pi$ is valid.
    For a deeper intuition, see~{\cite[Section 4]{CzerwinskiO21}}.

    We define the function $f:[1, k] \to [1, d]$ where $f(j) = i$ indicates that the $i$-th counter is subject to the $j$-th zero test. 
    In other words, $\vec{a}_{j} = \vec{a}_{j-1} - \vec{e}_{f(j)}$ holds for all $j\in [1,k]$.
    Formally, let $p_{j-1}(\vec{w}_{j-1})$ is last configuration observed in the copy $\Zz_{\vec{a}_{j-1}}$, for every $j \in [1,k]$.
    It remains to show, for each $j \in [1, k]$, that $\vec{w}_{j-1}[f(j)] = 0$ holds.

    Below we see how the $(d+1)$-th counter changes along the run $\rho$ and derive that the final value $\vec{y'}[d+1]$ is the sum of all the counter values at the moment when each zero test occurred in $\pi$.
    As we demand $\vec{y'}[d+1] = 0$ in the target configuration; it means
    that all the zero tested counters in $\pi$ were indeed zero.

    The run $\rho$ traverses through copies $\Zz_{\vec{a}_0}, \Zz_{\vec{a}_1}, \ldots, \Zz_{\vec{a}_k}$.
    Suppose that $\vec{v}_j \in \NN^{d+1}$ are the counter values at the first configuration in $\Zz_{\vec{a}_j}$ and $\vec{w}_j \in \NN^{d+1}$ are the counter values at the last configuration in $\Zz_{\vec{a}_j}$.
    Note that in the first copy $\vec{v}_0 = \vec{x}'$ and in the last copy $\vec{w}_k = \vec{y'}$.
    Since the transition between copies do not update the counters, we know that $\vec{w}_{j-1} = \vec{v}_j$ for each $j \in [1, k]$.
    Furthermore, by construction of $\Vv$, notice that in the copy $\Zz_{\vec{a}_j}$, we have that
    \begin{equation}\label{eq:one-copy}
        \vec{w}[d+1] - \vec{v}[d+1] = \vec{a}_j \cdot (\vec{w}[1;d] - \vec{v}[1;d]).
    \end{equation}

    By repeatedly applying~\Cref{eq:one-copy} to the segments of the run in each of the copies, we get that
    \begin{enumerate}
        \item[] $\vec{w}_0[d+1] - \vec{x}'[d+1] = \vec{a}_0\cdot(\vec{w}_0[1;d] - \vec{x}'[1;d])$,
        \item[] $\vec{w}_j[d+1] - \vec{v}_j[d+1] = \vec{a}_j\cdot(\vec{w}_j[1;d] - \vec{v}_j[1;d])$ for all $j \in [1, k-1]$, and
        \item[] $\vec{y}'[d+1] - \vec{v}_k[d+1] = \vec{a}_k\cdot(\vec{y}'[1;d] - \vec{v}_k[1;d])$.
    \end{enumerate}
    Given the overlap $\vec{w}_{j} = \vec{v}_{j+1}$, we know that 
    \begin{enumerate}
        \item[] $\vec{v}_1[d+1] - \vec{x}'[d+1] = \vec{a}_0\cdot(\vec{v}_1[1;d] - \vec{x}'[1;d])$,
        \item[] $\vec{v}_{j+1}[d+1] - \vec{v}_j[d+1] = \vec{a}_j\cdot(\vec{v}_{j+1}[1;d] - \vec{v}_j[1;d])$ for all $j \in [1, k-1]$, and
        \item[] $\vec{y}'[d+1] - \vec{v}_k[d+1] = \vec{a}_k\cdot(\vec{y}'[1;d] - \vec{v}_k[1;d])$.
    \end{enumerate}
    Summing all the above $k+1$ equations we arrive at
    \begin{equation}\label{eq:run-difference}
        \vec{y'}[d+1] - \vec{x}'[d+1] = 
        \vec{a}_k \cdot \vec{y'}[1;d] 
        - \left( 
        {\sum\limits_{j=0}^{k-1} (\vec{a}_{j+1}-\vec{a}_j) \cdot \vec{v}_{j+1}[1;d]}
        \right)
        - \vec{a}_0 \cdot \vec{x}'[1;d].
    \end{equation}
    Recall that by definition, $\vec{x}'[d+1] = \vec{a}_0\cdot\vec{x}'[1;d]$,
    $\vec{a}_k = \vec{0}$, and $\vec{y}'[d+1] = 0$.
    Then~\eqref{eq:run-difference} simplifies to
    \begin{equation*}
        \sum_{j=0}^{k-1} (\vec{a}_{j+1}-\vec{a}_j) \cdot \vec{v}_{j+1}[1;d] = 0.
    \end{equation*}
    Now, since $\vec{a}_{j+1} = \vec{a}_j - \vec{e}_{f(j+1)}$ we derive 
    \begin{equation*}
        \sum_{j=0}^{k-1} \vec{e}_{f(j+1)}\cdot\vec{v}_{j+1}[1;d]
        = \sum_{j=0}^{k-1} \vec{v}_{j+1}[f(j+1)]
        = \sum_{j=1}^{k} \vec{v}_j[f(j)]
        = 0.
    \end{equation*}
    By the above, and given that $\vec{v}_j \geq \vec{0}$, we conclude that, for each $j \in [1,k]$, that $\vec{v}_j[f(j)] = 0$ holds; this is exactly what was required.
    We can therefore conclude that $\config{s}{x} \reach \config{t}{y}$ in $\Zz$ if and only if $\config{s'}{\vec{0}} \reach \config{t_\vec{0}}{y'}$ in $\Vv$.

    Notice now that the construction of $\Vv$ preserves many properties of $\Zz$.
    First of all, if $\Zz$ is flat then, obviously, all the copies of $\Zz$ in $\Vv$ are also flat. 
    As we do not add any cycles between the copies, we know that $\Vv$ is flat as well.
    If $\Zz$ is a (simple) linear path scheme in which the zero-testing transitions do not lie on cycles, then not only is the number of zero tests fixed for each run in $\Zz$, but the exact order of which counters are zero-tested is fixed.
    In $\Vv$, we only leave the copies from the sequence of copies, which can actually be visited by a run.
    Therefore, there is only one sequence of copies of $\Zz$ that form $\Vv$.
    We now prune $\Vv$ so that we can guarantee it is a (simple) linear path scheme.
    All we need to do is remove any state in $\Vv$ which cannot be visited by any accepting run; this can be achieved by just performing a breadth first search on the underlying graph of $\Vv$.    
      
    Finally, observe that, even with pruning, $\Vv$ has size $\Oh((\size{\Zz} + \norm{\vec{x}}) \cdot (m+1)^d)$ and can be computed in that time.
    This follows directly from the construction.
    Indeed, there are $(m+1)^d$ copies of $\Zz$ and for each copy there are at most $\size{\Zz}$ many transitions in $\Vv$.
    There are an additional at most $\size{\Zz}$ many transitions on top of this: at most one from the one new initial state $s'$ to each of the copies.
\end{proof}

\section{NP-Hardness of Reachability in Unary Ultraflat 4-VASS}
\label{sec:4ultraflat}
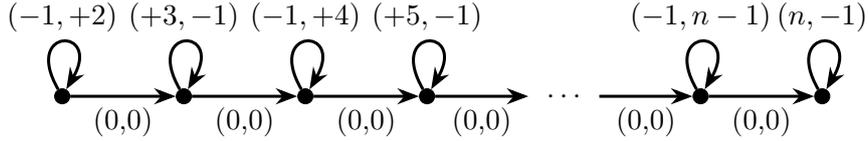
\begin{figure*}
	\centering
	\begin{tikzpicture}[scale = 0.8]
	\node[circle, fill = black, draw, inner sep = 0.7mm, minimum size = 1.4mm] (q1) at (0, 0) {};
	\node[circle, fill = black, draw, inner sep = 0.7mm, minimum size = 1.4mm] (q2) at (2, 0) {};
	\node[circle, fill = black, draw, inner sep = 0.7mm, minimum size = 1.4mm] (q3) at (4, 0) {};
	\node[circle, fill = black, draw, inner sep = 0.7mm, minimum size = 1.4mm] (q4) at (6, 0) {};
		\node (q5) at (8.25,0) {$\;\cdots\;$};
	\node[circle, fill = black, draw, inner sep = 0.7mm, minimum size = 1.4mm] (q6) at (10.5, 0) {};
	\node[circle, fill = black, draw, inner sep = 0.7mm, minimum size = 1.4mm] (q7) at (12.5, 0) {};
	
	\path[-Stealth, line width = 0.4mm] (q1) edge node[below] {(0,0)} (q2);
	\path[-Stealth, line width = 0.4mm] (q2) edge node[below] {(0,0)} (q3);
	\path[-Stealth, line width = 0.4mm] (q3) edge node[below] {(0,0)} (q4);
	\path[-Stealth, line width = 0.4mm] (q4) edge node[below] {(0,0)} (q5);
	\path[-Stealth, line width = 0.4mm] (q5) edge node[below] {(0,0)} (q6);
	\path[-Stealth, line width = 0.4mm] (q6) edge node[below] {(0,0)} (q7);

	\path[-Stealth, line width = 0.4mm, out = 120, in = 60, distance = 12mm] 
		(q1) edge node[above] {$(-1, +2)$} (q1);
	\path[-Stealth, line width = 0.4mm, out = 120, in = 60, distance = 12mm] 
		(q2) edge node[above] {$(+3, -1)$} (q2);
	\path[-Stealth, line width = 0.4mm, out = 120, in = 60, distance = 12mm] 
		(q3) edge node[above] {$(-1, +4)$} (q3);
	\path[-Stealth, line width = 0.4mm, out = 120, in = 60, distance = 12mm] 
		(q4) edge node[above] {$(+5, -1)$} (q4);
	\path[-Stealth, line width = 0.4mm, out = 120, in = 60, distance = 12mm] 
		(q6) edge node[above] {$(-1, n-1)$} (q6);
	\path[-Stealth, line width = 0.4mm, out = 120, in = 60, distance = 12mm] 
		(q7) edge node[above] {$(n, -1)$} (q7);			
\end{tikzpicture}
	\caption{
		An example ultraflat 2-VASS for ``weakly computing'' $n!$. 
		Suppose the initial counter values are set to $(1,0)$, then by maximally iterating all of the cycles, the final state can be reached with counter values $(n!, 0)$.
		This ultraflat VASS cannot be used to compute $n!$ since this configuration is not forced to be reached; instead, it is possible to just iterate the third cycle once and not iterate any other cycles and reach $(0, 4)$.
		The zero effect transitions between cycles make ultraflat VASS a particularly simple model of computation.
	}
        \vskip1ex
        \hrule
	\label{fig:ultraflat-example}
\end{figure*}

Ultraflat VASS are a further restricted kind of flat VASS.
They are simple linear path schemes in which the transitions between the cycles have zero effect. 
In other words, the counters can only be updated on the self-loops.
Ultraflat VASS were introduced by Leroux in~\cite{Leroux21} and make for another exceedingly weak model of computation, see an example in~\cref{fig:ultraflat-example}.
An \emph{ultraflat} $d$-VASS is a $d$-VASS $\Vv = (Q, T)$ such that $Q = \set{ q_1, \ldots, q_n }$ with $n = \size{Q}$, and for some vectors $\vec{a}_1, \ldots, \vec{a}_n \in \ZZ^d$,
\begin{equation*}
	T = \set{ (q_{j-1}, \vec{0}, q_j  : 2 \leq j \leq n } \cup \set{ (q_j, \vec{a}_j, q_j) : 1 \leq j \leq n }.
\end{equation*}

We answer an open problem posed by Leroux~{\cite[Section 4]{Leroux21}} by showing that reachability is \class{NP}-hard in 4-dimensional ultraflat unary VASS.

\begin{theorem}\label{thm:ultraflat-hardness}
	Reachability in unary ultraflat 4-VASS is \class{NP}-complete.
\end{theorem}

To achieve this, we extend the techniques and tricks developed for unary 3-SLPS.
The lower bound, again, comes from a reduction from SAT for which we reuse~\cref{lem:sat-encoding} (see the first part of~\cref{sec:hardness} for an overview of the reduction).
There are two challenges to overcome.
First, we need an ultraflat non-divisibility gadget.
Second, unlike the scenarios with unary and unitary SLPS, we need to modify the one-counter program \satprogram{\phi} (\cref{fig:lps-sat}) to make it ultraflat; Lines 4, 6, 8, and 10 are non-zero updates on transitions between cycles.

Our approach for non-divisibility assertions again requires some zero tests.
In~\cref{fig:ultraflat-non-div}, we detail an ultraflat unary 3-VASS with zero tests for non-divisibility (note that this implementation is needed since the unary 2-SLPS with zero tests for non-divisibility in~\cref{fig:unary-non-div} is not ultraflat).
We are, however, able to reuse the controlling counter technique (\cref{lem:controlling-counter}) for ultraflat VASS because the updates the additional counter receives are equal to a linear combination of the updates the three counters receive times the number of zero tests remaining. 
So, since the VASS is ultraflat, the additional counter will also have zero update on the transitions between cycles.
This maintains the property that the VASS is ultraflat.

\begin{figure}[ht!]
	\begin{algorithmic}[1]
		\Require \assert{x}{x}, \assert{y}{0}, \assert{z}{1}
		\Ffor{r}{1}{p-1}
			\State \LOOP \inc{x}{r}, \inc{y}{p+r}, \dec{z}{1}
		\EndFfor
		\State \zt{z} 
		\State \LOOP \dec{x}{p}, \inc{z}{p} 
		\State \zt{x}
		\State \LOOP \inc{x}{1}, \dec{z}{1}
		\State \zt{z}
		\Ffor{r}{1}{p-1} 
			\State \LOOP \dec{x}{r}, \dec{y}{p+r}, \inc{z}{1}
		\EndFfor
		\State \zt{y}
		\Ensure \assert{x}{x}, \assert{y}{0}, \assert{z}{1}
	\end{algorithmic}
	\caption{%
		The \nondiv{p} gadget implemented as a unary ultraflat 3-VASS with four zero tests for asserting that the initial value $v$ of counter $\var{x}$ is not divisible by the fixed positive integer $p$.
		The construction, which has size quadratic in $p$, uses three counters: the primary $\var{x}$ and ancillaries $\var{y}$ and $\var{z}$.
		From an initial configuration with $\var{x} = v$, $\var{y} = 0$, and $\var{z} = 1$, the final configuration with $\var{x} = v$, $\var{y} = 0$, and $\var{z} = 1$ can be reached if and only if $p$ does not divide $v$.
		\cref{clm:ultraflat-non-div} shows the correctness of this construction.
	}
        \vskip1ex
        \hrule
	\label{fig:ultraflat-non-div}
\end{figure}
\begin{claim}\label{clm:ultraflat-non-div}
	Let $p$ be a positive integer.
	In the unary ultraflat 3-VASS with zero tests in~\cref{fig:ultraflat-non-div}, from an initial configuration with counter values $\var{x} = v$, $\var{y} = 0$, and $\var{z} = 1$, the final state can be reached if and only if $p$ does not divide $v$.
	Moreover, in the case of reachability, the final configuration must have counter values $\var{x} = v$, $\var{y} = 0$, and $\var{z} = 1$.
	Furthermore, the unary ultraflat 3-VASS can be constructed in $\Oh(p^2)$ time and uses only four zero tests (which do not belong to cycles).
\end{claim}
\begin{proof}
	Suppose the final state is reachable from an initial configuration with counter values $\var{x} = v$, $\var{y} = 0$, and $\var{z} = 1$.
	For this to be true, each of the zero tests must have been successful.
	In order to pass the first zero test (Line 3), one of the first $p-1$ loops (prescribed by Line 1 and 2) must have been taken once.
	This means, after the first zero test, $\var{x} = v + r$, $\var{y} = p+r$, and $\var{z} = 0$.
	Looking ahead, to pass the second zero test (Line 5), there must exist a natural number $q \geq 1$ such that the next loop (Line 4) is taken $q$ many times, thus $v + r = q \cdot p$.
	Therefore, we can rearrange to see that $v = (q-1) \cdot p + (p-r)$, and since $p-r \in \set{1, \ldots, p-1}$, we know that $p$ does not divide $v$.

	Now, we will argue that when reaching the final state, the original counter values are restored.
	Currently, after the second zero test, the counter values are $\var{x} = 0$, $\var{y} = p+r$, and $\var{z} = v + r$.
	To pass the third zero test (Line 7), the prior loop (Line 6) must be taken maximally; the counter values will then be $\var{x} = v + r$, $\var{y} = p+r$, and $\var{z} = 0$.
	Now, since the fourth and final zero test (Line 10) is successful, the final $p-1$ loops (prescribed by Lines 8 and 9) must have been used.
	Notice that currently $\var{y} = p + r < 2p$.
	The update $\var{y}$ receives on any of these final loops is $-(p+r')$ for any $r' \in \set{1, \ldots, p-1}$; since $-(p+r') < -p$, at most one iteration of any of these loops can be taken.
	Given that the last zero test is passed, the loop with update \dec{y}{p+r} (that is, for the same~$r$ as previously) must have been taken; this also updates $\var{x}$ and $\var{z}$.
	Therefore, the final state is reached with counter values $\var{x} = v$, $\var{y} = 0$, and $\var{z} = 1$.

	Conversely, if $p$ does not divide $v$, there exist $q \in \NN$ and $r \in \set{1, \ldots, p-1}$ such that $v = q \cdot p + r$.
	We will use these values, $q$ and $r$, to construct a valid run.
	First, take one iteration of the $(p-r)$-th loop (prescribed by Lines 1 and 2).
	This reaches a configuration with counter values $\var{x} = v + p - r$, $\var{y} = 2p-r$, and $\var{z} = 0$; the first zero test (Line 3) is successfully passed.
	Now, since currently $\var{x} = v + p - r = q \cdot p + r + p - r = (q+1) \cdot p$, we can iterate the next loop (Line 4) $q+1$ many times.
	This leads to a configuration with counter values $\var{x} = 0$, $\var{y} = 2p-r$, and $\var{z} = v + p - r$; the second zero test (Line 5) is successfully passed.
	The following loop (Line 6) is iterated $v+p-r$ many times, so the counters then have values $\var{x} = v + p - r$, $\var{y} = 2p-r$, and $\var{z} = 0$; the third zero test (Line 7) is successfully passed.
	Among the $p-1$ remaining loops (prescribed by Lines 8 and 9), one iteration of the $(p-r)$-th loop is taken.
	Finally, the counter values reached are $\var{x} = v$, $\var{y} = 0$, and $\var{z} = 1$; indeed, the fourth and final zero test (Line 10) is successfully passed too.
\end{proof}

To overcome the second challenge, we provide an \ugadget{u} gadget that replaces non-zero updates \inc{x}{u} (on transitions between cycles, and where $u$ can be either positive or negative) with a series of carefully chosen cycles that must be executed in a specific way due to zero tests.

\begin{figure}[ht!]
	\begin{algorithmic}[1]
		\Require \assert{x}{v}, \assert{y}{0}, \assert{z}{1}
		\State \LOOP \inc{x}{u}, \inc{y}{1}, \dec{z}{1}
		\State \zt{z}
		\State \LOOP \dec{y}{1}, \inc{z}{1}
		\State \zt{y}
		\Ensure \assert{x}{v+u}, \assert{y}{0}, \assert{z}{1}
	\end{algorithmic}
	\caption{%
		The \ugadget{u} gadget that is a unary ultraflat 3-VASS with two zero tests.
		It is used to replace non-zero updates to counter $\var{x}$ on transitions between cycles.
		The construction, which has size linear in $u$, uses (the same) three counters $\var{x}$, $\var{y}$, and $\var{z}$.
		From an initial configuration with $\var{x} = v$, $\var{y} = 0$, and $\var{z} = 1$, the final configuration with $\var{x} = v + u$, $\var{y} = 0$, and $\var{z} = 1$ can be reached if and only if $v + u \geq 0$.
		\cref{clm:ultraflat-update} shows the correctness of this construction.
	}
        \vskip1ex
        \hrule
	\label{fig:ultraflat-update}
\end{figure}

\begin{claim}\label{clm:ultraflat-update}
	Let $u \in \ZZ$.
	In the unary ultraflat 3-VASS with zero tests presented in~\cref{fig:ultraflat-update}, from an initial configuration with $\var{x} = v$, $\var{y} = 0$, and $\var{z} = 1$, the final state can be reached if and only if $v + u \geq 0$.
	Moreover, in the case of reachability, the final configuration must have counter values $\var{x} = v + u$, $\var{y} = 0$, and $\var{z} = 1$.
	Furthermore, the unary ultraflat 3-VASS can be constructed in $\Oh(u)$ time and uses only two zero tests (neither of which belongs to any cycle).
\end{claim}
\begin{proof}
	Suppose the final state is reachable.
	This means that both zero tests were passed successfully.
	Given that initially $\var{z} = 1$, this means that both loops are iterated once.
	In particular, for the first loop to be iterated, $v + u \geq 0$ must hold ($\var{x}$ cannot have a negative value).
	The counter values reached at the final state are $\var{x} = v + u$, $\var{y} = 0$, and $\var{z} = 1$.

	Conversely, suppose that $v + u \geq 0$.
	Then to reach the final state, take the first loop once (this is possible since $v + u \geq 0$) and the second loop once.
	The final state is reached with counter values $\var{x} = v + u$, $\var{y} = 0$, and $\var{z} = 1$.
\end{proof}

We conclude this subsection with the proof that reachability in unary ultraflat VASS is \class{NP}-complete in dimension 4 and a conjecture that the same is true in dimension 3.

\begin{proof}[Proof of~\cref{thm:ultraflat-hardness}]
	We obtain the lower bound via a reduction from 3-CNF SAT; the \class{NP} upper bound is given by~\cref{thm:np}.

	We extend the reduction presented in~\cref{lem:sat-encoding} by implementing the non-divisibility assertions in polynomial time and replacing the non-zero updates on transitions between cycles (Lines 4, 6, 8, and 10 of~\cref{fig:lps-sat}) with copies of \ugadget{u} (\cref{fig:ultraflat-update}).
	As seen in~\cref{fig:ultraflat-non-div}, we can implement a non-divisibility assertion as a unary ultraflat 3-VASS that only uses four zero tests (see~\cref{clm:ultraflat-non-div} for its correctness).
	We remark that in order for this implementation to only use three counters (with zero tests), we need to be able to reuse ancillary counters $\var{y}$ and $\var{z}$ between the interleavings of \nondiv{p} and \ugadget{u} gadgets.
	This is possible thanks to the aligned pre and post conditions of both gadgets: both require (and ensure, respectively) $\var{y} = 0$ and $\var{z} = 1$.
	Otherwise, in between these gadgets, $\var{y}$ and $\var{z}$ are untouched.
	Accordingly, we obtain an instance of reachability in a unary ultraflat 3-VASS with zero tests that encodes the satisfiability of $\phi$.

	We can again use the controlling counter technique (\cref{lem:controlling-counter}) to obtain, in polynomial time, an equivalent instance of reachability in a unary ultraflat 4-VASS.
	Importantly, observe that this is only possible because the number of zero tests is polynomially bounded, which is the case since the zero tests only occur on transitions between cycles.
	Precisely, there are $m + \sum_{i=1}^n (p_i-2)$ many \nondiv{\cdot} gadgets, each with four tests, plus $2m + 2\cdot\sum_{i=1}^n (p_i-2)$ many \ugadget{\cdot} gadgets, each with two tests.
	Hence, reachability in unary ultraflat 4-VASS is \class{NP}-hard.
\end{proof}

\begin{conjecture}\label{thm:ultraflat-3-vass}
	Reachability in unary ultraflat 3-VASS is \class{NP}-hard.
\end{conjecture}

\section{Missing Proofs of~\cref{sec:hardness}}
\label{app:3slps}
\subsection{Proof of~\cref{lem:sat-encoding}}
\label{app:sat-encoding}
	As previously described, let $\phi$ be the given instance of CNF 3-SAT with variables $x_1, \ldots, x_n$, and let positive integers $p_1, \ldots, p_n$ be pairwise coprime.
	We choose $p_1, \ldots, p_n$ such that $\max\set{p_1, \ldots, p_n}$ is upper-bounded by a
    polynomial in~$n$.
    For example, by the prime number theorem, one can select the first $n$ primes: they are bounded above by $\Oh(n\log(n))$ and can be computed in $\Oh(n^{1+o(1)})$ time~\cite{AgrawalKS04}.  

	We assume, without loss of generality, that all clauses contain $3$~literals of distinct variables.
	We use the simple linear path scheme \satprogram{\phi} with non-divisibility checks that is presented as a counter program in~\Cref{fig:lps-sat}; its behaviour is largely self-explanatory.
	Suppose that, after the first loop, the value of $\var{x}$ is guaranteed to encode an assignment; we want to check whether this assignment is satisfying for the given instance~$\phi$.
	For each clause $C_j = (\ell_1 \lor \ell_2 \lor \ell_3)$, suppose $x_a$, $x_b$, and $x_c$ are the distinct variables of the literals $\ell_1$, $\ell_2$, and $\ell_3$, respectively.
	There exists exactly one assignment to these variables that falsifies this clause; this assignment corresponds to a combination of remainders modulo $p_a$, $p_b$, $p_c$, respectively.
	By the Chinese remainder theorem, there is a unique $r_j \in [0, p_a p_b p_c - 1]$ with these three remainders.
	Therefore, checking that $C_j$ is satisfied amounts to
	verifying that the value of the counter $\var{x}$ is not congruent to $r_j \bmod p_a p_b p_c$.

	Overall, the final state can be reached if and only if there exists an integer~$v$
	greater than (or equal to) the initial value of the counter~$\var{x}$ such that $v$ encodes an assignment to the Boolean variables $x_1, \ldots, x_n$, and this assignment satisfies each clause in $\phi$.
	Again by the Chinese remainder theorem, such an $v$ exists if and only if $\phi$ is satisfiable.

	Finally, we comment on the efficiency of this reduction.
	There are a linear number of non-divisibility assertions; precisely $m + \sum_{i=1}^n (p_i-2)$ many.
	The values $p_1, \ldots, p_n$; $q_1, \ldots, q_m$; and $r_1, \ldots, r_m$ are all at most $(\max\set{p_1, \ldots, p_n})^3$ and can be computed in polynomial time.
	Therefore, non-divisibility assertions can be implemented in polynomial time and \satprogram{\phi} can be constructed in polynomial time (with respect to $\size{\phi}$).

\subsection{Proof of~\cref{clm:unary-non-div}}
	Let $\var{x} = v$ and $\var{y} = 0$ be the initial counter values for the unary 2-SLPS with zero tests, presented in~\Cref{fig:unary-non-div}.
	First observe that after the execution of two initial updates to $\var{x}$ and $\var{y}$ (line 1), the invariant $\var{y} + \Delta \var x = p - 1$ is initialised. 
	As stated in~\Cref{fig:unary-non-div}, $\Delta \var{x}$ represents the difference between the current value of $\var{x}$ and its initial value~$v$.
	Following, all three loops (lines 2, 3, and 5)  leave the sum of the two counter values unchanged, thus leaving the invariant $\var{y} + \Delta \var{x} = p-1$ unchanged too.

	We will now prove that reachability is equivalent to non-divisibility of~$v$ by~$p$.
	Let $k$ denote the number of iterations of the first loop; observe that $k \in \set{ 0, 1, \ldots, p-2 }$.
	If the final state is reachable, then $v + 1 + k$ is divisible by~$p$.
	Since $1 + k < p$, it is true that $p$ does not divide $1 + k$, so it must also be true that $p$ does not divide $v$.
	Conversely, suppose $v$ is not divisible by~$p$, and in particular $v = q \cdot p + r$, where $q, r \in \NN$ and $0 < r \leq p - 1$.
	We can choose $k = (p - 1) - r$ iterations of the first loop; indeed $0 \leq k \leq p - 2$ permits a valid start to the run.
	Therefore, the zero test on~$\var{x}$ (line 4) can be passed, and so too can the subsequent zero test on~$\var{y}$ (line 6).

	It remains to prove that the initial value of the counter is properly restored at the end of the run.
	Before the final update to $\var{x}$ (line 7), the invariant $\var{y}+ \Delta\var{x} = p - 1$ holds.
	However, the value of $\var{y}$ is zero, so $\Delta\var{x} = p - 1$ at this moment.
	Hence, after the final update, \assert{x}{v} is restored. 
\qed

\subsection{Proof of~\cref{lem:fixed-updates}}
\label{app:fixed-updates}
	Let $u \geq 1$ be a fixed natural number.
	Let $\Uu$ be a $d$-SLPS with fixed updates in $\set{-u, -u+1, \ldots, u-1, u}$.
	We will construct a unitary $(d \cdot u)$-SLPS $\Vv$ that preserves the reachability relation of $\Uu$.

	For each of the $d$ counters in $\Uu$, we will have $u$ many counters into $\Vv$.
	Suppose $\var{x}$ is a counter in $\Uu$, suppose also that $\var{y}_1, \ldots, \var{y}_u$ are the corresponding counters in $\Vv$.
	The following will be invariant: 
	\begin{equation*}
		\var{x} = \var{y}_1 + \cdots + \var{y}_u.
	\end{equation*}
	Roughly speaking, the value of $\var{x}$ in $\Uu$ will be allowed to be spread freely across $\var{y}_1, \ldots, \var{y}_u$ in $\Vv$.
	Suppose there is a transition in $\Uu$ with update \inc{x}{t} for some $0 \leq t \leq u$, instead in $\Vv$ the equivalent transition has updates \increment{y}{1}{1}, $\ldots\,$, \increment{y}{t}{1}.
	Similarly, if there is a transition in $\Uu$ is labelled with updates \dec{x}{t} for some $0 \leq t \leq u$, instead in $\Vv$ the equivalent transition is labelled with updates \decrement{y}{1}{1}, $\ldots\,$, \decrement{y}{t}{1}. 

	As it stands, these modifications do not suffice.
	Consider, for example, the 1-SLPS with fixed updates in $\set{-2, 1, 0, 1, 2}$ (left), and the current modified unitary 2-SLPS (right).

	\vspace{0.1cm}
	\begin{minipage}{.5\textwidth}
	  	\begin{algorithmic}[1]
			\Require \assert{x}{0}
			\State \LOOP \inc{x}{1}
			\State \dec{x}{2}
		\end{algorithmic}
	\end{minipage}\vspace{0.1cm}
	\begin{minipage}{.5\textwidth}
	 	\begin{algorithmic}[1]
			\Require \assertt{y}{1}{0}, \assertt{y}{2}{0}
			\State \LOOP \increment{y}{1}{1}
			\State \decrement{y}{1}{1}, \decrement{y}{2}{1}
		\end{algorithmic}
	\end{minipage}
	Clearly these SLPSs are not equivalent: in the 1-SLPS, the final state can be reached (with any value in fact), and in the 2-SLPS, the final state is not reachable.

	To remedy this issue, we introduce the following gadget at each state (i.e. before and after every transition and self-loop).
	\begin{figure}
		\begin{algorithmic}[1]
			\State \LOOP \decrement{y}{2}{1}, \increment{y}{1}{1}
			\State \LOOP \decrement{y}{3}{1}, \increment{y}{1}{1}
			\State $\cdots$
			\State \LOOP \decrement{y}{u}{1}, \increment{y}{1}{1}
			\State \LOOP \decrement{y}{1}{1}, \increment{y}{2}{1}
			\State \LOOP \decrement{y}{1}{1}, \increment{y}{3}{1}
			\State $\cdots$
			\State \LOOP \decrement{y}{1}{1}, \increment{y}{u}{1}
		\end{algorithmic}
		\caption{%
			This gadget is designed to freely spread the value of $\var{y}_1 + \cdots + \var{y}_u$ across $\var{y}_1, \ldots, \var{y}_u$ so that negative updates can be handled.
		}
		\label{fig:counter-spreading}
	\end{figure}
	
	Since each of these loops preserves the invariant $\var{x} = \var{y}_1, \ldots, \var{y}_u$, the reachable values of the collection of counters $\var{y}_1, \ldots, \var{y}_u$ in $\Vv$ faithfully captures the reachable values of $\var{x}$ in $\Uu$.
\qed

\subsection{Unitary SLPS for Non-divisibility}
\label{app:unitary-non-div}
\begin{figure}[ht!]
	\begin{algorithmic}[1]
		\Require \assert{x}{v},
				 \assert{y}{0},
				 \assert{a_1}{0},
				 \assert{a_2}{0},
				 \assert{a_3}{0}.
		\Ffor{i}{1}{p-2} 
			\State \inc{y}{1}
		\EndFfor
		\State \inc{x}{1}  
		\State \LOOP \inc{x}{1}, \dec{y}{1} 
			\Comment{Guess $\Delta\var{x} \in \set{1, \ldots, p-1}$ for which $v + \Delta\var{x}$ is divisible by $p$.}

		\State \LOOP \inc{a_1}{1} 
		\Comment{Now, guess how many times $p$ divides $\var{x} = v+\Delta\var{x}$.}
		\Ffor{i}{1}{p} \Comment{Verify divisibility by multiplying that guess by $p$.} 
			\State \LOOP \dec{a_1}{1}, \inc{a_2}{1}, \inc{a_3}{1} 
			\Comment{Do this via repeated addition, storing the value in $\var{a_3}$.}
			\State \zt{a_1}
			\State \LOOP \inc{a_1}{1}, \dec{a_2}{1}
			\State \zt{a_2}
		\EndFfor
		\State \LOOP \dec{a_1}{1}
		\State \zt{a_1} 
		\Comment{Reset $\var{a}$ to zero for reuse.}
		\State \LOOP \dec{x}{1}, \dec{a_3}{1}, \inc{a_1}{1}
		\State \zt{x}, \zt{a_3} 
		\Comment{If the guesses were correct, $\var{x}$ and $\var{a_3}$ have equal value.}
		\State \LOOP \inc{x}{1}, \dec{a_1}{1}
		\State \zt{a_1} 
		\State \LOOP \inc{x}{1}, \dec{y}{1}
		\State \zt{y} 
		\Ffor{i}{1}{p-1}
			\Comment{Restore the original value $\var{x} = v$.}
			\State \dec{x}{1}
		\EndFfor
		\Ensure \assert{x}{v},
				\assert{y}{0},
				\assert{a_1}{0},
				\assert{a_2}{0},
				\assert{a_3}{0}.
	\end{algorithmic}
	\caption{%
		The \nondiv{p} gadget implemented as a unitary 5-SLPS with zero tests for asserting that the initial value $v$ of counter $\var{x}$ is not divisible by the fixed positive integer $p$.
		The construction, which has linear size in $p$, uses five counters: the primary~$\var{x}$ and four ancillary counters~$\var{y}$,~$\var{a_1}$,~$\var{a_2}$, and~$\var{a_3}$.
		From an initial configuration with $\var{x} = v$ and $\var{y}, \var{a_1}, \var{a_2}, \var{a_3} = 0$, the final configuration with $\var{x} = v$ and $\var{y}, \var{a_1}, \var{a_2}, \var{a_3} = 0$ can be reached if and only if $p$ does \emph{not} divide $v$.
		Note that $\Delta\var{x}$ is the change in the counter value of $\var{x}$.
	}
	\label{fig:unitary-non-div}
\end{figure}

\subsection{Proof of~\cref{clm:unitary-non-div}}
\label{app:unitary-non-div-correctness}
%\begin{proof}
	Let $\var{x} = v$, $\var{y} = \var{a_1} = \var{a_2} = \var{a_3} = 0$ be the initial counter values for the unitary 5-SLPS with zero tests presented in~\Cref{fig:unitary-non-div}.
	We will argue that reachability is equivalent to non-divisibility of $v$ by $p$.
	Overall, the SLPS first guesses a difference value $\Delta\var{x} \in \set{ 1, 2, \ldots, p-1 }$ (Lines 1 -- 4), then the SLPS proceeds to check that $\var{x} = v + \Delta\var{x}$ is divisible by $p$.
	This is due to the fact that $v$ is not divisible by $p$ if and only if such a $\Delta\var{x}$ exists for which $v + \Delta\var{x}$ is divisible by $p$.
	To achieve this, the quotient is guessed (Line 5) and immediately multiplied by $p$ (Lines 6 -- 10).
	If such a guess can be made, then the result, which is stored on counter $\var{a_3}$, will be equal to $v + \Delta\var{x}$, which is the current value of $\var{x}$.
	In such a scenario, the test for equality between $\var{x}$ and $\var{a_3}$ (Lines 11 -- 16) can succeed.
	Finally, now that $v + \Delta\var{x}$ has been verified to be divisible by $p$, the original value $\var{x} = v$ is restored (Lines 17 -- 20).
	\qed

\subsection{Proof of~\cref{lem:unitary-zero-tests}}
\label{app:unitary-zero-tests}
	This almost immediately follows from the proof of~{\cite[Lemma 2.6]{CzerwinskiO22}}.
	Roughly speaking,~{\cite[Lemma 2.6]{CzerwinskiO22}} states that for a unary $d$-VASS with zero tests $\Uu$ that uses at most $C$ many zero tests and whose counters are bounded by $B$, one can construct unary $(d+3)$-VASS $\Vv$ such that $\Run{\config{p}{\vec{0}}}{*}{\Uu}{\config{q}{\vec{0}}}$ if and only if $\Run{p'(B, 2C, 2BC, \vec{0}^d)}{*}{\Vv}{q'(B,\vec{0}^d+2)}$.
	In summary, the last $d$ counters of $\Vv$ will mimic the $d$ counters of $\Uu$ and the additional three counters $\var{b}$, $\var{c}$, and $\var{d}$ in $\Vv$ will be responsible for ensuring that the zero tests are simulated faithfully.
	To understand the core idea, consider one counter $\var{x}$ of the unary $d$-VASS $\Uu$, and suppose, initially that $\var{b} = B$, $\var{c} = 2C$, and $\var{d} = 2BC$.
	We will ensure that $\var{x} + \var{b} = B$ is invariant.
	At the beginning this is satisfied because, initially, $\var{x} = 0$.
	To maintain the invariant, whenever $\var{x}$ receives an update \inc{x}{u}, $\var{b}$ is updated by \dec{b}{u}, and whenever $\var{x}$ receives an update \dec{x}{u}, $\var{b}$ is updated by \inc{b}{u}.
	Given that that counters are bounded above by $B$, we know that $\var{b} \geq 0$ at all times.
	Then, instead of performing a zero test on $\var{x}$, the following gadget is inserted.
	\vspace{0.05in}\begin{algorithmic}[1]
		\State \LOOP \inc{x}{1}, \dec{b}{1}, \dec{d}{1}
		\State \LOOP \dec{x}{1}, \inc{b}{1}, \dec{d}{1}
		\State \dec{c}{2}
	\end{algorithmic}\vspace{0.05in}

	The idea is that, given that $\var{c}$ is initially set to $\var{c} = 2C$ and at the end of the run $\var{c}$ is required to be 0, we must traverse $C$ many of these gadgets.
	Furthermore, the greatest possible negative effect that $\var{d}$ can experience after one traversal of this gadget is $-2B$.
	That is because $\var{b} \leq B$ and $\var{x} \leq B$; indeed, it is possible for $\var{d}$ to lose $2B$ counter value if, initially $\var{x} = 0$ and, consequently, $\var{b} = B$.
	Most critically, given that $\var{d}$ is initially set to $\var{d} = 2BC$ and at the end of the run $\var{d}$ is required to be 0, and given that these gadgets are traversed (exactly) $C$ many times, it must be the case that $d$ is updated by $-2B$ counter value every time.
	Therefore, whenever this gadget is taken, we know that $\var{x} =0$, so the zero tests are simulated faithfully.
	
	There is a straightforward extension to this argument that allows for the $C$ many zero tests to be performed on any of the $d$ counters $\var{x}_1, \ldots, \var{x}_d$ in $\Uu$.
	Instead, the following gadget is inserted in place of a zero test 
	\vspace{0.05in}\begin{algorithmic}[1]
		\State \LOOP \increment{x}{1}{1}, \decrement{x}{2}{1}, \dec{d}{1}
		\State \LOOP \increment{x}{2}{1}, \decrement{x}{3}{1}, \dec{d}{1}
		\State $\cdots$
		\State \LOOP \increment{x}{d-1}{1}, \decrement{x}{d}{1}, \dec{d}{1}
		\State \LOOP \increment{x}{d}{1}, \dec{b}{1}, \dec{d}{1}
		\State \LOOP \inc{b}{1}, \decrement{x}{d}{1}, \dec{d}{1}
		\State \LOOP \increment{x}{d}{1}, \decrement{x}{d-1}{1}, \dec{d}{1}
		\State $\cdots$
		\State \LOOP \increment{x}{3}{1}, \decrement{x}{2}{1}, \dec{d}{1}
		\State \LOOP \increment{x}{2}{1}, \decrement{x}{1}{1}, \dec{d}{1}
	\end{algorithmic}\vspace{0.05in}
	For further details, see the discussion leading up to~{\cite[Lemma 2.6]{CzerwinskiO22}}.

	The are a few minor differences between~\cref{lem:unitary-zero-tests} and~{\cite[Lemma 2.6]{CzerwinskiO22}}, the first is that we wish to preserve the constant (unitary) updates and SLPS structure, as opposed to dealing with general VASS with zero tests (i.e. counter automata).
	The way zero tests are simulated is already just a sequence of single transition cycles with updates in $\set{-1, 0, 1}$, exactly as we wish.
	For the second difference, let us call $\var{b}$, $\var{c}$, and $\var{d}$  the ($d+1$)-st, ($d+2$)-nd, and ($d+3$)-rd counters of $\Vv$, respectively. 
	We require that $\var{b}$, $\var{c}$, and $\var{d}$ are all required to reach $0$ in the final configuration.
	As described in the proof of~{\cite[Lemma 2.6]{CzerwinskiO22}}, ensuring that $\var{d}$ reaches zero suffices to conclude that the zero tests were faithfully simulated, so to allow $\var{b}$ and $\var{c}$ counters to reach zero, we can append two unitary self-loops with updates \dec{b}{1} and \dec{c}{1} at the end of $\Vv$.

	We remark that the sum of $\var{x}_1 + \ldots + \var{x}_d$ (the $d$ counters in $\Uu$) only differs from the bounding counter $\var{b}$ by a constant amount.
	Therefore, assuming that the absolute value of the aggregate of any transition is bounded by $a$, then at most $a$ many counters can be incremented or decremented on any transition.
	For example, we may require a transition with updates: \inc{x_1}{1}, $\ldots\,$, \inc{x_a}{1}, \dec{b}{a}.
	Therefore, since this transition may belong to a self-loop, we cannot split it up into $a$ many updates.
	Instead, just as in the proof of~\cref{lem:fixed-updates}, to maintain the fact that cycles consist only of one transition in SLPSs, we will split the bounding counter $\var{b}$ into $a$ counters: $\var{b}_1, \ldots, \var{b}_a$.
	For any $t \leq a$, in place of \inc{b}{t} updates, we perform the updates \inc{b_1}{1}, $\ldots\,$, \inc{b_t}{1}; in place of \dec{b}{t}, we perform the updates \dec{b_1}{1}, $\ldots\,$, \dec{b_t}{1}.
	Then, in between all transitions, we add $2a$ self-loops, just like in~\Cref{fig:counter-spreading} for~\cref{lem:fixed-updates}.
	We note that this is why there is a different number of resulting counters between~{\cite[Lemma 2.6]{CzerwinskiO22}} and~\cref{lem:unitary-zero-tests} and; in~{\cite[Lemma 2.6]{CzerwinskiO22}} there is just one counter $\var{b}$ that is used to maintain the invariant and here, for~\cref{lem:unitary-zero-tests}, we use $a$ many counters $\var{b}_1, \ldots, \var{b}_a$.

	The absolute value of the aggregate effect of all transitions in the resulting SLPS $\Vv$ is at most one  because positive (resp. negative) updates to $\var{x}_1 + \ldots + \var{x}_d$ are always compensated by negative (resp. positive) updates to $\var{b}_1 + \ldots + \var{x}_a$.
	There are only two counters that we haven't considered: $\var{c}$ and $\var{d}$.
	Updates to $\var{c}$ only occur at the end of the series of $2d$ self-loops when simulating zero tests, the counter receives two consecutive \dec{c}{1} updates while no other counters are being updated.
	These transitions, therefore have an aggregate effect $-1$.
	Therefore, the only counter that remains is $\var{d}$; since it receives unitary updates, it can only change the aggregate effect of the other transitions by $-1$, $0$, or $+1$.
	Therefore, the absolute value of the aggregate effect of every transition in $\Vv$ is at most one.

	Finally, it remains to show size bounds on $\Vv$.
	After each transition, we need to inject the aforementioned sequence of $2s$ self-loops for freely spreading the value of the bounding counters around.
	Furthermore, each of (at most) $C$ zero tests in $\Uu$ is replaced by a sequence of $2d$ self-loops in $V$.
	Overall, $\Vv$ can be constructed in $\Oh(a\cdot\norm{\Uu} + d\cdot C)$ time.
	\qed

\subsection{The Ackermann Function and its Inverse}
\label{app:ackermann-function}

First, let us define a family of \emph{fast-growing} functions $G_i$ which we will show are comparable to the fast-growing functions defining the Ackermann function.
First, we let $G_1(n) = 2n$, and the subsequent functions are defined, for $i > 1$, inductively by composition,
\begin{equation*}
	G_{i+1}(n) \coloneqq \underbrace{G_i( G_i( \cdots G_i}_{\lfloor n/12 \rfloor \text{ times}}(1) \cdots ) ).
\end{equation*}
We let $A(n) = G_{(n-14)/4}(3)$; this function grows at the same asymptotic rate as the Ackermann function.
In particular, as we show in Claim~\ref{clm:functionA}, the inverse of the function $A$ is bounded by a constant times the inverse Ackermann function.

As they appear in~{\cite[Section 2]{CzerwinskiO21}}, the fast-growing functions defining the Ackermann function are as follows: $F_1(n) \coloneqq 2n$, and 
\begin{equation*}
	F_{i+1}(n) \coloneqq \underbrace{F_i( F_i( \cdots F_i}_{n\text{ times}}(1) \cdots ) ).
\end{equation*}
Accordingly, $\ackermann{n} = F_n(n)$ is the \emph{Ackermann function} and recall that $\inverseAckermann{n} = \iackermann{n}$ is the \emph{inverse Ackermann function}.
Note that these functions differ slightly from the original defined by Ackermann~\cite{Ackermann28}, however, for most applications the differences do not matter, this is discussed in~{\cite[Section 4]{Schmitz16}}, precise details can be found in~\cite{Ritchie65, Loeb70}.

\begin{claim}\label{clm:functionA}
	There is a constant $c \in \NN$ such that for all $n \in \NN$ the following inequality holds:
	\begin{equation*}
	A^{-1}(n) \leq c \cdot \alpha(n).
	\end{equation*}
\end{claim}
\begin{proof}
	Firstly, it is easy to show by induction that for any $n \geq 3$ we have $F_i(n) \geq F_{i-1}(n+1)$, therefore $F_{2n}(3) \geq F_n(n+3) \geq F_n(n) = \ackermann{n}$.
	Furthermore, it is also easy to prove $G_{12i}(n) \geq F_i(n)$ by induction.
	Then for any $n \geq 1$,
	\begin{equation*}
		A(110n) = G_{(110n-14)/4}(3) \geq G_{96n/4}(3) = G_{24n}(3) \geq F_{2n}(3) \geq \ackermann{n}.
	\end{equation*}
	Thus $A^{-1}(n) \leq 110 \alpha(n)$, which concludes the proof with $c = 110$.
\end{proof}

\subsection{Unitary SLPS Ackermann Generators of Ackermann Size}
\label{app:ackermann-generator}

The proof of~\cref{clm:unitary-triple} essentially follows the ideas presented in~\cite{CzerwinskiO21} and~\cite{Lasota22}; our proof is a modification of the proof in~{\cite[Theorem 8]{Lasota22}} which shows that for each $d \in \NN$ there is a VASS with $3d+2$ counters which essentially speaking realises the requirements of Claim~\ref{clm:unitary-triple}. 
We need to strengthen it in two ways: first, we require that the resulting VASS is an SLPS, and second, we require that the transitions have unitary updates. 
Both points force us to modify the proof, but the overall idea remains the same as~{\cite[Theorem 8]{Lasota22}}. 
However, we detail this proof from scratch for clarity.

We say that a $d$-SLPS $\Tt$ is a \emph{$B$-generator} if, from an initial configuration with counter values $\vec{0}^d$ and for every $C \in \NN$, the final state can only be reached with counter values $(\vec{0}^{d-3}, x, y, z)$, $x = B$, $y \geq C$, and $z = B \cdot y$.
With this definition, Claim~\ref{clm:unitary-triple} requires the existence of a $d$-dimensional $A(d)$-generator of size $\Oh(A(d))$.

Before we proceed further, observe that it is easy to design a $3$-SLPS with constant updates in $\set{-3, -2, -1, 0, 1, 2, 3}$ that is a $3$-generator.
\vspace{0.05in}\begin{algorithmic}[1]
	\Require \assert{x}{0}, \assert{y}{0}, \assert{z}{0}
	\State \inc{x}{3}
	\State \LOOP \inc{y}{1}, \inc{z}{3}
\end{algorithmic}\vspace{0.05in}
By applying~\cref{lem:fixed-updates} to this $3$-SLPS with constant updates in $\set{-3, -2, -1, 0, 1, 2, 3}$, we obtain a unitary $9$-SLPS that is a $3$-generator.

A key notion that was used in~\cite{CzerwinskiO21, Lasota22} is an $f$-amplifier for some function $f: \NN \to \NN$. 
Since we require an SLPS (as opposed to a VASS) with similar properties, we need to modify this notion somewhat; we introduce a second parameter to the definition of an amplifier.
For a function $f: \NN \to \NN$ and a constant $B \in \NN$ an SLPS is an $(f, B)$-\emph{amplifier} if,
from an initial configuration with counter values $(B, C, BC, \vec{0}^{d-3})$, the final state is reached with counter values $(\vec{0}^{d-3}, \,f(B), \,D, \,D\!\!\:\cdot\!\!\:f(B))$.
Moreover, for any $b \in \NN$, there exists a sufficiently large $C \in \NN$ such that a final configuration can be reached with $D \geq b$.
We call the first three and last three counters of the amplifier the \emph{input counters} and \emph{output counters}, respectively.
To prove~\cref{clm:unitary-triple}, it suffices to prove the following claim.

\begin{claim}\label{clm:amplifier}
	For each $d \in \NN$ and $B \in \NN$ there exists a unitary $(4d+8)$-SLPS of size $\Oh(G_d(B))$, which is an $(G_d, B)$-amplifier.
\end{claim}

Before proving~\cref{clm:amplifier} we show how one can use it to prove~\cref{clm:unitary-triple}.

\begin{proof}[Proof of~\cref{clm:unitary-triple}]
	By Claim~\ref{clm:amplifier}, there is a unitary $(4d+8)$-SLPS of size $\Oh(G_d(3))$ that is an $(G_d, 3)$-amplifier.
	Let $d' = 4d + 14$. 
	We define the required generator $d'$-SLPS $\Vv_{d'}$ by composing the previously detailed unitary 9-SLPS that is a 3-generator with the $(4d+8)$-SLPS that is a $(G_d, 3)$-amplifier.
	The three output counters of the $3$-generator coincide with the three input counters of the $(G_d, 3)$-amplifier; altogether the dimension of $\Vv_{d'}$ is $(4d+8)+6$.

	By definition if $\Vv$ reaches the final state with counter values $(\vec{0}^{d'-3}, x, y, z)$ then $x = G_d(3)$ and $z = G_d(3) \cdot y$. 
	The size of $\Vv_{d'}$ is $\Oh(G_d(3))$. 
	Now, notice that since $d = (d' - 14) / 4$, we know that $G_d(3) = A(d)$.
	Therefore, $\Vv_{d'}$ is indeed an $A(d)$-generator of size $\Oh(A(d))$, so by considering Claim~\ref{clm:unitary-triple} with dimension $d' \in \Oh(d)$, the $d'$-SLPS $\Vv_{d'}$ satisfies all the required conditions.
\end{proof}

Now, we proceed to prove~\cref{clm:amplifier}. 
\begin{proof}[Proof of~\cref{clm:amplifier}]
	We prove it by induction on $d$.

	Assume first that $d = 1$; recall that $G_1(n) = 2n$. 
	In this case, for every $B \in \NN$, the $(G_1, B)$-amplifiers is the same.
	The code in Figure~\ref{fig:F1-amplifier} describes the $(G_1, B)$-amplifier with input counters $\var{x_{in}}, \var{y_{in}}, \var{z_{in}}$ and output counters $\var{x_{out}}, \var{y_{out}}, \var{z_{out}}$.
	\begin{figure}[ht!]
		\begin{algorithmic}[1]
			\Require \assert{x_{in}}{B}, \assert{y_{in}}{C}, \assert{z_{in}}{BC}, \;\; \assert{x_{out}}{0}, \assert{y_{out}}{0}, \assert{z_{out}}{0}
			\State \LOOP \dec{x_{in}}{1}, \inc{x_{out}}{1}
			\State \LOOP \dec{y_{in}}{1}, \inc{y_{out}}{2}
			\State \LOOP \dec{z_{in}}{1}, \inc{z_{out}}{2}
			\Ensure \assert{x_{in}}{0}, \assert{y_{in}}{0}, \assert{z_{in}}{0}
		\end{algorithmic}
		\caption{The $(G_1, B)$-amplifier}
		\label{fig:F1-amplifier}
	\end{figure}
	It is easy to notice that the only possible values at the end are $\var{x_{out}} = B$, $\var{y_{out}} = 2C$ and $\var{z_{out}} = 2BC$. 
	By applying~\cref{lem:fixed-updates} to the $(G_1, B)$-amplifier, given that it has six counters and constant updates in $\set{-2, -1, 0, 1, 2}$, there is a unitary $12$-SLPS of size $\Oh(B)$ that is a $(G_1, B)$-amplifier.
	By Lemma~\ref{lem:fixed-updates} there is a unitary $12$-SLPS, which is an $(G_1, B)$-amplifier, as required for $d=1$. 

	For an induction step, let us assume that for all $B \in \NN$ there exist unitary $(4d+8)$-SLPSs that are $(G_d, B)$-amplifiers $\Vv_d^B$.
	We will proceed to construct a unitary $(4(d+1)+8)$-SLPS that is a $(G_{d+1}, B)$-amplifier $\Vv_{d+1}^B$, which is required in the statement of Claim~\ref{clm:amplifier}.
	For convenience, we write $\Vv_d^B[\var{x_{in}}, \var{y_{in}}, \var{z_{in}}, \var{x_{out}}, \var{y_{out}}, \var{z_{out}}]$ when we want to emphasise that $\var{x_{in}}, \var{y_{in}}, \var{z_{in}}$ are the input counters and $\var{x_{out}}, \var{y_{out}}, \var{z_{out}}$ are the output counters of the amplifier $\Vv_d^B$, respectively.
	
	We first construct a unitary SLPS with zero tests $\Pp_{d+1}^B$ that is a $(G_{d+1}, B)$-amplifier, which we present as a counter program in~\Cref{fig:higher-amplifier}.
	The input counters are $\var{x_{in}}, \var{y_{in}}, \var{z_{in}}$ and the output counters are $\var{x_{out}}, \var{y_{out}}, \var{z_{out}}$.
	We then use~\cref{lem:unitary-zero-tests} show that one can indeed obtain the required unitary SLPS $\Vv_{d+1}^B$ that is a $(G_{d+1}, B)$-amplifier.
	
	\begin{figure}[ht!]
		\begin{algorithmic}[1]
			\Require \assert{x_{in}}{A}, \assert{y_{in}}{B}, \assert{z_{in}}{A\cdot B}, \;\; \assert{x_{out}}{0}, \assert{y_{out}}{0}, \assert{z_{out}}{0}, \;\; \assert{x}{0}, \assert{y}{0}, \assert{z}{0}
			\State \inc{x_{out}}{1}
			\State \LOOP \inc{y_{out}}{1}, \inc{z_{out}}{1}
			\Ffor{i}{1}{B/12}
				\State $\Vv_{d}^{B'}[\var{x_{out}}, \var{y_{out}}, \var{y_{out}}, \var{x}, \var{y}, \var{z}] $	
				\Comment{Here, $B' = G_d^{(i-1)}(1)$.\hspace{1in}\textcolor{white}{}}
				\State \zt{x_{out}}
				\State \zt{y_{out}}
				\State \zt{z_{out}}
				\State \LOOP \dec{x}{1}, \inc{x_{out}}{1}
				\State \LOOP \dec{y}{1}, \inc{y_{out}}{1}
				\State \LOOP \dec{z}{1}, \inc{z_{out}}{1}	
				\State \zt{x}
				\State \zt{y}
				\State \zt{z}
			\EndFfor
			\Ensure \assert{x_{in}}{0}, \assert{y_{in}}{0}, \assert{z_{in}}{0}
		\end{algorithmic}
		\caption{A counter program is the unitary SLPS with zero tests $\Pp_{d+1}^B$ that is a $(G_{d+1}, B)$-amplifier.}
		\label{fig:higher-amplifier}
	\end{figure}

	We will first analyse the unitary SLPS with zero tests $\Pp_{d+1}^B$ in Figure~\ref{fig:higher-amplifier} as if counters $\var{x_{in}}$, $\var{y_{in}}$, and $\var{z_{in}}$ were not present; these three counters will be used to simulate the zero test (via Lemma~\ref{lem:unitary-zero-tests}).

	By the induction assumption, we know that $\Vv_d^{B'}$ (used in Line 4) is a unitary $(4d+8)$-SLPS (for any B' $\in \Nn$). 
	Notice that on top of these $4d+8$ counters, $\Pp_{d+1}^B$ only has $\var{x_{in}}$, $\var{y_{in}}$, and $\var{z_{in}}$.
	First, we argue an upper  bound on the size of $\Pp_{d+1}^B$.
	Notice that its size is dominated by the size of amplifiers $\Vv_d^{B}$ which is $\Oh(G_d(B))$.
	Therefore\footnote{For a function $f: \NN \to \NN$ we denote its $i$-th composition by $f^{(i)}$. Precisely, for all $k \in \NN$ we have $f^{(1)}(k) \coloneqq f(k)$ and for all $i > 1$ we have $f^{(i)}(k) \coloneqq f^{(i-1)}(f(k))$.}, the size in the $i$-th iteration inside the \texttt{FOR} loop is $G_d \circ G_d^{(i-1)}(1) = G_d^{(i)}(1) = G_{d+1}(12i)$.
	Observe that $G_{d+1}(n) \geq n \cdot G_{d+1}(n-1)$.
	Therefore, the sum of all the sizes of amplifiers (across all iterations of the \texttt{FOR} loop) is equal to $G_{d+1}(12) + G_{d+1}(24) + \ldots + G_{d+1}(12 \cdot B/12)$ and that is bounded by $2 \cdot G_{d+1}(B)$.

	Next, we focus on eliminating the zero tests from $\Pp_{d+1}^B$ to obtain the unitary SLPS $\Vv_{d+1}^B$.
	Observe first that in each iteration of the loop, six zero tests are performed; since \texttt{FOR} loop is iterated $B/12$ there are $B/2$ zero tests in total.
	Let $A$ be some number such that the sum of all the counters in $\Vv_{d}^B$ is bounded above by $A$. 
	Now we will use the triplet of counters $\var{x_{in}}$, $\var{y_{in}}$, and $\var{z_{in}}$.
	By Lemma~\ref{lem:unitary-zero-tests}, with the three counter values starting with initial values 
	$\var{x_{in}} = A$, $\var{y_{in}} = 2 \cdot B/2$, and $\var{z_{in}} = 2A \cdot B/2$, one can obtain
	Notice that every transition of $\Pp_{d+1}^B$ has an aggregate effect of absolute value at most two.
	Indeed, this is true by inspecting all of (explicitly written) transitions in Lines 1, 2, 8, 9, and 10 of~\Cref{fig:higher-amplifier}, and it will remain true inductively for $\Vv_d^{B'}$ (in Line 4). 
	Therefore the dimension of $V_{d+1}^B$ equals $(4d + 8) + (3-1) + 2 = (4d + 12) = (4(d+1)+8)$ as required.
	By Lemma~\ref{lem:unitary-zero-tests}, the size of $\Vv_{d+1}^B$ is bounded by the size of $\Oh(\norm{\Pp_{d+1}^B} + d \cdot B/2)$, which definitely is still at most $\Oh(G_{d+1}(B))$.

	It remains, most critically, to show that $\Vv_{d+1}^B$ is indeed a $(G_{d+1}, B)$-amplifier. 
	Let us inspect values of counters $\var{x_{out}}$, $\var{y_{out}}$, and $\var{z_{out}}$ after the $i$-the iteration of the \texttt{FOR} loop, let us denote these values $x_i$, $y_i$, and $z_i$, respectively.
	Also, let $x_0$, $y_0$, and $z_0$ be the values of $\var{x_{out}}$, $\var{y_{out}}$, and $\var{z_{out}}$ before the first iteration of the \texttt{FOR} loop. 
	We have $x_0 = 1$, $y_0 = C_0$, and $z_0 = C_0$ for some $C_0 \in \NN$. 

	In the first iteration, we use a $(G_d, 1)$-amplifier. 
	Due to the zero tests on Lines 5, 6, and 7, we know that the amplifier outputs on counters with values $\var{x} = G_d(1)$, $\var{y} = C_1$, and $\var{z} = G_d(1) \cdot C_1$, for some $C_1 \in \NN$.
	Now, the following loops (Lines 7, 8, and 9) and following zero tests (Lines 11, 12, and 13) transfer these values back to counters $\var{x_{out}}$, $\var{y_{out}}$, and $\var{z_{out}}$.
	So $x_1 = G_d(1)$, $y_1 = C_1$, and  $z_1 = G_d(1) \cdot C_1$.
	In a similar fashion, one can easily show, by induction, that $x_i = G_d^{(i)}(1), y_i = C_i, z_i = G_d^{(i)}(1) \cdot C_i$, for some arbitrarily large $C_i \in \NN$.
	This equality instantiated for $i = B/12$ shows that the final values are 
	$\var{x_{out}} = G_d^{(B/12)}(1)$, $\var{y_{out}} = C_d$, and $\var{z_{out}} = G_d^{(B/12)}(1) \cdot C_d$, for some arbitrarily large $C_d \in \NN$.
	We can therefore conclude that $\Vv_{d+1}^B$ is a $(G_{d+1}, B)$-amplifier, as required.
\end{proof}

\subsection{Proof of~\cref{thm:unitary-hardness}}
\label{app:unitary-hardness}

	As mentioned in the beginning of~\cref{sec:hardness}, and just like for the proof of~\cref{thm:3-lps-hardness}, we obtain the lower bound via a reduction from 3-CNF SAT; the \class{NP} upper bounds is given by~\cref{thm:np}.

	We again extend the reduction presented in~\cref{lem:sat-encoding} by implementing the non-divisibility assertions in polynomial time.
	As presented in~\Cref{fig:unitary-non-div}, we can implement a non-divisibility assertion as a unitary 5-SLPS that uses $\Oh(p)$ many zero tests (see~\cref{clm:unitary-non-div} for its correctness), where $n$ is the number of variables in the SAT instance.
	Note that between non-divisibility assertions, the four ancillary counters $\var{y}$, $\var{a}$, $\var{b}$, and $\var{c}$ are untouched; their values before and after each assertion are all zero.
	Accordingly, we obtain a unitary 5-SLPS with zero tests $\Uu$ in which reachability from the initial state $s$ with counter values $\vec{0}^5$ to a target state $t$ with counter values $\vec{0}^5$ is equivalent to the satisfiability of $\phi$.

	Now, we combine~\cref{lem:unitary-zero-tests} and~\cref{clm:unitary-triple} to simulate the zero tests.
	First, observe that in any run from $\vec{0}^5$, the sum of the counters is always bounded by an exponential value $B \in \NN$.
	That is true because the primary counter $\var{x}$ of $\Uu$ need not exceed the product of the first $n$ primes, where $n$ is the number of variables in $\phi$.
	Since $n \leq \norm{\Uu} \leq k$, we know that $B$ is exponential in the size of $\Uu$.
	Furthermore, notice that $C$, the number of zero tests is bounded above by the $\norm{\Uu} \leq k$ because the zero tests all lie on transitions between self-loops.
	Observe, by inspecting~\Cref{fig:lps-sat} and~\Cref{fig:unitary-non-div}, the absolute value of the aggregate effect of all transitions is at most one.
	Thus, by~\cref{lem:unitary-zero-tests}, we know can construct in $\poly{k}$ time, a unitary 8-SLPS $\Vv$ such that $\run{s(\vec{0}^5)}{*}{t(\vec{0}^5})$ in $\Uu$ if and only if $\run{s'(\vec{0}^5, B, 2C, 2BC)}{*}{t'(\vec{0}^5, 0, 0, 0)}$ in $\Vv$.

	The final obstacle is how to initialise the triple $\var{b} = B$, $\var{c} = 2C$, and $\var{d} = 2BC$ where $B$ is an exponential value, and $C \leq k$ is a polynomial value.
	For this, we use~\cref{clm:unitary-triple} by setting $d = \inverseAckermann{k}$.
	Accordingly, we obtain a $\inverseAckermann{k}$-SLPS $\Tt$ of size $\Oh(A(\inverseAckermann{k})) = \Oh(k)$ in which, from an initial configuration with counter values $\vec{0}^{\inverseAckermann{k}}$ a  final configuration with counter values $(\vec{0}^{\inverseAckermann{k}-3}, x, 2k, 2xk)$ can be reached for some $x \geq B$. 

	So, to conclude, we concatenate $\Vv$ to the end of $\Tt$ to create a $(\inverseAckermann{k}+9)$-SLPS for which $\vec{0}$ to $\vec{0}$ reachability coincides with the satisfiability of $\phi$.
	We remarked that the last three counters in $\Tt$ are exactly the seven counters $\var{b}$, $\var{c}$, and $\var{d}$ in $\Vv$ that are used to simulate zero tests.
	Therefore, reachability in $\inverseAckermann{k}$-VASS is \class{NP}-hard.
\qed

\section{Missing Proofs of~\cref{sec:polynomial}}
\label{app:decomposition}
\subsection{Proof of~\cref{lem:checkrun}}
\label{app:checkrun}
%\begin{proof}
    Clearly, if $\run{\vec{s}}{\pi}{\vec{t}}$, then $\zrun{\vec{s}}{\pi}{\vec{t}}$ holds and all configurations in the run $\run{\vec{s}}{\pi}{\vec{t}}$ are nonnegative so $\vec{a}_0^\pi, \vec{b}_0^\pi, \ldots, \vec{a}_k^\pi, \vec{a}_b^\pi \geq \vec{0}$ holds.

    Therefore, we assume that $\zrun{\vec{s}}{\pi}{\vec{t}}$ and $\vec{a}_0^\pi, \vec{b}_0^\pi, \ldots, \vec{a}_k^\pi, \vec{a}_b^\pi \geq \vec{0}$ hold; we would like to prove that $\run{\vec{s}}{\pi}{\vec{t}}$.
    For every $1 \leq i \leq k$, since $\vec{a}_i^\pi, \vec{b}_i^\pi \geq \vec{0}$, we know that $\run{\vec{a}_{i}}{\alpha_i}{\vec{b}_i}$ is a run. 
    Therefore, it suffices to prove that, for every $1 \leq i \leq k$, $\run{\vec{b}_{i-1}}{\beta_i^{n_i}}{\vec{a}_i}$ is also a run. 
    Precisely, $\run{\vec{b}_{i-1}}{\beta_i^{n_i}}{\vec{a}_i} = (\vec{b}_{i-1} + j\cdot\beta_i)_{j=0}^{n_i}$.
    Given that, in this run, we only iterate the vector $\beta_i$, for each coordinate, the sequence of values observed is monotonic; the first counter is either non-decreasing or non-increasing and the second counter is either non-decreasing or non-increasing
    From this, we conclude that $\run{\vec{b}_{i-1}}{\beta_i^{n_i}}{\vec{a}_i}$ is a run if and only if $\vec{b}_{i-1}, \vec{a}_i \in \NN^2$, which we assumed.
    Therefore, as all of the subruns are themselves runs, we know that $\vec{s} = \run{\vec{a}_0}{\alpha_0 \beta_1^{n_1} \alpha_1 \cdots \alpha_{k-1} \beta_k^{n_k} \alpha_k }{\vec{b}_k} = \vec{t}$ is indeed a run.
\qed%\end{proof}

\subsection{Proof of~\cref{clm:zero-effect-shifting}}
\label{app:zero-effect-shifting}
%\begin{proof}
    For (1), given that $e_1 \cdot \beta_{i_1} + e_2 \cdot \beta_{i_2} + e_3 \cdot \beta_{i_3} = \vec{0}$, it must be true that $\eff{\pi} = \eff{\rho}$.
    It immediately follows that $\zrun{\vec{s}}{\pi}{\vec{t}}$ if and only if $\zrun{\vec{s}}{\rho}{\vec{t}}$.

    For (2), given condition (ii) of~\cref{def:shifting}, we know that $\rho(0,i_1-1] = \pi(0,i_1-1]$ and $\rho(i_3,k] = \pi(i_3, k]$.
    Now, consider the midpoints ($\vec{a}_0^\pi, \vec{b}_0^\pi, \ldots, \vec{a}_k^\pi, \vec{b}_k^\pi$ and $\vec{a}_0^\rho, \vec{b}_0^\rho, \ldots, \vec{a}_k^\rho, \vec{b}_k^\rho$) of the two $\ZZ$-runs $\zrun{\vec{s}}{\pi}{\vec{t}}$ and $\zrun{\vec{s}}{\rho}{\vec{t}}$, respectively.
    Given that $\rho(0,i_1-1] = \pi(0,i_1-1]$, we know that $\vec{a}_j^\pi = \vec{a}_j^\rho$ and $\vec{b}_j^\pi = \vec{b}_j^\rho$ for all $j \in \set{1, \ldots, i_1-1}$; given that $\rho(i_3,k] = \pi(i_3, k]$, we know that $\vec{a}_j^\pi = \vec{a}_j^\rho$ and $\vec{b}_j^\pi = \vec{b}_j^\rho$ for all $j \in \set{i_3, \ldots, k}$.
\qed%\end{proof}

\subsection{Proof of~\cref{clm:shifting-subclaim}}
\label{app:shifting-subclaim}
%\begin{proof}
    Given that $\rho(0,i-1] = \pi(0,i-1]$, we know that $\vec{a}_s^\pi = \vec{a}_s^\rho$ for all $s \in \set{0, 1, \ldots, i-1}$.
    Thus it suffices to analyse the midpoints $\vec{a}_i^\rho, \vec{b}_i^\rho, \ldots, \vec{a}_{j-1}^\rho, \vec{a}_{j-1}^\rho$.

    If $\beta_i$ is zero on a coordinate, then, for each of the midpoints, the value of that coordinate is unchanged by the shift.
    In other words and more precisely, if there exists $\iota \in \set{1, 2}$ such that $\beta_i[\iota] = 0$, then $\vec{a}^\rho_s[\iota] = \vec{a}^\pi_s[\iota]$ and $\vec{b}^\rho_s[\iota] = \vec{b}^\pi_s[\iota]$, for all $i \leq s < j$.
    Since $\run{\vec{s}}{\pi}{\vec{t}}$, we know that in this case, $\vec{a}^\rho_i[\iota], \vec{b}^\rho_i[\iota], \ldots, \vec{a}^\rho_{j-1}[\iota], \vec{b}^\rho_{j-1}[\iota] \geq 0$.

    We will now check that the value of other coordinates of all the midpoints are nonnegative.
    On the remaining coordinate $\kappa = 3-\iota$, or both coordinates if both coordinates of $\beta_i$ are nonzero, we know that the value decreases based on the effect of the additional $\beta_i^x$ and subsequent subpath $\alpha_i \cdots \beta_{j-1}^{n_{j-1}} \alpha_{j-1}$.
    Since $\abs{x} \leq N$ and for all $i < s < j$, $n_s < N$, the maximal decrease on such a coordinate is bounded by 
    \begin{equation*}
        N(j-i)\cdot\size{\Vv} + N\cdot\size{\Vv} \leq Nk\cdot\size{\Vv} + N\cdot\size{\Vv} \leq N\cdot\size{\Vv}^2 + N\cdot\size{\Vv} \leq B.
    \end{equation*}
    Given that $\vec{a}^\pi_i[c'] \geq B$, we therefore conclude that $\vec{a}^\rho_i[c'], \vec{b}^\rho_i[c'], \ldots, \vec{a}^\rho_{j-1}[c'], \vec{b}^\rho_{j-1}[c'] \geq 0$.

    Together, regardless of whether $\vec{a}_i^\pi$ is far or, for some $\iota \in \set{1, 2},\, \beta_i[\iota] = 0$ and $\vec{a}_i[3-\iota] \geq B$, we know that for both coordinates $\kappa \in \set{1,2}$, $\vec{a}^\rho_i[\kappa], \vec{b}^\rho_i[\kappa], \ldots, \vec{a}^\rho_{j-1}[\kappa], \vec{b}^\rho_{j-1}[\kappa] \geq 0$.
    Therefore, $\vec{a}^\rho_i, \vec{b}^\rho_i, \ldots, \vec{a}^\rho_{j-1}, \vec{b}^\rho_{j-1} \geq (0,0)$.
    Hence, following $\rho(0,j-1]$ from the initial configuration $\vec{s}$ yields a run.
    This is equivalent to observing that $\run{\vec{s}}{\rho(0,j-1]}{\vec{v}}$ for some $\vec{v} \geq \vec{0}$.
\qed%\end{proof}

\subsection{Proof of~\cref{clm:case1-shifting}}
\label{app:case1-shifting}
%\begin{claimproof}
	First, by only considering the prefix of the shifted path $\rho$ before the cycle $\beta_b$, we know that $\run{\vec{s}}{\rho(0,b-1]}{\vec{b}^\rho_{b-1}}$ is indeed a run via~\cref{clm:shifting-subclaim} (by setting $i = a$, $j = b$, and $x = x_1$).

	Second, by~\Cref{eq:case1} and~\cref{clm:zero-effect-shifting}, we also know that the midpoints observed after the shift are unchanged (as well as before).
	Therefore, $\vec{a}_b^\rho = \vec{a}_b^\pi$.
	Since $\vec{a}_b^\pi \geq \vec{0}$, we know that $\run{\vec{b}^\rho_{b-1}}{\beta_b^{n_b-x_2}}{\vec{a}_b^\rho}$ is a run.
	We also know that the suffix of the paths after $\beta_b$ are unchanged; so $\rho(b,k] = \pi(b,k]$.
	Accordingly, we deduce that $\run{\vec{a}_b^\rho}{\rho(b,k]}{\vec{t}}$.

	Altogether,
	\begin{equation*}
	    \vec{s} \xrightarrow{\rho(0,b-1]} \vec{b}^\rho_{b-1} \xrightarrow{\beta_b^{n_b-x_2}} \vec{a}_b^\rho \xrightarrow{\rho(b,k]} \vec{t}.
	    \vspace{-0.25in}
	\end{equation*}
\qed
%\end{claimproof}

\subsection{Proof of~\cref{clm:case2-later-cycle}}
\label{app:case2-later-cycle}
%\begin{claimproof}
    Suppose for the sake of contradiction, that there does not exist $c \in I_{\pi[i,j)}$ such that $c > b$.
    This argument is very similar to the argument proving the existence of $b$.
    In this case, we know that $I_{\pi[i,j)} \cap \set{a+1, \ldots, j} = \set{b}$.
    Therefore, for all $s \in \set{a+1, \ldots, j}\setminus\set{b}$, $n_s < N$.
    So, we deduce that
    \begin{equation*}
        \sum_{s \in \set{a+1, \ldots, j}\setminus\set{b}} n_s 
            < \sum_{s \in \set{a+1, \ldots, j}\setminus\set{b}} N 
            < N(j-a) \leq Nk \leq N\cdot\size{\Vv}.
    \end{equation*}
    This implies that $\pi[i,j)$ is essential.
    However, this contradicts an opening assumption that $\pi[i,j)$ is not essential.
    Therefore, there exists $c \in I_{\pi[i,j)}$ such that $c > b$.
\qed%\end{claimproof}

\subsection{Proof of~\cref{clm:case2run}}
\label{app:case2run}
%\begin{claimproof}
    First, by only considering the prefix of the shifted path $\rho$ before the cycle $\beta_b$, we know that $\run{\vec{s}}{\rho(0, b-1]}{\vec{b}^\rho_{b-1}}$ is a run via~\cref{clm:shifting-subclaim} (by setting $i = a$, $j = b$, and $x = x_1$).

    Next, we can lower bound $\vec{a}_b^\rho$ based on the lower bound given in this case $\vec{a}^\pi_b \geq (B,B)$.
    The only difference between $\pi$ and $\rho$ until the midpoints $\vec{a}_b^\pi$ and $\vec{a}_b^\rho$ are reached is that $\rho$ has $x_1$ iterations of $\beta_a$ and $x_2$ iterations of $\beta_b$ added.
    Recall that $\abs{x_1}, \abs{x_2} \leq N$.
    Therefore, $\vec{a}_b^\rho - \vec{a}^\pi_b \geq (- N(\beta_b+\beta_a), -N(\beta_b+\beta_a) \geq (-N\cdot\size{\Vv}, -N\cdot\size{\Vv})$.
    Since $\vec{a}^\pi_b \geq (B,B)$, we deduce that $\vec{a}^\rho_b \geq (N\cdot\size{\Vv}^2 + N\cdot\size{\Vv}, N\cdot\size{\Vv}^2 + N\cdot\size{\Vv})$.
    Note that since $\vec{a}^\rho_b \geq \vec{0}$, $\run{\vec{b}^\rho_{b-1}}{\beta_b^{n_b+x_2}}{\vec{a}^\rho_b}$ is also a run.

    We can now also use this lower bound on $\vec{a}^\rho_b$ to argue that the next part of the run, that is $\run{\vec{a}^\rho_b}{\rho(b,c-1]}{\vec{b}^\rho_{c-1}}$, is indeed a run.
    Recall that $\rho(b,c-1] = \alpha_b\,\beta_{b+1}^{n_{b+1}}\,\cdots\,\beta_{c-1}^{n_{c-1}}\,\alpha_{c-1}$ and, for each $b < s < c$, $n_s < N$ (which means that $n_s \leq N-1$).
    For each coordinate $\iota \in \set{1,2}$, 
    \begin{align*}
        \eff{\rho(b,c-1]}[\iota] 
        & = \eff{\alpha_b\,\alpha_{b+1}\,\cdots\alpha_{c-1}}[\iota] + \eff{\beta_{b+1}^{n_{b+1}} \, \cdots \, \beta_{c-1}^{n_{c-1}}}[\iota] \\
        & \geq -\size{\Vv} - (N-1)(\eff{\beta_{b+1}}[\iota] + \cdots + \eff{\beta_{c-1}}[\iota]) \\
        & \geq -\size{\Vv} - (N-1)\cdot\size{\Vv} \\
        & = -N\cdot\size{\Vv}.
    \end{align*}
    Now, since $\vec{a}^\rho_b \geq (N\cdot\size{\Vv}^2 + N\cdot\size{\Vv}, N\cdot\size{\Vv}^2 + N\cdot\size{\Vv})$, we know that all of the following midpoints are nonnegative $\vec{a}^\rho_b, \vec{b}^\rho_b, \ldots, \vec{a}^\rho_{c-1}, \vec{b}^\rho_{c-1} \geq (N\cdot\size{\Vv}^2, N\cdot\size{\Vv}^2) \geq \vec{0}$.
    Thus $\run{\vec{a}^\rho_b}{\rho(b,c-1]}{\vec{b}^\rho_{c-1}}$ is indeed a run.

    Now, given~\Cref{eq:case2} and by~\cref{clm:zero-effect-shifting}, we know that $\vec{a}^\rho_c = \vec{a}^\pi_c$.
    Since $\vec{a}^\pi_c \geq \vec{0}$, we know that $\run{\vec{b}^\rho_{c-1}}{\beta_c^{n_c-x_3}}{\vec{a}^\rho_c}$ is a run.
    We also know that suffix of the paths after $\beta_c$ are unchanged; so $\rho(c,k] = \pi(c,k]$ are the same, so $\run{\vec{a}^\rho_c}{\rho(c,k]}{\vec{t}}$ is a run as well.

    Altogether, 
    \begin{equation*}
        \vec{s} \xrightarrow{\rho(0,b-1]} 
        \vec{b}^\rho_{b-1} \xrightarrow{\beta_b^{n_b+x_2}}
        \vec{a}^\rho_b \xrightarrow{\rho(b,c-1]}
        \vec{b}^\rho_{c-1} \xrightarrow{\beta_c^{n_c-x_3}}
        \vec{a}^\rho_c \xrightarrow{\rho(c,k]} \vec{t}.
        \vspace{-0.25in}
    \end{equation*}
\qed
%\end{claimproof}

\subsection{Proof of~\cref{clm:case3-later-cycle}}
\label{app:case3-later-cycle}
%\begin{claimproof}
    First, recall that, for all $s \in \{a+1, \ldots, b-1\}$, $n_s < N$.
    Given the opening assumption that $\pi[i,j)$ is not essential, we can deduce the following lower bound on $\sum_{s \in \set{b+1, \ldots, j}} n_s$.
    \begin{align}\label{eq:bj-bound}
        \sum_{s \in \set{a+1, \ldots, j}\setminus\set{b}} n_s 
        & > B + 2N\cdot\size{\Vv}^2 + N\cdot\size{\Vv} + \size{\Vv} \nonumber\\
        \implies \sum_{s \in \set{b+1, \ldots, j}} n_s 
        & > B + 2N\cdot\size{\Vv}^2 + N\cdot\size{\Vv} + \size{\Vv} - \sum_{s \in \set{a+1, \ldots, b-1}} n_s \nonumber\\
        & > B + 2N\cdot\size{\Vv}^2 + N\cdot\size{\Vv} + \size{\Vv} - N\cdot(b-a-1) \nonumber\\
        & \geq B + 2N\cdot\size{\Vv}^2 + \size{\Vv}
    \end{align}

    Now, let $C_{<0} = \set{s \in \set{b+1, \ldots, j} : \beta_s[1] < 0}$ and $C_{\geq0} = \set{s \in \set{b+1, \ldots, j} : \beta_s[1] \geq 0}$.
    We will now argue that $\sum_{s \in C_{\geq0}} n_s \geq N\cdot\size{\Vv}$.
    Assume, for sake of contradiction, that $\sum_{s \in C_{\geq0}} n_s < N\cdot\size{\Vv}$.
    Given~\Cref{eq:bj-bound} and the fact that $C_{<0}$ and $C_{\geq0}$ partition $\set{b+1, \ldots, j}$, we deduce that $\sum_{s \in C_{<0}} n_s > B + N\cdot\size{\Vv}^2 + \size{\Vv}$.
    We can therefore obtain the following upper bound on $\eff{\alpha_b\,\beta_{b+1}^{n_{b+1}}\,\cdots\,\alpha_{j-1}\,\beta_j^{n_j}}[1]$.
    \begin{align*}
        \eff{\alpha_b\,\beta_{b+1}^{n_{b+1}}\,\cdots\,\alpha_{j-1}\,\beta_j^{n_j}}[1]
        & = \eff{\alpha_{b+1}\,\cdots\,\alpha_{j-1}}[1] + \eff{\beta_{b+1}^{n_{b+1}}\,\cdots\,\beta_j^{n_j}}[1] \\
        & \leq \size{\Vv} + (-1)\sum_{s\in C_{<0}} n_s + \size{\Vv}\sum_{s \in C_{\geq0}} n_s \\
        & < \size{\Vv} - B - N\cdot\size{\Vv}^2 - \size{\Vv} + N\cdot\size{\Vv}^2 \\
        & = - B
    \end{align*}

    So, we have deduced that $\eff{\alpha_b\,\beta_{b+1}^{n_{b+1}}\,\cdots\,\alpha_{j-1}\,\beta_j^{n_j}}[1] < -B$.
    Since $\vec{a}^\pi_b[1] < B$ and that
    \begin{equation*}
        \run{\vec{a}^\pi_b}{\alpha_b\,\beta_{b+1}^{n_{b+1}}\,\cdots\,\alpha_{j-1}\,\beta_j^{n_j}}{\vec{a}^\pi_j},
    \end{equation*}
    we would deduce that $\vec{a}^\pi_j[1] < 0$.
    However, this is contradictory to the fact that $\run{\vec{s}}{\pi}{\vec{t}}$ is a run.
    We therefore conclude that $\sum_{s \in C_{\geq0}} n_s \geq N\cdot\size{\Vv}$.

    Now, given that $\sum_{s \in C_{\geq0}} n_s \geq N\cdot\size{\Vv}$ and $\abs{C_{\geq0}} < k \leq \size{\Vv}$, we deduce, using the pigeonhole principle, that there exists $c \in C_{\geq0}$ such that $n_s \geq N$.
    In other words, there exists $c \in I_{\pi[i,j)}$ such that $c > b$ and $\beta_c[1] \geq 0$.
\qed%\end{claimproof}

\subsection{Proof of~\cref{clm:case3-shifting}}
\label{app:case3-shifting}
%\begin{claimproof}
    Given that $a$, $b$, $x_1$, and $x_2$ are defined identically in Case 2 and Case 3, we know that $\run{\vec{s}}{\rho(0,b)}{\vec{a}^\rho_b}$ is a run.
    See~\cref{clm:case2run} for the details.
    Furthermore, given~\Cref{eq:case3}, and that $\rho$ is obtained by shifting $\pi$ at $a,b,c$ by $x_1,x_2,x_3$, we know that $\run{\vec{a}^\rho_c}{\rho(c,t)}{\vec{t}}$ is a run as well.
    
    Therefore, in this case, it suffices to argue that $\run{\vec{a}^\rho_b}{\rho(b,c)}{\vec{a}^\rho_c}$ is indeed a run.
    We will split the analysis into two parts by arguing that the two counters remain nonnegative separately.

    For the first counter, recall that $\beta_c[1] \geq 0$ and $x_3 > 0$.
    Notice that $x_1\cdot\beta_a[1] + x_2\cdot\beta_b[1] = x_3\cdot\beta_c[1]$ (\Cref{eq:case3}); so $x_1\cdot\beta_a[1] + x_2\cdot\beta_b[1] \geq 0$.
    Thus, $\vec{a}^\rho_b[1] = \vec{a}^\pi_b[1] + x_1\cdot\beta_a[1] + x_2\cdot\beta_b[1] \geq \vec{a}^\pi_b[1]$.
    Thus, $\vec{a}^\rho_b[1] \geq \vec{a}^\pi_b[1]$, $\vec{b}^\rho_b[1] \geq \vec{b}^\pi_b[1]$, $\ldots$\,, $\vec{a}^\rho_{c-1}[1] \geq \vec{a}^\pi_{c-1}[1]$, $\vec{b}^\rho_{c-1}[1] \geq \vec{b}^\pi_{c-1}[1]$, and $\vec{a}^\rho_{c}[1] = \vec{a}^\pi_{c}[1]$.
    In other words, the value of the first counter in the run over $\rho(b,c)$ is at least as great as the value of the first counter in the run over $\pi(b,c)$.

    As for the second counter, we will first prove that $\vec{a}^\pi_{b+1}[2], \ldots, \vec{a}^\pi_{c-1}[2] \geq (N+1)\cdot\size{\Vv}$ before using these lower bounds to argue that the second counter remains nonnegative over $\run{\vec{a}^\rho_b}{\rho(b,c)}{\vec{a}^\rho_c}$.
    Suppose, for sake of contradiction, that there exists $d \in \set{b+1, \ldots, c-1}$ such that $\vec{a}^\pi_d[2] < (N+1)\cdot\size{\Vv}$.
    Recall that $\vec{a}^\pi_b[2] \geq C = \size{\Vv}^2\cdot(B + (2N+3)\cdot\size{\Vv})$ and that $\run{\vec{a}^\pi_b}{\pi(b,d)}{\vec{a}^\pi_d}$, we therefore can deduce that
    \begin{align*}
        \eff{\pi(b,d)}[2] 
        & = \eff{\alpha_b\,\beta_{b+1}^{n_{b+1}}\,\cdots\,\alpha_{d-1}\,\beta_d^{n_d}}[2] \\
        & = \vec{a}^\pi_d[2] - \vec{a}^\pi_b[2] \\
        & < (N+1)\cdot\size{\Vv} - \size{\Vv}^2\cdot(B + (2N+3)\cdot\size{\Vv}) \\
        & \leq -\size{\Vv}^2\cdot(B + (N+2)\cdot\size{\Vv})
    \end{align*}
    \begin{align*}
        \implies \eff{\beta_{b+1}^{n_{b+1}} \, \cdots \, \beta_d^{n_d}}[2] 
        & < -\size{\Vv}^2\cdot(B + (N+2)\cdot\size{\Vv}) - \eff{\alpha_b \, \cdots \, \alpha_{d-1}}[2] \\
        & \leq -\size{\Vv}^2\cdot(B + (N+2)\cdot\size{\Vv}) + \size{\Vv} \\
        & \leq  -\size{\Vv}^2\cdot(B + (N+1)\cdot\size{\Vv})
        .
    \end{align*}
    Observe that, for all $s \in \set{b+1, \ldots, d}$, $\eff{\beta_s}[2] \leq \size{\Vv}$ and $d-b \leq k \leq \size{\Vv}$. 
    Thus, by the pigeonhole principle, there exists $s \in \set{b+1, \ldots, d}$ such that $n_s > B + (N+1)\cdot\size{\Vv}$.

    Now, we will analyse the effect of $\pi(b,s)$ on the first counter.
    Since $n_s > B + (N+1)\cdot\size{\Vv} > N$, we know that $s \in I_{\pi[i,j)}$.
    Furthermore, given that $s < c$, we know that $\eff{\beta_s}[1] < 0$ (by the minimality of $c$).
    Therefore, $\eff{\beta_s^{n_s}}[1] \leq B + (N+1)\cdot\size{\Vv}$.
    We will now argue that $\vec{a}^\pi_s[1] < 0$. 
    Recall that $\vec{a}^\pi_b < B$, observe that 
    \begin{align*}
        \vec{a}^\pi_s[1] 
        & = \vec{a}^\pi_b[1] + \eff{\pi(b,s)}[1] \\
        & = \vec{a}^\pi_b[1] + \eff{\alpha_b \, \beta_{b+1}^{n_{b+1}} \, \cdots \, \alpha_{s-1} \, \beta_s^{n_s}}[1] \\
        & = \vec{a}^\pi_b[1] + \eff{\alpha_b \, \cdots \, \alpha_{s-1}}[1] + \eff{\beta_{b+1}^{n_{b+1}} \, \cdots \, \beta_s^{n_s}}[1] \\
        & < B + \size{\Vv} + \eff{\beta_{b+1}^{n_{b+1}} \, \cdots \, \beta_{s-1}^{n_{s-1}}}[1] + \eff{\beta_s^{n_s}}[1] \\
        & \leq B + \size{\Vv} + \eff{\beta_{b+1}^{n_{b+1}} \, \cdots \, \beta_{s-1}^{n_{s-1}}}[1] - (B + (N+1)\cdot\size{\Vv}).
    \end{align*}
    Now, again by the minimality of $c$, we know that for all $s \in \set{b+1, \ldots, s-1}$, that if $\beta_s[1] \geq 0$, then $n_s < N$; this means that $\eff{\beta_{b+1}^{n_{b+1}} \, \cdots \, \beta_{s-1}^{n_{s-1}}}[1] < N\cdot\size{\Vv}$.
    \begin{equation*}
        \vec{a}^\pi_s[1] < B + \size{\Vv} + N\cdot{\size{\Vv}} - (B + (N+1)\cdot\size{\Vv}) = 0
    \end{equation*}
    This creates a contradiction, since $\run{\vec{s}}{\pi}{\vec{t}}$ is a run.
    We therefore conclude that $\vec{a}^\pi_{b+1}[2], \ldots, \vec{a}^\pi_{c-1}[2] \geq (N+1)\cdot\size{\Vv}$.

    Now, for all $s \in \set{b+1, \ldots, c-1}$, 
    \begin{align*}
        \vec{a}^\rho_s[2]
        & = \vec{a}^\pi_s[2] + x_1\cdot\beta_a[2] + x_2\cdot\beta_b[2] \\
        & \geq (N+1)\cdot\size{\Vv} - N\cdot\size{\Vv} \\
        & = \size{\Vv}.
    \end{align*}
    Furthermore, for all $s \in \set{b+1, \ldots, c-1}$, given that $\run{\vec{a}^\rho_s}{\alpha_s}{\vec{b}^\rho_s}$, we know that $\vec{b}^\rho_s[2] = \vec{a}^\rho_s[2] + \eff{\alpha_s}[2] \geq \size{\Vv} - \size{\Vv} = 0$.
    Together, $\vec{a}^\rho_{b+1}[2], \vec{b}^\rho_{b+1}[2], \ldots, \vec{a}^\rho_{c-1}[2], \vec{b}^\rho_{c-1}[2]$; the second counter remains nonnegative.
    
    This concludes the proof, because we have verified that $\run{\vec{a}^\rho_b}{\rho(b,c)}{\vec{a}^\rho_c}$ is a run.
\qed%\end{claimproof}

\subsection{Proof of~\cref{lem:nobigpoints}}
\label{app:nobigpoints}
%\begin{proof}
    If $\pi[i,j)$ is essential, or $\pi$ is not $C$-safe over $[i,j)$, then $\rho = \pi$ immediately satisfies this lemma.
    Therefore, let us assume that $\pi[i,j)$ is not essential and that $\pi$ is $C$-safe over $[i,j)$.
    
    Then, via~\cref{lem:iteration}, we can obtain a path $\rho$ such that $\run{\vec{s}}{\rho}{\vec{t}}$, $\rho(0,i-1] = \pi(0,i-1]$, $\rho(j,k] = \pi(j,k]$, and $\rho_1[i,j) \prec \pi[i,j)$.
    Now, if $\rho[i,j)$ is essential, or $\rho$ is not $C$-safe over $[i,j)$, then $\rho$ satisfies this lemma.
    Therefore, let us assume that $\rho[i,j)$ is not essential and that $\rho$ is $C$-safe over $[i,j)$.
    Let us define $\rho_0 = \pi$ and $\rho_1 = \rho$.
    Then, by applying~\cref{lem:iteration} to $\rho_1$, we obtain a path $\rho_2$ such that $\run{\vec{s}}{\rho_2}{\vec{t}}$, $\rho_2(0,i-1] = \rho_1(0,i-1]$, $\rho_2(j,k] = \rho_1(j,k]$, and $\rho_2[i,j) \prec \rho_1[i,j)$.

    By repeating this argument over and over again, by assuming that each of the obtained paths does not contain essential subpaths and are $C$-safe over $[i,j)$, we can obtain a sequence of paths $(\rho_1, \rho_2, \ldots)$ such that, for all $s \geq 1$, $\run{\vec{s}}{\rho_s}{\vec{t}}$, $\rho_s(0,i-1] = \rho_{s-1}(0,i-1]$, $\rho_s(j,k] = \rho_{s-1}(j,k]$, and $\rho_s[i,j) \prec \rho_{s-1}[i,j)$.
    
    Given that $\prec$ is the reverse lexicographic order, like the lexicographic order, it is a well-order. 
    We therefore know that the sequence of paths obtained must terminate after some finite number $t \in \NN$ of steps $(\rho_1, \rho_2, \ldots, \rho_t)$.
    It is therefore the case that, for $\rho_t$, there does not exist a path $\rho_{t+1}$ such that $\run{\vec{s}}{\rho_{t+1}}{\vec{t}}$, $\rho_{t+1}(0,i-1] = \rho_{t}(0,i-1]$, $\rho_{t+1}(j,k] = \rho_t(j,k]$, and $\rho_{t+1}[i,j) \prec \rho_t[i,j)$.

    Finally, by applying~\cref{lem:iteration} to $\rho_t$, we must obtain a path $\rho'$ that satisfies conditions (a) or (b) instead.
    In other words, there exists a path $\rho'$ such that $\run{\vec{s}}{\rho}{\vec{t}}$, $\rho'(0,i-1] = \rho_t(0,i-1]$, $\rho'(j,k] = \rho_t(j,k]$ and either $\rho'[i,j)$ is essential, or $\rho'$ is not $C$-safe over $[i,j)$.
    We conclude this proof by observing that $\pi(0, i-1] = \rho_1(0,i-1] = \ldots = \rho_t(0,i-1] = \rho'(0,i-1]$ and that $\pi(j,k] = \rho_1(j,k] = \ldots = \rho_t(j,k] = \rho'(j,k]$.
\qed%\end{proof}

\subsection{Proof of~\cref{clm:progressive}}
\label{app:progressive}
%\begin{claimproof}
    Suppose, for sake of contradiction, that there exists consecutive indices $i, j \in I$ such that $\vec{a}^\pi_i$ and $\vec{a}^\pi_j$ are not progressive.
    Without loss of generality, suppose that both $\vec{a}^\pi_i$ and $\vec{a}^\pi_j$ are horizontal.
    This means that
    \begin{equation*}
        \vec{a}^\pi_i[1], \vec{a}^\pi_j[1] > C \text{\; and \;} \vec{a}^\pi_i[2], \vec{a}^\pi_j[2] < B.
    \end{equation*}
    Additionally, before proceeding to find a contradiction, we will use the following equation several times
    \begin{align}\label{eq:bj-setup}
        \vec{a}^\pi_j 
        & = \vec{a}^\pi_i + \eff{\alpha_i \, \beta_{i+1}^{n_{i+1}} \, \cdots \, \alpha_{j-1} \, \beta_j^{n_j} } \\
        & = \vec{a}^\pi_i + \eff{\alpha_i \, \cdots \, \alpha_{j-1}} + \eff{\beta_{i+1}^{n_{i+1}} \, \cdots  \, \beta_{j-1}^{n_{j-1}}} + \eff{ \beta_j^{n_j} }. \nonumber
    \end{align}

    We will now argue that regardless of whether $\beta_j[2] < 0$, $\beta_j[2] > 0$, or $\beta_j[2] = 0$, we will arrive at a contradiction.

    First we consider the scenario when $\beta_j[2] < 0$.
    Since $i$ and $j$ are consecutive indices in $I$, we know that, for all $s \in \set{ i+1, \ldots, j-1 }$, $n_s < N$.
    Thus, $\eff{\beta_{i+1}^{n_{i+1}} \, \cdots  \, \beta_{j-1}^{n_{j-1}}}[2] < N\cdot\size{\Vv}$; we also know that $\eff{\alpha_i \, \cdots \, \alpha_{j-1}}[2] < \size{\Vv}$.
    Combining these facts with~\Cref{eq:bj-setup}, we deduce that
    \begin{align*}
        \vec{a}^\pi_j[2] 
        & < \vec{a}^\pi_i[2] + \size{\Vv} + N\cdot\size{\Vv} + \eff{ \beta_j^{n_j} }[2] \\
        & < B + (N+1)\cdot\size{\Vv} - (B + (N+1)\cdot\size{\Vv}) \\
        & = 0.
    \end{align*}
    This contradicts the fact that $\run{\vec{s}}{\pi}{\vec{t}}$.

    Now, consider the scenario when $\beta_j[2] > 0$.
    This time, using~\Cref{eq:bj-setup}, we deduce that
    \begin{align*}
        \vec{a}^\pi_j[2] 
        & \geq \vec{a}^\pi_i[2] - \size{\Vv} - N\cdot\size{\Vv} + \eff{ \beta_j^{n_j} }[2] \\
        & \geq 0 - (N+1)\cdot\size{\Vv} + B + (N+1)\cdot\size{\Vv} \\
        & = B.
    \end{align*}
    However, this contradicts the assumption that $\vec{a}^\pi_j$ was horizontal.

    Now, suppose that $\beta_j[2] = 0$.
    Given that $\vec{a}^\pi_j[1] > C \geq B$, we determine that $\vec{a}^\pi_j$ is a shiftable midpoint.
    This cannot be the case, since $j \in I \sset\set{q+1, \ldots, a-1}$ implies that $j < a$.
    Recall that $a \in \set{q+1, q+2,  \ldots, k}$ is the minimal index such that $\vec{a}^\pi_a$ is a shiftable midpoint. 
    This concludes the proof.
\qed%\end{claimproof}

\subsection{Proof of~\cref{clm:mixed}}
\label{app:mixed}
%\begin{claimproof}
    For (1), suppose for sake of contradiction that there exists $j \in I$ such that $n_j \geq B + (N+1)\cdot{\size{\Vv}}$ and that $\beta_j$ is positive.
    Then, simply, because $\vec{b}^\pi_{j-1} \geq \vec{0}$ and 
    \begin{equation*}
        \run{\vec{b}^\pi_{j-1}}{\beta_j^{n_j}}{\vec{a}^\pi_j},
    \end{equation*}
    we deduce that, for $\iota \in \set{1,2}$, $\beta_j[\iota] > 0 \implies \vec{a}^\pi_j[\iota] \geq B + (N+1)\cdot{\size{\Vv}} \geq B$.
    This means that $\vec{a}^\pi_j$ is a shiftable midpoint.
    This cannot be the case, since $j \in I \sset\set{q+1, \ldots, a-1}$ implies that $j < a$.
    Recall that $a \in \set{q+1, q+2, \ldots, k}$ is the minimal index such that $\vec{a}^\pi_a$ is a shiftable midpoint. 

    For (2), suppose for sake of contradiction that there exists $i, j \in I$ such that $n_j \geq B + (N+1)\cdot{\size{\Vv}}$, $i < j$, and $\beta_j$ is negative.
    In fact, let $i \in I$ be the greatest index such that $i < j$; in other words, $i$ and $j$ are consecutive indices in $I$.
    This means that
    \begin{align}\label{eq:bj-setup-2}
        \vec{a}^\pi_j 
        & = \vec{a}^\pi_i + \eff{\alpha_i \, \beta_{i+1}^{n_{i+1}} \, \cdots \, \alpha_{j-1} \, \beta_j^{n_j} } \\
        & = \vec{a}^\pi_i + \eff{\alpha_i \, \cdots \, \alpha_{j-1}} + \eff{\beta_{i+1}^{n_{i+1}} \, \cdots  \, \beta_{j-1}^{n_{j-1}}} + \eff{ \beta_j^{n_j} }. \nonumber
    \end{align}

    By~\cref{clm:progressive}, the pair of vectors $\beta_i$ and $\beta_j$ are progressive.
    Without loss of generality, we shall assume that $\vec{a}_i$ is horizontal and $\vec{a}_j$ is vertical; in other words,
    \begin{equation*}
        \vec{a}^\pi_i[1], \vec{a}^\pi_j[2] > C \text{\; and \;} \vec{a}^\pi_i[2], \vec{a}^\pi_j[1] < B
    \end{equation*}

    Like in the proof of~\cref{clm:progressive}, we will argue that regardless of whether $\beta_j[2] < 0$, $\beta_j[2] > 0$, or $\beta_j[2] = 0$, we will arrive at a contradiction.

    First, suppose $\beta_j[2] < 0$.
    Then, using~\Cref{eq:bj-setup-2},
    \begin{align*}
        \vec{a}^\pi_j[2] 
        & < \vec{a}^\pi_i[2] + \size{\Vv} + N\cdot\size{\Vv} + \eff{\beta_j^{n_j}} \\
        & < \vec{a}^\pi_i[2] + (N+1)\cdot\size{\Vv} - (B + (N+1)\cdot\size{\Vv}) \\
        & = 0.
    \end{align*}
    This contradicts the fact that $\run{\vec{s}}{\pi}{\vec{t}}$.

    Now, suppose $\beta_j[2] > 0$.
    Using~\Cref{eq:bj-setup-2}, we deduce that
    \begin{align*}
        \vec{a}^\pi_j[2] 
        & \geq \vec{a}^\pi_i[2] - \size{\Vv} - N\cdot\size{\Vv} + \eff{\beta_j^{n_j}} \\
        & \geq 0 - (N+1)\cdot\size{\Vv} + (B + (N+1)\cdot\size{\Vv}) \\
        & = B.
    \end{align*}
    This contradicts the fact that $\vec{a}^\pi_j$ is horizontal.

    Lastly, to conclude the proof, if $\beta_j[2] = 0$, then
    \begin{align*}
        C < \vec{a}^\pi_j[2] 
        & \leq \vec{a}^\pi_i[2] + \size{\Vv} + N\cdot\size{\Vv} + \eff{\beta_j^{n_j}} \\
        & < B + (N+1)\cdot\size{\Vv} + 0 \\
        & < C.
    \end{align*}
\qed%\end{claimproof}

\section{Missing Proofs of~\cref{sec:algorithm}}
\label{app:algorithm}
\subsection{Proof of~\cref{lem:close-reachability}}
\label{app:close-reachability}
%\begin{proof}[Proof of~\cref{lem:close-reachability}]
    We will split this proof into three cases depending on whether $\vec{s}[1]$ and $\vec{s}[2]$ are bounded above or below by $Q+P$, where $Q$ is a value that is polynomially in $\size{\Vv}$.
    For clarity, one should think that $Q$ is a large (yet still polynomial) value in $\size{\Vv}$, we will remark on a bound for $Q$ at the end of this proof.

    \paragraph*{Case 1.} When $\vec{s}[1], \vec{s}[2] < P+Q$.

    Roughly speaking, in this case, since both the counter values in initial and target configurations are polynomially bounded, we can use the non-deterministic log-space (\class{NL} $\sset$ \class{P}) algorithm for reachability in unary 2-VASS~\cite{EnglertLT16} to decide, in polynomial time, whether $\run{\vec{s}}{*}{\vec{t}}$.
    See~\cref{claim:ranko}.

    \paragraph*{Case 2.} When $\vec{s}[1], \vec{s}[2] \geq P+Q$.
    
    Intuitively, since both counters start at `large' values, we know that the target values of both counters are also `large'.
    Loosely speaking, we can therefore can treat this instance of reachability as an instance of reachability in a unary encoded 2-SLPS with integer counters; here the nonnegativity condition is not imposed.
    
    In fact, we can encode the problem as an instance of Integer Linear Programming (ILP) with two-dimensional constraints. 
    Precisely, let $k$ be the number of self-loops and $\Delta$ be maximum of the absolute value of any update (to either counter). 
    Now, for every $i \in {1, \ldots, k}$, let $A_i \in \set{-\Delta, \ldots, \Delta}^2$ be the update of the $i$-th loop; note that $A_i = \beta_i$.
    Then, let $A \in
    \{-\Delta,\ldots,\Delta\}^{2 \times k}$ be the matrix consisting of these updates.
    Then, we encode the problem as the following instance of ILP:
    \begin{align}\label{eq:ilp-ew}
        \set{ \vec{x} \in \mathbb{N}^k : A\,\vec{x} = \vec{t'} },
    \end{align}
    where $\vec{t'} = \vec{t} - \vec{s}$. 
    The task is to decide if the above set is empty.

    Observe that the number of constraints is constant. 
    Eisenbrand and Weismantel show that such an instance of ILP can be solved in time $\poly{\Delta \cdot k \cdot \norm{\vec{t'}}}$~\cite[Theorem 2.1]{DBLP:journals/talg/EisenbrandW20}. 
    Since our 2-SLPS is encoded in unary, we know that $\Delta$ is linearly bounded, precisely $\Delta \leq \size{\Vv}$.
    Additionally, by the initial assumption, we know that $\norm{\vec{t'}} \leq P$. 
    Hence, this instance of ILP can be solved in polynomial time.
    
    Finally, we need to show that the answer to reachability corresponds to the solution to the above instance of ILP. 
    Eisenbrand and Weismantel also show that if there exists a solution to~\Cref{eq:ilp-ew}, then there also exists a solution $\vec{y} \in \mathbb{N}^k$ such that $\norm{\vec{y}} \leq \poly{k \cdot \Delta \cdot \norm{\vec{t'}}}$~\cite[Theorem 3.3]{DBLP:journals/talg/EisenbrandW20}.
    Note that the linear relaxation of~\Cref{eq:ilp-ew} that is used in \cite[Theorem 3.3]{DBLP:journals/talg/EisenbrandW20} has $\ell_1$-norm bounded by $\poly{\norm{\vec{t'}}_\infty}$ which is polynomial.
    From such a solution $\vec{y}$, we can construct a path $\pi = \epath{\vec{y}[1], \ldots, \vec{y}[k]}$ that witnesses reachability $\run{\vec{s}}{\pi}{\vec{t}}$.
    Since $\poly{k \cdot \Delta \cdot \norm{\vec{t'}}} \leq \poly{\size{\Vv}^2 \cdot P}$, we know that $\pi$ has length in $P + \size{\Vv}$.
    We note that, in particular, the $\run{\vec{s}}{\pi}{\vec{t}}$ is valid because the counters are decreased (at most) by a value that is at most polynomial and because, by the opening assumption of this case, $\vec{s}[1], \vec{s}[2] \geq P+Q$ for some $Q$ that is appropriately chosen to be large enough.
    
    \paragraph{Case 3.} Either $\vec{s}[1] \geq P+Q$ and $\vec{s}[2] < P+Q$ or $\vec{s}[2] < P+Q$ and $\vec{s}[2] \geq P+Q$. 
    
    Without loss of generality, assume that $\vec{s}[1] \geq P+Q$ and $\vec{s}[2] < P+Q$; the other scenario is symmetric.
    Since $\norm{\vec{t}-\vec{s}} \leq P$, we know that $\abs{\vec{t}[1] - \vec{s}[1]} \leq P$, therefore $\vec{t}[1] \geq Q$.
    Roughly speaking, since the first counter starts and ends with large values, we need not be concerned about this counter dropping below zero; however, since the second counter starts and ends with small values, we need to be concerned about the nonnegativity condition on this counter.
    Therefore, intuitively speaking, we can suppose that the first counter is not subject to the nonnegativity condition.

    Accordingly, we shall define a one-counter automaton over a fixed alphabet $\Sigma = \set{a,b}$ that, in some way, captures the behaviour of the unary 2-SLPS $\Vv$ when the first counter is not subject to the nonnegativity condition.
    We use the definition from~{\cite[Section 2]{AtigCHKSZ16}}, as later in the proof we will invoke a result on the representation of Parikh image of languages of one-counter automata~\cite{AtigCHKSZ16}. (Standard) formal definition of the semantics of these automata can be found in the same paper.
    
    Syntactically, a one-counter automaton is a quintuple $\Aa = (Q', \Sigma, \delta, q_0, F)$.
    The set~$Q'$ of states of our automaton will contain the states of the 2-SLPS (and some additional states, as will be described later when defining $\delta$).
    The alphabet is fixed to be $\Sigma = \set{a,b}$.
    The initial state $q_0$ is the same as the state of the initial configuration $\vec{s}$.
    The set of final states $F \sset Q'$ is the singleton set $F = \set{q_f}$ where $q_f$ is the same as the state of the target configuration $\vec{t}$.
    
    The transition function $\delta$ is defined using the transitions of the 2-SLPS as follows.
    Suppose there is a transition $(p,(x,y),q)$ in the 2-SLPS. We will split this transition into parts such that the counters only receive incremental ($+1$), decremental ($-1$), or zero updates; such a split will require additional intermediate states to be introduced into $Q'$.
    Furthermore, the second counter of the 2-SLPS will be mimicked by the one counter of $\Aa$ and the first counter of the 2-SLPS will be mimicked by the letters read by $\Aa$.
    Precisely, if a transition increments the first counter of $\Vv$, then there is an `$a$' on the corresponding transition in $\Aa$; if a transition decrements the first counter of $\Vv$, then there is a `$b$' on the corresponding transition in $\Aa$; and if a transition does not increment or decrement the first counter of $\Vv$, there will be two sequential transitions in $\Aa$ that read an `$a$' and a `$b$', respectively.
    To give a concrete example, the transition $(p, (5, -3), q)$ in the unary 2-SLPS $\Vv$ will become the sequence of transitions $(p, a, -1, p_2), (p_2, a, -1, p_3), (p_3, a, -1, p_4), (p_4, a, 0, p_5), (p_5, a, 0, q)$ in the one-counter automaton~$\Aa$.

    The idea for the second counter of $\Vv$ is that, since the second counter in $\Vv$ initially starts with value $\vec{s}[2] < P + Q$, and since $\norm{\vec{t}-\vec{s}} \leq P$, we know that the counter much end with a value $\vec{t}[2] < 2P + Q$.
    Since both $\vec{s}[2]$ and $\vec{t}[2]$ are polynomially bounded, it is therefore important that the nonnegativity condition is monitored, hence the counter in $\Aa$ mimics the second counter in $\Vv$.
    Since it is assumed that runs from $\Aa$ start from the initial configuration $q_0(0)$ and words are accepted by $\Aa$ if a final state in $F$ is reached with zero counter value (in other words, the final configuration must be $q_f(0)$), we need to make one additional modification to $\Aa$.
    We prepend, to the initial state, a sequence of transitions that increment the counter so that the starting counter value is $\vec{s}[2]$; we also append, to the final state, a sequence of transitions that decrement $\vec{t}[2]$ from the counter so that the final state must be reached with exactly $\vec{t}[2]$ on the counter.

    The idea for the first counter of $\Vv$ is that, since we are ignoring the nonnegativity condition of the first counter, we only care for the final counter value to differ from the initial counter value by precisely $\vec{t}[1] - \vec{s}[1] \in \ZZ$.
    Therefore, we only need to determine if $\Aa$ accepts a word such that the number of `$a$'s read minus the number of `$b$'s read is equal to $\vec{t}[1] - \vec{t}[2]$.
    For a set of words $W$ over an alphabet $\Sigma$, the \emph{Parikh image} $\parikh{W} \sset \NN^\Sigma$ is the set
        \begin{equation*}
        \parikh{W} \coloneqq
        \{ 
            \vec{v} \in \NN^\Sigma
            : 
            \text{some $w \in W$
                  contains, for each $\sigma \in \Sigma$, exactly $\vec{v}[\sigma]$ occurrences of $\sigma$}
        \}.
        \end{equation*}
    With this definition in hand, we need to check, in polynomial time, if there exists a vector $\vec{v} \in \parikh{\lang{\Aa}}$ such that $\vec{v}[a] - \vec{v}[b] = \vec{t}[1] - \vec{s}[1]$ and for which $\vec{v}[a], \vec{v}[b] \leq P+Q$.
    The bound $\vec{v}[a], \vec{v}[b] \leq P+Q$, that we require, is used to inform us that the length of the path $\pi$ in $\Aa$ such that $\run{q_0(\vec{s}[2])}{\pi}{q_f(\vec{t}[2]})$ is bounded above by $P+Q$.
    Now, since $\vec{s}[1] \geq P + Q$, we therefore know that the corresponding run in $\Vv$ that witnesses reachability from $\vec{s}$ to $\vec{t}$ remains nonnegative.

    The main result of~{\cite[Section 4]{AtigCHKSZ16}} gives an algorithm that constructs, for a given one-counter automaton over a fixed alphabet, a non-deterministic finite automaton (NFA) whose language has the same Parikh image.
    Instead of invoking this result, we invoke several lemmas that give us a semilinear representation of this Parikh image, for our automaton~$\Aa$ over~$\Sigma = \set{a,b}$.
    Using the terminology of that paper, $\Aa$ is a simple OCA.
    We omit the definitions of semilinear and linear sets; as $|\Sigma| = 2$, all vectors in these sets will be two-dimensional (think $\NN^2$).
    By~{\cite[Lemma 15]{AtigCHKSZ16}}, we obtain a representation of the semilinear set $\parikh{\lang{\Aa}}$ (``semilinear representation'') that has polynomial size.
    More precisely, the number of linear sets in the union, the number of basis vectors, and the number of period vectors, as well as all of the components in said vectors, are all polynomially bounded.
    By using~{\cite[Lemma 16]{AtigCHKSZ16}}, this description can be computed in polynomial time.
    Since we care to observe a vector $\vec{v} \in \parikh{\lang{\Aa}}$ such that $\vec{v}[a] - \vec{v}[b] = \vec{t}[1] - \vec{s}[1]$, we can define the (semi)linear set $R = \set{\vec{v} \in \NN^\Sigma : \vec{v}[a] - \vec{v}[b] = \vec{t}[1] - \vec{s}[1]}$.
    We next use the handy fact that the intersection of two semilinear sets with polynomial descriptions is itself semilinear and has a polynomial description too~{\cite[Theorem 6]{ChistikovH16}}.
    Importantly for our purposes, when we measure the description size here, the maximum component of each vector in the semilinear representation (in the notation of~\cite{ChistikovH16}: the \emph{norm}) is bounded from above by a polynomial.
    Therefore, if $\parikh{\lang{\Aa}}\cap R \neq \emptyset$, then there exists $\vec{v} \in\parikh{\lang{\Aa}}\cap R$ such that $\vec{v}$ is polynomially bounded (on both components $\vec{v}[a]$ and $\vec{v}[b]$).

    Now we know that if there exists $\vec{v} \in\parikh{\lang{\Aa}}$ such that $\vec{v}[a] - \vec{v}[b] = \vec{t}[1] - \vec{s}[1]$, then there exists such a $\vec{v}$ that is polynomially bounded and since $\Sigma$ is fixed, there are at most polynomially many such vectors $\vec{v}$.
    Finally, it remains to enumerate every possible polynomially bounded vector $\vec{v}$ such that $\vec{v}[a] - \vec{v}[b] = \vec{t}[1] - \vec{s}[1]$ and, by~{\cite[Lemma 17]{AtigCHKSZ16}}, determine whether $\vec{v} \in \parikh{\lang{\Aa}}$ in polynomial time.

    Finally, we remark that the integer~$Q$ depends on the polynomial bound on $\vec{v} \in \parikh{\lang{\Aa}}\cap R$ (if $\parikh{\lang{\Aa}}\cap R \neq \emptyset$).
    This bound itself depends on the size of semilinear description of $\parikh{\lang{\Aa}}$ (as discussed above, this is polynomial in the size of $\Aa$; recall that the size of $\Aa$ is itself polynomial in the size of $\Vv$); the size of the semilinear description of $R$; and the polynomial bounds obtained when intersecting two semilinear sets with polynomial-size descriptions.
\qed%\end{proof}

\subsection{Proof of~\cref{lem:anyrun}}
\label{app:anyrun}
%\begin{proof}
    This lemma follows from~\cref{pro:simplify}, if there is a run from $\vec{a}_i$ to $\vec{a}_j$, then there is a path $\pi$ that is a simplified reachability witness (\cref{def:simple-witness}) for  $\run{\vec{a}_i}{*}{\vec{a}_j}$.
    However, such a run does not guarantee that cycles $\beta_{i_1}$ and $\beta_{i_2}$ specifically are taken at least $M$ times; hence $\run{S'}{*}{T}$ (instead of $\run{S}{*}{T}$).
\qed%\end{proof}

\subsection{Proof of~\cref{clm:short}}
\label{app:short}
\begin{figure}[ht!]
        \centering
        \includegraphics[width=0.8\textwidth]{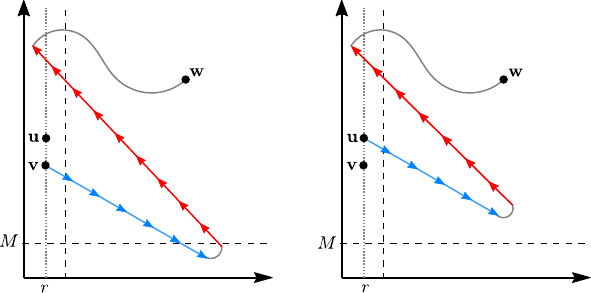}
        \caption{
            This figure accompanies the proof of~\cref{clm:short}. 
            Blue (\textcolor{blue}{$\boldsymbol\searrow$}) and red (\textcolor{red}{$\boldsymbol\nwarrow$}) cycles are taken large number of times. 
            In that case, we have two pairs of counter values $\vec{u} = (r, u)$ and $\vec{v} = (r, v)$; the goal is to show that if $\vec{w}$ is reachable from $\vec{v}$ then $\vec{w}$ is also reachable from $\vec{u}$.
            We know that the vector $\vec{u} - \vec{v} = (m \cdot h, 0)$, for some $m \in \NN$, can be represented as a combination of blue and red cycles. 
            Hence, given a run from $\vec{v}$ to $\vec{w}$, we can construct a run from $\vec{u}$ to $\vec{w}$ by removing this combination.
            Note that this operation only moves the run higher, so there is no concern about the second counter observing a sub-zero value.
            We do, however, have to be careful in some extreme cases since the
            grey segment of the run (between blue cycle iterations and red cycle
            iterations) may decrease the value of the first component. But this is easily dealt with since its length is at most $\polyflat$.
        }
        \label{fig:clm:short}
    \end{figure}

%\begin{claimproof}
    See~\Cref{fig:clm:short} for the sketch of the idea behind this proof.

    We begin by noting that the index $i$ in $(r,v)_i$ and $(r, u)_i$ is just to emphasise that this is the $i$-th midpoint; recall that $Z_r \sset Y_x \sset C_2$ and $C_2$ is a set of counter values obtained from configurations in $\Ss(i,i_1,i_2)$.

    Let $\vec{u}_i = (r, u)_i$ and $\vec{v}_i = (r, v)_i$.
    We denote $S_1 = (\vec{u}_i, \beta_{i_1}, \beta_{i_2})$ and $S_2 = (\vec{v}_i, \beta_{i_1}, \beta_{i_2})$.
    By assumption, we have $S_1, S_2 \in \Ss(i, \beta_{i_1}, \beta_{i_2})$.

    Let $T = (\vec{w}_j, \beta_{i_{3}}, \beta_{i_4}) \in \Ss$ for some $i < j$. We will prove that if $\run{S_2}{*}{T}$ then $\run{S_1'}{*}{T}$ for some $S_1' = (\vec{u}_i, \beta_{i_1'}, \beta_{i_2'})$. 
    Given that $\Ss$ is complete, we know that $S_1' \in \Ss$. 
    Therefore, and by~\cref{lem:anyrun}, it suffices to show that there is a run $\run{\vec{u}_i}{*}{\vec{w}_j}$.
    
    Given that $\run{S_2}{*}{T}$, there exists a path $\pi$ such that $\run{\vec{v}_i}{\pi}{\vec{w}_j}$, where $\pi = \epath{n_{i+1}, \ldots, n_j}$. 
    Recall that $u, v \in Z_r$, so there exists $m \in \NN$ such that $v = u + m \cdot h$. Also recall that $x_1\cdot\beta_{i_1} + x_2\cdot\beta_{i_2} = (h,0)$. 
    We define $\rho$ to be the path that is obtained by $\pi$ shifting at $i_1, i_2$ by $-x_1\cdot m, -x_2\cdot m$.
    We will prove that $\run{\vec{u}_i}{*}{\vec{w}_j}$. 
    Since $\eff{\rho} = \eff{\pi}$, we only need to check that the number of times each cycle is taken in $\rho$ is nonnegative and that no midpoint configuration drops below zero. 
    We start by showing that $n_{i_1} > M\cdot\size{\Vv}^2 + m \cdot x_1 \geq 0$ and $n_{i_2} > M\cdot\size{\Vv}^2 + m \cdot x_2 \geq 0$. 
    
    We will argue that $n_{i_1}, n_{i_2} \geq \frac{v-M}{\size{\Vv}}$.
    Indeed, recall that $\abs{\beta_{i_1}[1]}, \abs{\beta_{i_1}[2]}, \abs{\beta_{i_2}[1]}, \abs{\beta_{i_1}[2]} \leq \size{\Vv}$.
    Since $\beta_{i_1}$ and $\beta_{i_2}$ are the cycles that are used to change between $M$-slim midpoint configurations, we know that $n_{i_1} \cdot \size{\Vv} + M \geq v$ and $n_{i_2} \cdot \size{\Vv} + M \geq v$, which proves the inequalities.
    
    Next, recall that $h > 0$. Then,
    \begin{align*}
             m \leq m h = u - v \leq v\left(1 + \frac{1}{2\cdot\size{\Vv}^2}\right) - \frac{M}{2\cdot\size{\Vv}^2} - v = \frac{(v - M)}{2\cdot\size{\Vv}^2}.
    \end{align*}
    
    Also, recall that $2M \cdot \size{\Vv}^3 + M \leq v$, which is equivalent to the inequality $M \cdot \size{\Vv}^2 \leq \frac{v - M}{2\cdot\size{\Vv}}$. 
    Finally, recall that $x_1, x_2 \leq \size{\Vv}$; therefore,
    \begin{multline*}
        \min\set{n_{i_1}, n_{i_2}}
        \geq \frac{v - M}{\size{\Vv}} 
        = \frac{v - M}{2\cdot\size{\Vv}} + \frac{v - M}{2\cdot\size{\Vv}} \\
        \geq M\cdot\size{\Vv}^2 + \size{\Vv} \cdot \frac{v - M}{2\cdot\size{\Vv}^2}
        \geq M\cdot\size{\Vv}^2 + \max\set{x_1\cdot m, x_2\cdot m}.
    \end{multline*}

    It remains to show that no midpoint configurations $\vec{a}_s$, $\vec{b}_s$ drop below zero. 
    Note that because $\eff{\rho} + m(\eff{\beta_{i_1}^{x_1}} + \eff{\beta_{i_2}^{x_2}}) = \eff{\pi}$, the suffixes and prefixes of $\pi$ and $\rho$ are the same.
    Precisely, $\rho(0,i_1-1] = \pi(0,i_1-1]$ and $\rho(i_2, k) = \pi(i_2, k)$.
    So the only midpoint configurations that differ are between $i_1$ and $i_2$.
    Recall that $\beta_{i_1}$ is a mixed vector; there is a coordinate $\iota \in \set{1,2}$ such that $\beta_{i_1}[\iota] < 0$.
    Without loss of generality suppose that $\beta_{i_1}[1] < 0$ and $\beta_{i_1}[2] > 0$.
    On the first coordinate, the counter values are increased by the shift from $\pi$ to $\rho$.
    Therefore, it only remains to check the second counter.
    Given the definition of dynamic programming transition and that that $\pi$ is the path witnessing reachability of the configurations between the tuples $S_2$ and $T$, we know that $n_{i_1}+1, \ldots, n_{i_2}-1 < M$; this means that $n_{i_1}+1, \ldots, n_{i_2}-1 \leq M-1$.
    Therefore, we know that
    \begin{align*}
        \eff{\alpha_{i_1} \, \beta_{i_1+1}^{n_{i_1+1}} \, \cdots \, \alpha_{i_2-2} \, \beta_{i_2-1}^{n_{i_2-1}}}[2]
        & = \eff{\alpha_{i_1} \, \cdots \, \alpha_{i_2-2}}[2] + \eff{\beta_{i_1+1}^{n_{i_1+1}} \, \cdots \, \beta_{i_2-1}^{n_{i_2-1}}}[2] \\
        & \geq -\size{\Vv} - (M-1)\cdot\size{\Vv}^2 \\
        & \geq -M\cdot\size{\Vv}^2.
    \end{align*}
    Recall that $n_{i_1} \geq M\cdot\size{\Vv}^2 + x_1\cdot m$ and since $\beta_{i_1}[2] > 0$, so we know that (before the shift) $\vec{a}_{i_1}[2] \geq M\cdot\size{\Vv}^2 + x_1\cdot m$.
    Therefore, and so this coordinate cannot drop below zero after the shift.
%\end{claimproof}

\subsection{Proof of~\cref{clm:long}}
\label{app:long}
%\begin{claimproof}
    We need $P_1$ to be great enough such that for all $v \geq P_1$, we have $v \cdot (1 + \frac{1}{2\cdot\size{\Vv}^2}) - M\cdot\size{\Vv}^2 \geq P_1 (1 + \frac{1}{4\cdot\size{\Vv}^2})$; it is clear that $P_1$ is polynomial in $\size{\Vv}$. 
    Towards a contradiction, suppose there exist $v_1, v_2, \ldots, v_{P_2} \in Z_r$ such that $v_1 \geq P_1$ and, for all $i \in \set{1, \ldots, m_2-1}$, $v_i < v_{i+1}$.
    We get that $v_{P_2} \geq P_1 \cdot (1 + \frac{1}{4\cdot\size{\Vv}^2})^{P_2}$.
    
    Let $c$ be the constant bound from~\cref{rem:bounds}. 
    Recall that $(1 + \frac{1}{x})^x \ge 2$ for $x > 0$. 
    Thus for $P_2 \ge (4\cdot\size{\Vv}^2) \cdot c(\size{\Vv} + \bitsize{\vec{s}} + \bitsize{\vec{t}})^3$, we get $v_{P_2} \ge P_1 > 2^{c(n + \bitsize{\vec{s}} + \bitsize{\vec{t}})^3}$.
    This contradicts~\cref{rem:bounds}.
    Therefore, we know that $\abs{\set{v \in Z_r : v \geq P_1}} \leq P_2$, and since $Z_r \sset \set{0, 1, \ldots, P_1-1} \cup \set{v \in Z_r : v \geq P_1}$, we know that $\abs{Z_r} \leq P_1 + P_2$.
\qed%\end{claimproof}

\subsection{Proof of~\cref{lem:important-sets}}
\label{app:important-sets}
%\begin{proof}
    See~\Cref{fig:alg} on~\cpageref{fig:alg} for some intuition behind the proof. 

    Note that $\Ss_0 = \set{(\vec{a}_{0}, \beta_{i_1}, \beta_{i_2}) : 1 \leq i_1 < i_2 \leq k \text{ or } i_1 = i_2 = \infty}$ is trivially computable in polynomial time. 
    Suppose we have computed all sets $\Ss_0, \ldots \Ss_{i-1}$; we now want to compute $\Ss_i$.
    Let $S_i = (\vec{a}_i, \beta_{i_1}, \beta_{i_2}) \in \Ss_i'$ and let $S_j = (\vec{a}_j, \beta_{j_1}, \beta_{j_2}) \in \Ss_j$ be such that $\run{S_j}{}{S_i}$, for some $j < i$.
    Then, according to \cref{def:transition}, let $\pi = \epath{n_{j+1}, \ldots, n_i}$ be a path such that $\run{\vec{a}_j}{\pi}{\vec{a}_i}$. 
    We will consider two cases depending on whether $j < j_1 < j_2 \leq k$ or $j_1 = j_2 = \infty$.

    \paragraph*{Case 1.} When $j < j_1 < j_2 \leq k$, we know that $\beta_{j_1}, \beta_{j_2} \neq \emptyset$.

    We consider two scenarios, depending on whether there exists $c \in \QQ$ such that $\beta_{j_1} = c\cdot\beta_{j_2}$.
    If there does not exist $c \in \QQ$ such that $\beta_{j_1} = c\cdot\beta_{j_2}$, then the lemma follows directly from~\cref{lem:bound-on-configurations}. 
    
    Otherwise, if there does exist $c \in \QQ$ such that $\beta_{j_1} = c\cdot\beta_{j_2}$, then since $\vec{a}_{j_1}$ and $\vec{a}_{j_2}$ are $M$-slim and since the pair of cycles $\beta_{j_1}$ and $\beta_{j_2}$ are opposite, then it must be the case that $n_{j_1} \cdot \beta_{j_1} + n_{j_2} \cdot \beta_{j_2} \leq (2M, 2M)$. 
    Since, for all $s \in \set{j+1, \ldots, i}\setminus\set{j_1,j_2}$, $n_s < M$, we know that $\eff{\pi}[1],\eff{\pi}[2] \leq 2M + M\cdot\size{\Vv}^2$ on both coordinates. 
    It is, therefore, the case that we require to check whether $\run{\vec{a}_j}{*}{\vec{a}_i}$ for vectors $\vec{a}_j \in \set{\vec{a}_i + \vec{u} : \vec{u} \in [0,2M + M\cdot\size{\Vv}^2]\times[0,2M + M\cdot\size{\Vv}^2]}$.
    It is therefore true, that in this case, $\norm{\vec{a}_i - \vec{a}_j} \leq 2M + M\cdot\size{\Vv}^2$.
    Also, observe that there are at most $(2M+M\cdot\size{\Vv}^2 + 1)^2$ many such vectors $\vec{a}_j$.
    This means that we can apply~\cref{lem:close-reachability} at most polynomially many times.

    \paragraph*{Case 2.} When $j_1 = j_2 = \infty$, we know that $\beta_{j_1} = \beta_{j_2} = \emptyset$.

    Here we consider two scenarios, depending on whether $j \leq k$ or $j = k+1$. 
    If $j \le k$ then by~\cref{def:transition} only short runs are allowed. 
    Then, like in the second scenario of the previous case, there are only polynomially many configurations to consider and each of these configurations differs by a polynomial amount on each counter.
    This, again, can be verified in polynomial time using~\cref{lem:close-reachability}. 
    
    Now, suppose $j = k+1$. 
    Then there exists a $\pi = \epath{n_{i+1},\ldots,n_{k}}$ and three indices $i_1, i_2, i_3 \in \set{i+1, \ldots, k}$ such that $i_1 < i_2 < i_3$ and, for all $s \in \set{i+1, \ldots, k} \setminus \set{s_1,s_2, s_3}$, $n_s \leq M$.
    To verify a candidate path $\pi$, we need to consider all candidates for indices $s_1, s_2, s_3 \in \set{i+1, \ldots, k}$ for which there less than $k^3 \leq \size{\Vv}^3$ many choices. 
    The effect of $\pi$ (on both counters) in-between indices $i$, $i_1$, $i_2$, $i_3$, and $k$ is polynomially bounded; precisely, for all $\tau \in \set{\pi(i,i_1-1], \pi(i_1,i_2-1], \pi(i_2,i_3-1], \pi(i_3,k)}$, 
    \begin{equation*}
        (-M\cdot\size{\Vv}^2, -M\cdot\size{\Vv}^2) \leq \eff{\tau} \leq (M\cdot\size{\Vv}^2, M\cdot\size{\Vv}^2).
    \end{equation*}
    Then one needs to decide if there exists a run with such effects (between indices $i$, $i_1$, $i_2$, $i_3$, and $k$); since each of the effects of each path is polynomially bounded, there are only polynomial many combinations; so this verified polynomial time.

    Finally, we only need to verify whether there are $n_{i_1}$, $n_{i_2}$ and $n_{i_3}$ for which $\pi$ exists.
    Given that one can simply encode an instance of reachability in a (simple) linear path scheme as an instance of ILP that uses a constant number of variables for each cycle in the (S)LPS.
    Since only a constant number of cycles ($\beta_{i_1}$, $\beta_{i_2}$, and $\beta_{i_3}$) do not have their exponents fixed, the instance of ILP will only have a constant number of variables.
	Such an ILP can be solved in polynomial time by algorithms in, e.g.,~\cite{Lenstra83,Kannan83,ReisR23}.
    This concludes the second case.
%\end{proof} 
\end{appendix}

\end{document}